\documentclass[submission, Phys]{SciPost}
\usepackage{graphicx,epstopdf}              
\usepackage{mathtools}                        
\usepackage{amsfonts,amsthm,amssymb,latexsym,amsmath}
\usepackage{mathrsfs}
\usepackage{bbm}
\usepackage[capitalize]{cleveref}
\usepackage{bm}
\usepackage{units}  
\usepackage{braket}  
\usepackage{textcomp} 
\usepackage{hhline}
\usepackage{multirow}   
\usepackage{makecell}
\usepackage{tabularx}
\usepackage{hyperref}
\usepackage{tikz}
\usepackage{tikz-3dplot}
\usepackage{wasysym}
\usepackage{colortbl}
\usepackage{pgfplots}
\usepackage[normalem]{ulem}
\usepackage{subcaption}

\usepackage{framed}
\definecolor{shadecolor}{gray}{0.95}

\usetikzlibrary{external}
\usetikzlibrary{decorations.markings}
\usetikzlibrary{decorations.pathreplacing}
\usetikzlibrary{decorations.text}
\usetikzlibrary{arrows,calc,external,shapes.geometric}

\tikzstyle{A}=[circle,draw=red!50,fill=red!20,thick]
\tikzstyle{R}=[circle,draw=blue!50,fill=blue!20,thick]
\tikzstyle{U}=[rectangle,draw=green!50,fill=green!20,thick]
\tikzstyle{V}=[rectangle,draw=orange!50,fill=orange!20,thick]

\def\ZZ{\mathbb Z}

\definecolor{newblue}{RGB}{112,178,255}
\definecolor{neworange}{RGB}{255,204,112}
\definecolor{blue2}{RGB}{120,0,255}
\definecolor{red2}{RGB}{255,0,120}
\definecolor{green2}{RGB}{0,130,130}

\hypersetup{
colorlinks = true,
linkcolor=magenta,
citecolor=[rgb]{0.15,0.65,0.13},
urlcolor = [rgb]{0.25, 0.41, 0.88}}

\newcommand{\notA}{\textrm{$\bar A$}}
\newcommand{\AnotA}{\textrm{$A\bar A$}}

\makeatletter
\def\l@subsection#1#2{}
\def\l@subsubsection#1#2{}
\makeatother

\begin{document}

\begin{center}{\Large \textbf{Quasi-Phases of Quantum Matter} 
}\end{center}

\begin{center}
Nick G. Jones\textsuperscript{1}, Paul Fendley\textsuperscript{2}, Nathanan Tantivasadakarn\textsuperscript{3}, Ruben Verresen\textsuperscript{4}
\end{center}

\begin{center}\small
{\bf 1} St John’s College and Mathematical Institute, University of Oxford, UK\\
{\bf 2} All Souls College and Rudolf Peierls Centre for Theoretical Physics, University of Oxford, UK\\
{\bf 3} C. N. Yang Institute for Theoretical Physics, Stony Brook University, USA\\
{\bf 4} Pritzker School of Molecular Engineering, University of Chicago, USA
\end{center}

\begin{center}
\today
\end{center}

\section*{Abstract}
Within quantum many-body phase diagrams, continuous transition lines can appear and then terminate. However, it has remained unclear whether the neighboring regimes can be meaningfully described as distinct phases, akin to water and vapour, since any sharp characterization would have to go beyond the usual notion of phases of matter. We answer this in the affirmative by introducing quasi-phases: gapped regimes near a critical point that are connected in the broader phase diagram, but not by any gapped path accessible from small perturbations of the critical theory. The physical consequences include interface modes between regimes in the same phase of matter. More striking still, describing both the extended critical line and the gapped path connecting the regimes requires a bigger multicritical parent theory. In spin-chain examples with continuous terminating lines with central charge $c=1$, we find multicritical $c=\tfrac{3}{2}$ and $c=\tfrac{9}{5}$ orbifolds describing so-called `quotient symmetry protected' and `representation symmetry protected' topological phenomena. While one-dimensional symmetry-protected topological phases are labeled by projective representation classes, these quasi-phases can have interface modes carrying higher-dimensional linear representations. We turn this into a constructive principle through a weakening of the Lieb–Schultz–Mattis anomaly. Supporting distinct quasi-phases makes an unnecessary critical point \emph{infrared-necessary}, even if it is unnecessary at the ultraviolet scale. This condition is more restrictive than simply having a codimension-one critical line, revealing richer structure within phases of matter than their conventional classification suggests.

\tableofcontents

\section{Introduction \label{sec:intro}}
Much of the history of condensed matter physics is focused on identifying phases of matter and the transitions between them. In some cases, there are complete classifications, including the case of gapped phases of matter in 1+1D spin chains with an on-site symmetry $G$\cite{Fidkowski11,Turner11,Chen11,Schuch11,Chen13}. These are labeled by the unbroken symmetry acting on the ground state, and a projective symmetry class. Non-trivial projective classes correspond to symmetry-protected topological (SPT) phases \cite{Gu09,Senthil15}, with the projective representation protecting degenerate edge modes. If two symmetry-preserving Hamiltonians share the same class, then the Hamiltonians can be connected by a $G$-symmetric adiabatic path, while if it differs then they must be separated by a phase transition where the gap closes. 

\textbf{Unnecessary criticality.} $\;$ This programme has been so successful that generic quantum criticalities of codimension one (e.g.,\  extended lines in a two-dimensional phase diagram) that are \textit{not} implied by such a classification were dubbed \emph{unnecessary} \cite{BiSenthil19}. Such transitions can persist over an extended region but eventually terminate, providing continuous analogues of the liquid-vapour transition in water, and have been studied in Refs.~\cite{MoudgalyaPollmann15,Anfuso07,Jian20,Verresen21,He24,Prakash24,Senthil2024,Prakash25,Zhang25,Zhang26}. 
Such terminating critical surfaces of codimension one locally separate  neighboring gapped regions in the same phase of matter, even though the regions are not labeled by different global topological invariants. This raises the question of whether there is nevertheless some physical distinction between these two sides, similar to how liquid and vapour have clearly contrasting physical properties. The immediate challenge is whether such a question can be ascribed meaning: is there a sharp notion of how the neighboring gapped regions can look distinct, despite ultimately being in the same gapped phase? Indeed, a key motivation of the present work is to make this meaningful.

\textbf{SPT-like phenomena in the `trivial' phase.} $\;$ Coming from the opposite direction, examples have emerged of gapped phases with SPT-like properties despite corresponding to a trivial projective class. In particular, for these ground states, the classification tells us that there must exist a symmetric gapped path to a featureless product state, but nevertheless parametrically robust features have been observed (including edge modes and entanglement degeneracies). In fact, simple interpolations between such spin chains and a trivial Hamiltonian have given rise to continuous phase transitions, apparently providing natural examples of the aforementioned unnecessary criticality. However, to the best of our knowledge, there is currently no available general explanation for why this should be so, i.e., why such continuous phase transitions arise.

In special instances, one can explain the appearance of such 
criticality by a separation of energy scales \cite{Verresen21}. Even though the symmetry class associated to the gapped phase is a genuine linear representation (as acting on the edge modes, say), it may act projectively if we integrate out certain massive degrees of freedom. Hence, as long as certain symmetry sectors remain at much higher energy scales than others, two regimes of the same phase can look distinct, separated by a phase transition. Since the process of integrating out gapped symmetry sectors reduces the faithful symmetry action to a quotient group, the associated physics has been called \emph{quotient symmetry protected topological (QSPT) phenomena} \cite{Verresen21}. Even here we note that no precise definition of a \emph{QSPT phase} has been given; indeed it is not clear whether this can be made precise, explaining why the word `phenomena' rather than `phase' was used. In contrast, the \textit{transition} does have a robust characterization: the same quotienting mechanism leads to an emergent anomaly \cite{Metlitski18}, which can only be removed by either spontaneously breaking the symmetry or bringing down new gapless modes.

More mysterious are SPT-like phase diagrams where one cannot appeal to any such separation of energy scales. This already occurs when taking the smallest non-Abelian protecting group, $S_3$. In Ref.~\cite{OBrien20} an $S_3$-symmetric spin chain was introduced whose natural boundary conditions host a two-dimensional irreducible linear representation, and whose direct interpolation to a trivial phase leads to a continuous phase transition. Due to the appearance of the edge modes with a non-trivial linear representation, one side of the phase transition was dubbed a \emph{representation symmetry protected topological} (RSPT) phase. However, the classification tells us that the two sides of the transition should be connected by a symmetric gapped path, although no explicit path has been studied. In particular, it remains unclear how stable the continuous critical point is, and whether there is a general mechanism for its appearance. RSPT phenomenology appears also with integer-spin Affleck-Kennedy-Lieb-Tasaki (AKLT) states \cite{Pollmann12} (see also \Cref{app:AKLT}), and odd-$N$-state Onsager-integrable chiral clock models \cite{Jones25}. 

\textbf{Quasi-phases as a unifying theme.} $\;$ In this work we resolve these tensions and open questions by focusing our attention on connectivity of gapped phases in the vicinity of critical points. We provide a sharp reason for why certain types of criticality that are unnecessary from the SPT classification are, in fact, necessary. We dub such behavior {\em `infrared-necessary' (IR-necessary) criticality}, and will explain how it separates neighboring gapped \emph{quasi-phases}. Quasi-phases\footnote{The term quasi-phases has previously been used to describe regions separated by sharp but analytic `pseudo-transitions', e.g., in finite-temperature one-dimensional chains and in spin ice\cite{timonin2011spin,de_Souza_2018} (in some cases originating from a zero-temperature phase transition \cite{Krokhmalskii_2021}). We use the term in a different sense, since in our phase diagrams we have genuine critical points separating the two regions.}
are in the same bulk phase as classified by topological invariants, but can be distinguished sufficiently close to a critical point. We show how QSPT and RSPT phenomena naturally arise in this way, giving them a sharp meaning. The ensuing physical properties then can `radiate out' further away in the phase diagram, akin to the role of a finite-temperature critical fan \cite{Sachdev11}.

We explain the key ideas of quasi-phases in Section~\ref{sec:conceptual}, where we also outline how the subsequent sections unpack these ideas in exemplar phase diagrams. We synthesize the intuition gained into a broader framework in Section~\ref{sec:framework}.

\section{Quasi-phases: key idea and outline}
\label{sec:conceptual}
Let us consider gapped phases for concreteness. The usual equivalence relation is that two ground states of local gapped Hamiltonians are in the same phase if and only if they are connected by a path of local gapped Hamiltonians. In other words, we look at path-connectedness. Sometimes the phase diagram becomes richer when one imposes more structure: a celebrated instance is that of SPT phases, where restricting to \emph{symmetric} paths suddenly distinguishes parts of the phase diagram which would have been otherwise connected.

Here we introduce and study a new kind of restriction that also leads to a rich structure beyond the conventional classification. We explore phases of matter \emph{relative} to a given quantum critical point. More precisely, we define a natural equivalence relation for gapped phases by considering path-connectedness under the requirement that the path traverses {\em only those states obtained under the renormalization group (RG) for perturbations of some initial critical point}. More colloquially speaking: even if two gapped states obtained by different perturbations of a quantum critical system should strictly belong to the same phase of matter according to the usual classification, the critical point might be `unaware' of their connectivity. If this is the case, we say that such states are in distinct \emph{quasi-phases}.

While a critical transition between such quasi-phases may be unnecessary from the SPT classification, considering only RG flows can make a transition unavoidable. Indeed, at a generic unnecessary criticality, the single relevant perturbation can flow to two gapped states depending on the choice of sign (since they have codimension one, see \cref{sec:intro}). These states automatically \textit{seem} distinct in the vicinity of the critical point. This is, however, not necessarily so: while being connected by a homogeneous path of nearby gapped phases is \emph{sufficient} to be in the same quasi-phase, it is not \emph{necessary}. Indeed, as we make more precise in \cref{sec:framework}, a sufficient and necessary condition is to look at whether the neighborhood of the critical point can describe a smooth interface between the two gapped phases. Such an interface can exist even at an unnecessary criticality with one relevant operator---a second operator that is irrelevant in the bulk can be dangerously irrelevant at the interface, gapping it out. 

Our work thus identifies a strict \emph{subclass} of generic unnecessary criticalities, where the nearby gapped phases genuinely look distinct in its vicinity. We label such criticality IR-necessary. Interesting examples of IR-necessary criticality naturally arise at the QSPT and RSPT transitions discussed in \cref{sec:intro}. Indeed, since the existence of a smooth interface (arising from a nearby critical point) is a faithful diagnostic of being in the same quasi-phase, distinct quasi-phases naturally give rise to \emph{interface modes}. If lattice boundary conditions are sufficiently natural, one can often interpret these as low-energy interfaces to the trivial vacuum, thereby explaining the edge modes observed in various examples. Note that once such edge modes appear, the robustness of irreducible representations (irreps) means that they will persist over a finite range of parameter space, hence they do not require fine-tuning. Conversely, as we will discuss in \cref{sec:frameworkconstruct}, analyzing interfaces can be used to \emph{construct} lattice models whose nearby phase diagrams host quasi-phases.

One particularly interesting consequence of having a critical point whose nearby phase diagram hosts distinct quasi-phases is that it corresponds to one region of a bigger parent theory.
Indeed, the broader phase diagram can host a gapped path between the distinct quasi-phases, which by definition cannot be captured by an IR-necessary critical point. One needs a bigger universality class that on the one hand can flow to the starting critical point, and on the other hand describe the gapped path. The study of quasi-phases is thus not only a means to extend the notion of SPT-like and other interesting gapped phenomena, but also to provide a guideline to search for new quantum critical systems. In contrast, unnecessary critical lines that are not IR-necessary can simply terminate without requiring a bigger parent theory.

To broaden the accessibility of our work, we postpone the formal definition and framework of quasi-phases to \cref{sec:framework} and instead use the bulk of the paper to arrive at these ideas and their conclusions organically by studying illustrative examples. Although these various examples will have quite distinct microscopic structure, in each case we naturally arrive at a parent theory which possesses a critical point described by a conformal field theory (CFT) \cite{Ginsparg1988} built in a very particular fashion. Namely, it is described by an \emph{orbifold CFT} comprised of two CFTs coupled together via gauging a diagonal symmetry. One of the CFTs describes the phase transition between the distinct quasi-phases in the original phase diagram, whereas the different gapped regimes arrived at by perturbing the other CFT determine whether or not there is a path in the full phase diagram connecting the two quasi-phases. Thus while the quasi-phases are distinct relative to the initial starting CFT, the broader phase diagram includes gapped paths corresponding to different flows from the parent orbifold theory.

\textbf{Outline of paper.} $\;$ Since many examples of quasi-phases studied in this work are reminiscent of conventional SPT phases, \cref{sec:bondalt} starts by reviewing one of the simplest settings of two distinct 1+1D SPT phases and their continuous phase transition, described by a compact boson CFT. This example also serves as a limiting case for both QSPT and RSPT models, where we deform a genuine SPT transition by either explicitly breaking or enhancing the SPT-protecting symmetry (see also discussion of the `failed SPT' premise in Ref.~\cite{Prakash23}).
As mentioned in the introduction, QSPT phenomena are the simplest setting for quasi-phase physics, due to the emergent anomaly. In \cref{sec:QSPT} we outline the QSPT physics of the bond-alternating spin-1 XX chain with single-ion anisotropy. This model has a $D_4$ symmetry that acts as a projective quotient in the large anisotropy limit. We derive a physically motivated field-theoretic description of the parent orbifold theory, consisting of an Ising CFT coupled to the free boson. In \cref{sec:RSPT} we discuss RSPT phenomena in a perturbed spin-$\tfrac12$ XX chain with $S_3$ symmetry, where we uncover a phase diagram controlled by an orbifold of the Potts CFT with the free boson. In \cref{sec:potts,sec:pottsRSPT} we consider first a pair of symmetry-breaking quasi-phases and then RSPT physics in Hamiltonians related to the 3-state Potts model, making connections to the Hamiltonians of Ref.~\cite{OBrien20}. Finally, we outline our broader framework in \cref{sec:framework} and discuss possible connections to non-contractible loops of gapped Hamiltonians, i.e., generalized Thouless charge pumps.

\section{SPT physics and criticality in the spin-$\tfrac12$ bond-alternating XXZ chain}\label{sec:bondalt}
In this section we recall the bond-alternating XXZ chain, which is one of the simplest models containing two distinct SPT phases separated by a phase transition. Its $O(2)=U(1)\rtimes \mathbb{Z}_2$ symmetry protects this distinction. In later sections we show how it arises both in QSPT and RSPT contexts. 
Namely, in \cref{sec:QSPT} we show it appears as a limiting case of a spin-1 chain that exhibits QSPT phenomena; its symmetry is a quotient of the symmetry group of the full model.
Then in \cref{sec:RSPT} we consider a perturbation of this model that breaks the $U(1)$ down to a $\mathbb{Z}_3$ subgroup. As a consequence, the topological distinction between the two SPT phases disappears, giving rise to RSPT phenomena.

QSPT and RSPT constitute two different classes of \emph{short-range entangled} (SRE) quasi-phases \cite{Kitaev_talk,Gaiotto_2019,Kapustin20}. In 1+1D spin chains, SRE phases coincide with \emph{invertible phases}, where the state has a corresponding inverse state such that stacking the two can be symmetrically deformed to a symmetric product state. Such systems have unique ground states on a periodic chain.

\subsection{Projective symmetry fractionalization}
\label{sec:xxproj}
The spin-$\tfrac12$ bond-alternating XXZ chain Hamiltonian with open boundary conditions is\footnote{The square in the bond alternation is not the typical parameterization but is convenient below.}
\begin{align} \label{eq:HXX}
H_{\delta XXZ} = - \sum_{n=1}^{L-1} \left( 1 + (-1)^n \delta \right)^2\Big( X_n X_{n+1} + Y_n Y_{n+1} -\Delta Z_n Z_{n+1} \Big)\ ,
\end{align}
with $\delta$ parameterizing the degree of staggering. Fixing $\Delta$\,=\,0 yields the bond-alternating XX chain $H_{\delta XX}$, which is exactly solvable by Jordan-Wigner transformation to free fermions \cite{Affleck_Les_Houches}. Fixing $\delta$\,=\,0 as well makes it gapless, with the continuum limit described by the $c$\,=\,1 compactified free boson, described in \cref{sec:boson} below. For $\delta\ne 0$, $H_{\delta XX}$ is gapped. Tuning away from $\Delta$\,=\,0 we lose free-fermion solvability, but the $\delta$\,=\,0 case remains gapless for $-1< \Delta\leq1$ and the phase structure is unchanged.

The symmetries of $H_{\delta XXZ}$ of interest to us are the $U(1)$ generated by $\tfrac{1}{2}\sum_n Z_n$ and the $\mathbb{Z}_2$ generated by $\mathcal{C}=\prod_n X_n$. Together they form an $O(2) = U(1) \rtimes \mathbb Z_2$. It is well known that the transition at $\delta$\,=\,0 separates two distinct SPT phases \cite{Verresen17}, as a bosonic analogue of the SSH chain \cite{Su79}. The minimal symmetry group needed to protect the SPT physics is $\ZZ_2\times\ZZ_2 \subset O(2)$, generated by $\mathcal{C}$ along with $P=\prod_n Z_n$. It is convenient to take the number of sites $L$ to be even and to define a two-site unit cell, e.g., $\{2n-1,2n\}$. $P$ and $\mathcal{C}$ then act as a product of linear representations of $\ZZ_2\times\ZZ_2$ on each unit cell. Similarly, $O(2)$ acts linearly on these unit cells.

Taking $\delta<0$ results in a trivial phase while $\delta>0$ corresponds to a non-trivial SPT phase. This behavior can already be gleaned from the completely staggered limits $\delta = \pm 1$, there the resulting ground states are 
\begin{align}
\begin{tikzpicture}
\node at (0,0){
    \begin{tikzpicture}
    \node at (3.5/2,0.7) {$\delta=-1$};
    \foreach \x in {0,1,2,3} {
        \filldraw[color=black!12] (\x+0.25,0) ellipse (0.4 and 0.2);
    \foreach \x in {0,1,2,3,4,5,6,7}
        \filldraw (\x/2,0) circle (1pt);
    };
    \end{tikzpicture}
};
\node at (6,0){
    \begin{tikzpicture}
    \node at (3.5/2,0.7) {$\delta=1$};
    \foreach \x in {0,1,2} {
        \filldraw[color=black!12] (\x+0.25+0.5,0) ellipse (0.4 and 0.2);
    \foreach \x in {0,1,2,3,4,5,6,7}
        \filldraw (\x/2,0) circle (1pt);
    };
    \end{tikzpicture}
};
\end{tikzpicture}
\label{dimerpic}
\end{align}
where each oval dimer labels a Bell pair $\tfrac{1}{\sqrt{2}}({\ket{\uparrow \downarrow} + \ket{\downarrow \uparrow}})$. While in the trivial phase all spins are paired, at $\delta=1$ those at the boundary are decoupled, resulting in a fourfold ground-state degeneracy. These `edge modes' persist for any $\delta>0$, albeit with a dressing which decays exponentially into the bulk. (They are the usual edge modes of the SSH chain \cite{Su79}, and can be calculated by fermionizing \eqref{eq:HXX}.) 
While the symmetry generators $P$ and $\mathcal{C}$ commute within each two-site unit cell, they anticommute on each of the boundary spins when $\delta >0$ (see below). In other words, the $\mathbb Z_2 \times \mathbb Z_2$ acts in a non-trivial projective representation\footnote{Linear representations of a group $G$ acting on Hilbert space correspond to unitary matrices that satisfy $V(g)V(h)=V(gh)$. Projective representations satisfy $V(g)V(h)=e^{i \phi(g,h)} V(gh)$. We allow gauge transformations $V(g) \rightarrow e^{i \chi(g)} V(g)$, and non-trivial projective representations are those that are not gauge-equivalent to a linear representation. Mathematically, non-trivial projective representations correspond to classes in $H^2(G,U(1))$. That the two commuting $\mathbb{Z}_2$ group elements anticommute in this representation cannot be removed by a gauge transformation. \label{footnote:reps}} on the boundary. Such behavior cannot change without a bulk phase transition, and is characteristic of an SPT phase. The phase diagram of $H_{\delta XXZ}$ for $-1< \Delta\leq1$ is thus
\begin{center}
\begin{tikzpicture}
\draw[thick,->] (-4,0) -- (4,0) node[anchor= west]{$\delta$};
\node (node1) at (-0,-0.) {$\bullet$}; 
\node (node2) at (0.,0.5) {$c=1$}; 
\node (node2b) at (0.,-0.5) {$0$}; 
\node (node3) at (2.,-0.5) {SPT}; 
\node (node4) at (-2.,-0.5) {Trivial}; 
\node (node1) at (5,-0.) {.}; 
\end{tikzpicture}
\end{center}
Below we describe how this one-parameter phase diagram embeds in larger phase diagrams, and the codimension of the criticality.

The projective representations are also apparent in the entanglement. 
Given an entanglement cut between two unit cells in the bulk of $H_{\delta XXZ}$, the fractionalization of on-site symmetries onto the Schmidt basis gives rise to different SPT classes \cite{Pollmann10,Chen11,Schuch11,Pollmann12}. In particular, for linear on-site representations, this fractionalization can be projective, and the cohomology class of this projective representation classifies the SPT phase. 
Since for $\delta=-1$ the ground state is a product state over unit cells, the symmetry fractionalization (on the unique Schmidt state) is trivial. In contrast, the cut between the unit cells for $\delta=1$ crosses the Bell pair and so results in fractionalization. In
the Schmidt basis $\{\ket{\uparrow},\ket{\downarrow}\}$ the $U(1)$ symmetry $e^{-i\frac{\theta}{2} (Z_{2j-1}+Z_{2j})}$ fractionalizes as $V_{\theta}=e^{-i\frac{\theta}{2} \sigma_z}$ at $\delta$\,=\,1. Likewise, on-site charge conjugation $X_{2j-1}X_{2j}$ fractionalizes as $V_\mathcal{C}=\sigma_x$ on the Schmidt basis. 
The $O(2)$ acts projectively here, since the fractionalized $V_{2\pi}=-\mathbb{I}$. For the $\ZZ_2\times\ZZ_2$ subgroup, we have $V_\pi=-i \sigma_z$ which anticommutes with $V_\mathcal{C}=\sigma_x$. This implies a non-trivial projective class and is the exact analogue of the boundary action discussed above. Under perturbation away from $\delta$\,=\,1, the projective representation must survive until there is a bulk transition, indicating SPT behavior throughout the phase. In \cref{app:pumpxx} we give a complementary charge-pump argument that there is a relative SPT class between $\delta=\pm1$ \cite{Jones25b}. 

\subsection{The compactified free-boson description of the critical point}\label{sec:boson}
At $\delta$\,=\,0, \cref{eq:HXX} reduces to the usual XXZ spin chain, described in the scaling limit by a compactified free boson\cite{Affleck_Les_Houches}. We fix our conventions for this field theory throughout the paper in terms of $\Phi(x)$ and its conjugate field $\tfrac{1}{2\pi}\partial_x \Theta(x)$ obeying $[\Phi(x),\partial_y \Theta(y)] = 2\pi i \delta(x-y)$. The Hamiltonian is given by
\begin{align} \label{eq:compactboson}
H_{K} = \frac{1}{2\pi} \int \left( \frac{1}{4K} (\partial_x \Theta)^2 + K (\partial_x \Phi)^2 \right) \mathrm dx ~~,
\end{align}
where the marginal parameter $K$ depends on $\Delta$. For the XX model $\Delta$\,=\,0, it takes its free-fermion value $K$\,=\,1. This free-boson CFT has central charge $c$\,=\,1 and contains primary fields labeled by integers $m$ and $\widehat{m}$, with scaling dimensions
\begin{align}
    e^{\pm im \Phi}:\quad \Delta_m = \frac{m^2}{4K}\ .\qquad\quad e^{\pm i\widehat{m}\Theta}:\quad \Delta_{\widehat{m}} = \widehat{m}^2 K \ .
\label{Deltaboson}
\end{align}
 On the fields, our $O(2)$ acts as follows
\begin{alignat}{2}
\Phi &\overset{U(\alpha)}{\longrightarrow} \Phi+ \alpha\ ,  \qquad&\Phi &\overset{\mathcal{C}}{\rightarrow} -\Phi\ ,\nonumber\\
\Theta&\overset{U(\alpha)}{\longrightarrow} \Theta\ ,  &  \Theta&\overset{\mathcal{C}}{\rightarrow}- \Theta \ .
\end{alignat}
Moreover, translation takes $e^{\pm i\widehat{m}\Theta}\rightarrow (-1)^{\widehat{m}}e^{\pm i\widehat{m}\Theta}$ and leaves $\Phi$ invariant. 
These symmetry actions can be inferred from standard correspondences between lattice operators and continuum fields \cite{Affleck1987,Affleck_Les_Houches,Sachdev11}:
\begin{align} 
X_n \pm i Y_n\ & \sim\ e^{\pm i \Phi}  +\cdots \nonumber \\
\label{eq:XXXcontinuumlattice}
Z_n\ & \sim\  \partial_x \Theta\ +\ (-1)^n \sin \Theta +\cdots\\
-X_n X_{n+1} - Y_n Y_{n+1} + \Delta Z_n Z_{n+1} & \sim\  1\ +\ (\partial_x \Theta)^2 +(\partial_x \Phi)^2\ +\ (-1)^n \cos \Theta +\cdots \ ,
\nonumber
\end{align}
where we set all prefactors to one.

The bond alternation from turning on $\delta$ in Eq.~\eqref{eq:HXX} generates a term $\cos \Theta$ in the effective Hamiltonian. It
is the most relevant perturbation permitted by the $O(2)$ symmetry; while the operators $\cos(m\Phi)$ for integer $m$ are in the CFT, they are not invariant under the $U$(1) symmetry and so cannot appear in the Hamiltonian. Moreover, $\sin(n\Phi)$ and $\sin(\widehat{n}\Theta)$ are not $\mathcal{C}$ invariant. For $\delta<0$ ($\delta >0$) the perturbation causes a flow to a ground state with $\Theta$\,=\,0 ($\Theta=\pi$), resulting in the trivial (SPT) phase.  These two phases are distinct because there is no symmetric gapped path local to the field theory that connects the two ground states. Namely, a path in the space of perturbations $\cos(\hat{m}\Theta)$ can have a unique vacuum only at $\Theta=0,\pi$, because all other values occur in charge-conjugate pairs. The two are therefore not connected by any path induced by perturbations by homogeneous operators; see also \cref{sec:framework}. 

Relaxing charge conjugation, a gapped path connecting the two is obtained using the two-parameter perturbation $\lambda \cos \Theta + \mu \sin \Theta$---we can then continuously shift $\Theta$. If we instead relax the $U(1)$ symmetry but keep $\mathcal C$, a more subtle gapped path connecting the vacua comes from the perturbation $\lambda \cos \Theta + \mu \cos \Phi$; since $\Theta$ and $\Phi$ do not commute, we cannot track a semiclassical value of $\Theta$. For $K>1/8$, $\cos(\Phi)$ induces a flow to a unique vacuum, this could allow a gapped path, and indeed this theory is gapped\footnote{This is most easily understood at the $SU(2)$-invariant value $K=\tfrac12$, where we can rotate $\cos(\Phi)$ into $\sin(\Theta)$  and recover the first path. For $K<1/8$ see \cref{sec:irrelevant}.} for all $\lambda, \mu$ allowing us to tune through $\lambda =0$ \cite{Lecheminant02}. We will return to such paths below when we break $U(1)$ down to a subgroup.

\section{QSPT phenomena in a spin-1 chain}\label{sec:QSPT}

In this section, we turn to the bond-alternating spin-1 XX chain with single-ion anisotropy \cite{Schulz86,Chen03,Shiraishi25}. We will see that large anisotropy energetically freezes out $S^z=0$, recovering the spin-$\tfrac12$ chain of the previous section. In this low-energy regime, we find two different SPT phases. However, unfreezing the third degree of freedom reveals that the low-energy symmetry is a quotient, and the full UV symmetry group removes the topological distinction between the phases. Here we investigate gapped paths between the two phases and the unusual phase diagram that results.

We first introduce the spin-1 Hamiltonian
\begin{align}
H_{\delta D} = - \sum_n \left[ \left( 1 + (-1)^n \delta \right)\left( S^x_n S^x_{n+1} + S^y_n S^y_{n+1}\right) + D \left( S^z_n \right)^2 \right] \label{eq:spin1XX}
\end{align}
where we label the states on each site in the $S^z$-diagonal basis as
\begin{align}
     S^z\ket{0} \equiv 0, \quad S^z\ket{\uparrow} \equiv \ket{\uparrow}, \quad S^z\ket{\downarrow} \equiv -\ket{\downarrow} \ ,
\end{align}
and in this basis define
\begin{align}
\label{eq:spin1operators}
    S^x =&\frac{1}{\sqrt{2}}\begin{pmatrix}
        0 & 1&1\\
        1& 0& 0\\
        1 & 0& 0
    \end{pmatrix}, &
    S^y =&\frac{1}{\sqrt{2}}\begin{pmatrix}
        0 & i&-i\\
        -i& 0& 0\\
        i & 0& 0
    \end{pmatrix}, &
    S^z =&\begin{pmatrix}
        0 & 0&0\\
        0& 1& 0\\
        0 & 0& -1
    \end{pmatrix}.
\end{align}

The symmetries of Eq.~\eqref{eq:spin1XX} of relevance are the $U(1)$ symmetry generated by $\sum_n S^z_n$, as well as $\pi$-rotations around the three principal axes, $\mathcal R^i = \prod_n e^{\pi i S^i_n}$ for $i=x,y,z$. Notably $\mathcal R^z$ is the order-two element in $U(1)$ and $\mathcal R^x$ acts as a charge conjugation symmetry that negates the $U(1)$ charge. Together they make up $O(2) = U(1) \rtimes \mathbb Z_2$. In this section, we also consider its discrete subgroup $D_4 = \mathbb Z_4 \rtimes \mathbb Z_2 \subset O(2)$. Remarkably, the Hamiltonian \eqref{eq:spin1XX} has a non-local $SU(2)$ symmetry for all values of $(D,\delta)$\cite{Kitzawa03}, with the $U(1)$ a local subgroup. This symmetry is non-invertible; the generators satisfy the $SU(2)$ algebra but do not generate the full group. It affects certain microscopic features of the phase diagram but not our high-level arguments.

\begin{figure}\centering
\includegraphics[]{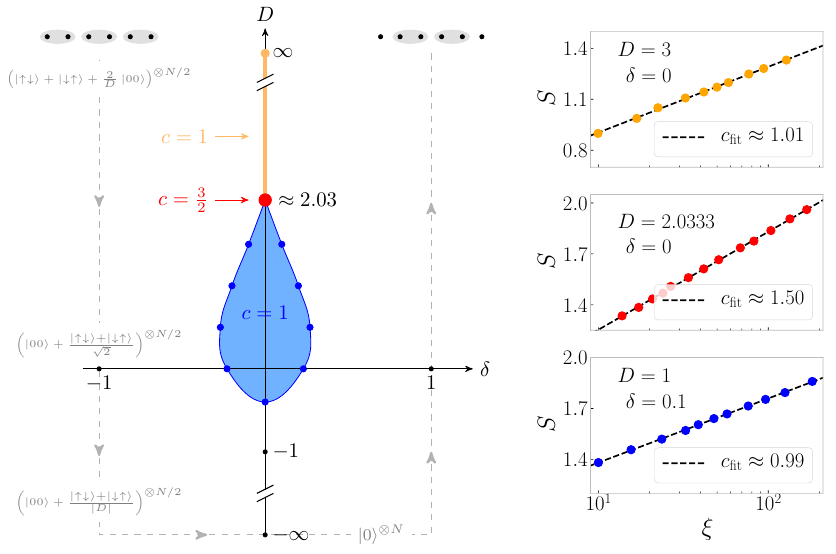}
    \caption{\small Left: Ground state phase diagram of the spin-1 chain in Eq.~\eqref{eq:spin1XX}. For $D \gg 1$, the two bond alternations are \emph{quasi-phases} separated by a continuous phase transition.
    The latter can be gapped out by tuning through the multicritical point with central charge $c=\frac{3}{2}$. Right: central charges for the critical regimes.}
    \label{fig:XXsingleionDMRG}
\end{figure}

The local $O(2)$ or $D_4$ symmetry can in principle protect the non-trivial Haldane SPT phase (as realized by, e.g., the spin-1 Heisenberg chain \cite{Pollmann10}), but we will show that \cref{eq:spin1XX} does not host this non-trivial phase. Nevertheless, we explain that when $D$ is large and positive, the spin-1 chain \eqref{eq:spin1XX} will have two distinct \emph{quasi-phases} for $\delta>0$ and $\delta<0$, separated by \emph{IR-necessary criticality}. In particular, we obtain the rather rich and peculiar phase diagram shown in \cref{fig:XXsingleionDMRG}, which we will gradually unpack and explain over the course of this section.

\subsection{Recovering the SPT in the strong-field limit}
We start by exploring the $D \to +\infty$ limit, where the two quasi-phases become genuinely distinct SPT phases. This limit is described by an effective spin-$\tfrac12$ chain since the local condition $\left(S^z_n\right)^2 = 1$ requires all low-energy states are comprised of $\{ \ket{S^z_j=1} \equiv \ket{\uparrow}, \ket{S^z=-1} \equiv \ket{\downarrow} \}$. The effective Hamiltonian can be written using the Pauli operators\footnote{If one prefers a definition of $Z_n$ which is unitary on the spin-1 Hilbert space, one can take $Z_n = i e^{-i \frac{\pi}{2} S^z_n} = \exp \left( i \pi \frac{1-S^z_n}{2} \right)$.} $Z_n = S^z_n$, $X_n = -e^{\pi i S^x_n}$, and $Y_n=iX_n Z_n$, whose action in this subspace exactly coincides with the usual Pauli matrices. Then second-order perturbation theory (see Appendix~\ref{app:pert}) yields the effective Hamiltonian in the low-energy Hilbert space \cite{Chen03}
\begin{align} \label{eq:Heff}
H_{\rm eff} = - \frac{1}{4D}\sum_n \left( 1 + (-1)^n \delta \right)^2\left( X_n X_{n+1} + Y_n Y_{n+1} - Z_n Z_{n+1} \right) + \mathcal O \left(D^{-3} \right).
\end{align}
This Hamiltonian is precisely that of the spin-$\tfrac12$ bond-alternating Heisenberg model \eqref{eq:HXX} with $\Delta$\,=\,1, displaying the SPT physics discussed in \cref{sec:bondalt}. For the $O(2)$ symmetry discussed above, tuning $\delta$ takes us between two different SPT phases \emph{for this symmetry group}. For $\delta$\,=\,0 it is unitarily equivalent to the Heisenberg antiferromagnet, as can be seen by applying $\prod_{n} Z_{2n}$. The $SU(2)$ is local in this reduction, as the non-local dressing present in the full Hilbert space acts as a constant on the low-energy subspace. In the field theory \eqref{eq:compactboson}, the $SU(2)$-invariant point corresponds to $K=\tfrac12$. 

\subsection{From phase to quasi-phase}
We have seen that in the $D \to + \infty$ limit, the two bond alternations in Eq.~\eqref{eq:spin1XX} ($\delta>0$ and $\delta<0$) are distinct phases of matter. Here we show that once $D$ is allowed to be finite, there exists a path between the two phases without closing the gap.

The two distinct SPT phases are protected by the $\mathbb Z_2 \times \mathbb Z_2$ symmetry generated by  $\prod_n Z_n$ and $\prod_n X_n$.
In terms of the original spin-1 variables, these two symmetry generators correspond to $\sqrt{{\mathcal R}^z}= \prod_j e^{i\frac{\pi}{2} S^z_j}$ and $\mathcal{R}^x= \prod_j e^{i\pi S^x_j}$ respectively (recall we take $L$ even). In the larger Hilbert space, the former does not square to one, but instead to the $\pi$-rotation $\mathcal R^z = \prod_j e^{i \pi S^z_j} = (-1)^{n_0}$, where $n_0$ is the number of sites occupied by the state $\ket{0}$. 
In the full spin-1 Hilbert space the corresponding symmetry is therefore enhanced to
$D_4 = \mathbb Z_4 \rtimes \mathbb Z_2$.
The generator $\sqrt{{\mathcal R}^z}$ acts as $\prod_n i Z_n$ in the spin-$\tfrac12$ Hilbert space with $n_0$\,=\,0, and the full $D_4$ symmetry of the spin-1 chain reduces to the \emph{quotient} symmetry $D_4/\mathbb Z_2 \cong \mathbb Z_2 \times \mathbb Z_2$ as $D \to +\infty$.\footnote{If we consider the full $O(2)$ symmetry of the spin-$\tfrac12$ chain, we have $V(\alpha) \equiv e^{i \alpha \sum_n {Z_n}/{2}} = e^{i{\alpha} \sum_n S^z_n/2}$ in the original spin-1 variables, such that $V(2\pi)$ only \emph{looks} like an identity operator in the low-energy theory, but actually acts like $\mathcal R^z = \prod_n e^{i \pi S^z_n}$ on the high-energy variables. In conclusion, the apparent `SPT' phase in the large $D$ limit is protected by $O(2)/ \mathbb Z_2$. Alternatively, one can say that the UV spin-1 $O(2)$ is the double cover of the quotient $O(2)$, where the former is denoted Pin$^+(2)$.} 

Crucially, the non-trivial \emph{projective} representation of $\mathbb Z_2 \times \mathbb Z_2$ (which appeared, e.g., on the boundary of the spin-$\tfrac12$ chain for $\delta>0$) is actually a \emph{linear} two-dimensional representation of this $D_4$. This means that from the perspective of the spin-1 chain, the two bond-alternations are \emph{not} distinct SPT phases. Indeed, it is known that extending symmetry groups (by embedding a system into a bigger Hilbert space) can trivialize SPT phases 
\cite{WangWenWitten18,Prakash18,Tachikawa20,Thorngren20,Thorngren21}.
Nevertheless, Ref.~\cite{Verresen21} pointed out how various SPT phenomena (such as edge modes and phase transitions) remain parametrically robust. Since these phenomena were protected by the low-energy quotient symmetry, they were dubbed quotient symmetry protected topological (QSPT) phenomena.
\subsection{Gapped path connecting the quasi-phases}\label{subsec:paths}
The fact that the $D_4$ symmetry of the spin-1 chain is not sufficient to distinguish these two phases can be seen most clearly by constructing a {gapped $D_4$-symmetric path between the two distinct bond-alternations}. This is indicated in the phase diagram in \cref{fig:XXsingleionDMRG}. We start with $\delta=-1$, where the Hamiltonian \eqref{eq:spin1XX} decouples into a series of two-spin-1 systems. Hence, it is sufficient to solve 
\begin{align}
H_{12} \equiv -2\big(S^x_1 S^x_2 + S^y_1 S^y_2) - D \big(S^z_1)^2 - D \big(S^z_2)^2.
\end{align} This Hamiltonian has a gap $\sqrt{D^2+8}-\max \{ 2,D \}>0$, and its ground state is:
\begin{align}
\ket{\psi}_{12} \propto \ket{\uparrow \downarrow} + \ket{\downarrow \uparrow} + \frac{\sqrt{D^2+8} - D}{2} \ket{00}.
\end{align}
Hence, the ground state of the full system $\ket{\Psi} = \bigotimes_{n}\ket{\psi}_{2n-1,2n}$ smoothly interpolates from $\left(\frac{\ket{\uparrow \downarrow} + \ket{\downarrow \uparrow}}{\sqrt{2}}\right)^{\otimes N/2}$ in the $D \to + \infty$ limit to $\ket{0}^{\otimes N}$ in the $D \to - \infty$ limit; see \cref{fig:XXsingleionDMRG}.  Since in this regime the ground state is independent of bond-alternation, we can freely tune $\delta$ from  $-1 $ to  $1$, after which we can ramp back into the spin-$\tfrac12$ chain regime.

As apparent in \cref{fig:XXsingleionDMRG}, this gapped path is rather lengthy. Indeed, in this phase diagram the criticality extends from $D=\infty $ to negative $D$. In this way, it allows us to see that the two bond-alternations can be thought of as distinct \emph{quasi-phases} with differing physical properties.  
The key observation here is to take open boundary conditions. Along the portion of the path with $\delta$\,=\,1, the leftmost spin has `edge' Hamiltonian $H_{\rm edge} = - D (S^z_1)^2$. Thus for $D>0$ the ground state is (at least) twofold degenerate because of this edge mode. The edge mode is gapped out for $D<0$, giving a unique ground state. 

The degenerate states $S_1^z=\pm1$ form a two-dimensional linear irrep of $D_4$, and so can only disappear by driving a level crossing to a trivial representation (a `boundary phase transition').  Including additional $D_4$-symmetric boundary perturbations can only change the location of this transition, not remove it altogether. There will always be an open region with $\delta>0$ and large $D$ 
where the edge mode and the ensuing degeneracy are present. Thus while it is possible to connect the two phases without closing the gap, there is still a sharp (boundary) transition between the two.
Analyzing these edge modes is not the primary focus of the present work; see Ref.~\cite{Verresen21} for a detailed study of edge modes due to QSPT phenomenology. 

Note also that while for the edge mode we see a boundary transition, for the bulk system we will see this reflected in the entanglement spectrum. Indeed the entanglement spectrum will be degenerate at infinite $D$, due to the SPT projective representation of the quotient group. When we leave the quotient we are able to introduce entanglement singlets without closing the bulk gap, and we will see an entanglement transition. (The entanglement transition and boundary transition do not generically coincide in the larger phase diagram, but may do in the vicinity of the critical point \cite{Verresen21}.) 

Crucially, the parametric stability in this case can be traced back to energetic stability. More precisely, whenever there is a large energy scale associated to the excitations carrying charge under the $\mathbb Z_2$ subgroup, the low-energy theory should be governed by the effective quotient symmetry. This means that such an energetic separation of scales can protect the degeneracies in the edge mode and entanglement spectrum. Operationally, this is captured by the large energy terms suggesting that perturbation theory is controlled. Indeed, as long as the low-energy theory is well-described by a spin-$\tfrac12$ chain, the Hilbert space \textit{cannot} accommodate a unique boundary state or non-degenerate entanglement spectrum. Correspondingly, breaking the edge mode signals that we have entered a non-perturbative regime, where perturbation theory can no longer be controlled. This is in sharp contrast to the RSPT example in the next section: although it also exhibits parametrically robust features, it is not controlled by any given energy scale.

\subsection{Phase diagram}
We now turn to a study of the full ground-state phase diagram of the Hamiltonian in Eq.~\eqref{eq:spin1XX}. We have shown that the distinct $\delta>0$ and $\delta<0$ SPT phases at infinite $D$ become quasi-phases, as we lower $D$, that can be connected without closing the gap. Moreover, the two quasi-phases differ in their boundary physics, at least for a regime of sufficiently large $D$. 
The path connecting the two quasi-phases crosses the $\delta$-axis at a sufficiently negative value of $D$, while the SPT protection occurs as $D\to\infty$. Something interesting must therefore happen in the middle!

We first analyze what happens along the $\delta$\,=\,0 line. 
Due to the exact $SU(2)$ symmetry \cite{Kitzawa03}, so long as we remain in a $c=1$ criticality, this pins $K=\tfrac12$. 
For large $D$, we can perturb around the bosonic effective field theory \eqref{eq:compactboson}. Other terms may appear in the effective Hamiltonian arising from higher orders in perturbation theory in $1/D$. The only relevant operator in the free-boson theory invariant under the local $O(2)$ symmetry is $\cos(\Theta)$, which when added to the effective action corresponds to taking $\delta\ne 0 $.
As discussed in \cref{sec:boson}, no matter what (relevant, marginal or irrelevant) operators are generated in the effective action, we cannot have a gapped path in the vicinity of this theory that connects $\Theta=0$ to $\Theta=\pi$. Hence, the gapless line extends to some finite $D$. The operator $\cos(2\Theta)$ is symmetric and thus it will be included in the effective low-energy Hamiltonian at $\delta$\,=\,0 and large but finite $D$; if it were (marginally) relevant this could lead to a first-order line. It is a marginal operator at this value of $K$, but we observe that the sign of the generated coupling means it is marginally irrelevant and the $c=1$ criticality has some finite extent. To allow for the gapped path, this must terminate at some point, and we observe numerically that this is via a $c=3/2$ multicritical point into an extended gapless $c=1$ phase. Indeed, this model has a rich history \cite{Schulz86} and was most recently studied by Shiraishi and Nomura \cite{Shiraishi25} where they found indications of a multicritical point described by a CFT with central charge $c=\tfrac32$, although a precise characterization of the field theory was not achieved. See also related discussion in \cite{Prakash23,Mondal23,Prakash24}. We will discuss the field theory description in detail below; first we review our numerical results. 

We analyzed the behavior on the critical line numerically using the infinite density matrix renormalization group (iDMRG) \cite{White92,Hauschild24a,Hauschild24b}, as displayed in 
\cref{fig:XXsingleionDMRG}. Although the two bond-alternating patterns for large $D$ are connected in the low-$D$ regime via the gray dashed line (derived in \cref{subsec:paths}), a rich landscape of criticality emerges. This is numerically diagnosed by `entanglement scaling', whereby one gradually increases the bond dimension of the matrix product state and tracks how the correlation length $\xi$ and bipartition entanglement entropy $S$ diverge; for a CFT this should scale as $S \sim \frac{c}{6} \log \xi$ where $c$ is the central charge \cite{Calabrese2004,Pollmann09}. We see in \cref{fig:XXsingleionDMRG} that the translation-invariant chain gives a central charge $c=1$ for large $D$, consistent with our solvable limit in Eq.~\eqref{eq:Heff}. As we decrease $D$, this criticality persists (orange line) until we hit a multicritical point with $c={3}/{2}$ at $D \approx 2.0333$ (red dot) (consistent with recent observations in Ref.~\cite{Shiraishi25}). For smaller $D$, an extended gapless phase ($c=1$) opens up, which eventually gaps out via a Berezinskii-Kosterlitz-Thouless (BKT) transition.

\subsection{Field theory}

\subsubsection{Arriving at \texorpdfstring{$c=\tfrac32$}{c=3/2} from large $D$}
The multicritical point and the extended gapless region in \cref{fig:XXsingleionDMRG} can be understood in depth from a more general field-theory treatment. 
One can view the $S^z_n$\,=\,0 as vacancies in a critical spin-$\tfrac12$ XXZ chain.
The  vacancies themselves can become critical at some value of $D$, and so cause a phase transition. A classic example of such is the appearance of a tricritical point in the Ising model with vacancies, at which the critical line becomes a first-order transition \cite{Blume1971}.

We find no first-order transition, but rather that the multicritical point is a {\em deconfinement} transition of the sort considered in \cite{Shi2011,Serna2017}. Namely, vacancies are created in local pairs via the terms in the Hamiltonian taking $\ket{+-}$ or $\ket{-+}$ to $\ket{00}$. At large positive $D$, these vacancies have a much higher energy and so can be integrated out. At small values, they can break the local pairs and proliferate throughout the chain, i.e., are deconfined. The boson effective theory must be at minimum amended, as one can no longer integrate out the spins with $S^z_n$\,=\,0.  

It is most natural to describe these $U(1)$-neutral degrees of freedom via an Ising field. Indeed, $e^{i\pi S^z_n}=\pm 1$ is frozen at $-1$ as $D\to\infty$, but fluctuates for $D$ finite. This $\mathbb{Z}_2$ degree of freedom is not part of the free-boson field theory. The simplest way to take it into account is to include an Ising field $\sigma(x)$ in the effective field theory along with the boson $\Phi$ (where the neutral field $e^{i\pi S^z_n}$ corresponds to $\varepsilon$ that drives the Ising transition). If both the boson and the emergent Ising degrees of freedom are critical, the ensuing CFT indeed has central charge $c=\tfrac32$. Such an effective field theory describes the low-energy physics of an integrable chain described below in \eqref{eq:extendedspin1}, that is smoothly connected to our model 
\cite{OBrien20,OBrien20b}.

To identify the lattice operators with those in the field theory, we need to understand how the Ising and bosonic degrees of freedom are coupled. To illuminate the situation, we consider the field-theory description of the lattice operators $S^{\pm}_n$. 
These operators do not arise in the effective spin-$\tfrac12$ description at $D\to\infty$, as they couple the $\uparrow,\downarrow$ states to the $S^z$\,=\,0 states. They therefore must involve the Ising degrees of freedom. 
As follows from \eqref{eq:XXXcontinuumlattice} (and the identification of states in the low-energy space), the charge-$\pm 2$ operator $(S^{\pm}_n)^2$ is identified with $e^{\pm i\Phi}$ in the field theory. The operators $e^{\pm i \Phi/2}$ of dimension $1/8$ have charge $\pm 1$, and indeed are \emph{not present} in the $SU(2)_1$ field theory describing the $D\to\infty$ limit. 
We therefore can identify\begin{align}
S_n^\pm\ \sim\ \sigma\, e^{\pm i\frac12 \Phi}
\label{Ssigma}
\end{align}
in the field theory. As the operator product of two $\sigma$ fields contains the identity, the product of each operator with itself does indeed contain $e^{\pm i \Phi}$ as necessary. Moreover, when the Ising degrees of freedom are gapped, we can neglect their fluctuations and replace $\sigma$ in \eqref{Ssigma} with its expectation value. When the $\sigma$ degrees of freedom are in a disordered phase, $\braket{\sigma}\simeq 0$, ensuring that this operator does not appear in the CFT. 

To fully identify the coupling between Ising and boson theories, we note that $\sigma$ must be even under both $\mathcal{R}^z$ and the $\mathcal{R}^x$ spin-flip symmetry, as it is $U(1)$-neutral. The Ising field theory however has a $\mathbb{Z}_2$ symmetry $\mathcal{R}_{\rm I}:\sigma\to-\sigma$ that has no obvious analogue in the lattice model.  The operators $\sigma e^{\pm i\frac12 \Phi}$ are odd under $\mathcal{R}^z$ as desired, but also odd under $\mathcal{R}_{\rm I}$. Since the latter symmetry does not appear in the lattice model, it is natural then to remove it from the field theory by gauging it. We therefore declare the diagonal $\mathbb{Z}_2$ generated by $\mathcal{R}_{\rm I}\mathcal{R}^z$ to be 
\begin{alignat}{2}
\mathrm{gauge~symmetry:} \qquad\sigma ~&{\rightarrow} ~- \sigma \qquad    &
\tfrac\Phi2~ &{\rightarrow} ~\tfrac\Phi2 +\pi \ . \label{eq:gaugeZ2}
\end{alignat}
Then $\sigma\, e^{\pm i\frac12 \Phi}$ is gauge-invariant and appears in the CFT. In the background we seem to have a new elementary bosonic field $\tilde{\Phi}=\Phi/2$ that is $2\pi$ periodic so that the vertex operators are well defined. This would appear to imply that our field $\Phi$ should really be understood as $4\pi$ periodic. But note that after gauging we should characterize the large $D$ regime by an exponentially decaying two-point function  $\langle \sigma(x) \sigma(y)\rangle$. Then our gauge symmetry recovers the $2\pi$ periodicity of $\Phi$ in this limit. 

The effective field-theory Hamiltonian of \eqref{eq:Heff} with $\delta\ne 0$ contains $\cos\Theta$, but this operator is not local with respect to that in \eqref{Ssigma} (although $\cos(2\Theta)$ is). The $\Theta$ vertex operators must therefore also couple to the Ising degrees of freedom in a non-trivial way. The Ising field theory contains a disorder field $\mu$, Kramers-Wannier dual to $\sigma$. Duality requires that it also be odd under $\mathcal{R}_{\rm I}$. While $\mu$ is non-local with respect to $\sigma$, gauge invariance requires coupling the latter to the $\Phi$ field as in \eqref{Ssigma}. Thus if we couple $\mu$ to $\Theta$ in a similar fashion, we obtain operators that are mutually local.  
Namely, the operators $\mu e^{\pm i \Theta}$ are local with respect to both $e^{2i\widehat{m}\Theta}$ and $\sigma e^{i(m+\frac12) \Phi}$. 

We thus find that for $D$ finite, the additional Ising degrees of freedom result in more operators in the spectrum than simply those in \eqref{Deltaboson}. The gauge-invariant ones coupling the two theories are 
\begin{align}
    \sigma^m e^{i \frac{m}2 \Phi},\quad \mu^{\widehat{m}} e^{\pm i\widehat{m}\Theta}\ ,
\label{sigmamu}
\end{align}
where $m$ and $\widehat{m}$ are integers. These operators remain mutually local even when the Ising degrees of freedom become critical. 
The other identifications in \eqref{eq:XXXcontinuumlattice} (again, setting coefficients to one) generalize to 
\begin{align}
\begin{aligned}
S^z_n & \sim  \partial_x \Theta + (-1)^n \mu \sin\Theta + \cdots\cr
S^x_n S^x_{n+1} + S^y_n S^y_{n+1}+D(S^z_n)^2 &\sim 1+ (\partial_x \Theta)^2 +(\partial_x \Phi)^2+(-1)^n \mu\cos \Theta + \mathcal{H}^{\rm Ising}+\cdots \ .
\label{SSmu}
\end{aligned}
\end{align}
The Ising Hamiltonian density $\mathcal{H}^{\rm Ising}$ depends on the coupling $D$. For large values of $D$, it is gapped, and so does not affect the critical behavior. The effective spin-$\tfrac12$ chain behavior \eqref{eq:Heff} arises when $
\mu$ has long-range order, so that \eqref{SSmu} reduces to \eqref{eq:XXXcontinuumlattice}. The Ising degrees of freedom in the large-$D$ phase must in effect be in their disordered phase, where creating vacancies costs energy. Indeed, $S^\pm$ will have exponentially decaying two-point correlation functions, and not appear in the effective CFT for $\delta=0$. Thus we obtain the correct $D\to\infty$ limit simply by stipulating that the Ising degrees of freedom are gapped and in the disordered phase, for large separations $\braket{\sigma(x)\sigma(y)}=0$ and $\braket{\mu(x)\mu(y)}\ne 0$.

As we lower $D$, the fluctuations of the Ising degrees of freedom increase, and eventually become critical. Our numerics show that this happens at $D \approx 2.0333$ and $\delta=0$. The ensuing Ising criticality contributes an additional $c_{\rm I}=\tfrac12$ to the central charge \cite{Ginsparg1988}, thus accounting for the $c=1+\tfrac12 =\tfrac32$ observed in our numerics. The critical field theory can be written in terms of free-boson and Ising CFTs with the allowed operators constrained as in \eqref{Ssigma}, i.e., operators with odd-integer charge under the $U(1)$ are glued to those with odd charge under $\mathcal{R}_I$. Such behavior is that of the `super-affine' line of $c=\tfrac32$ CFTs, obtained by taking an `orbifold' of the boson and Ising theories \cite{Dixon88}. In the current parlance, this orbifold corresponds precisely to gauging $\mathcal{R}_{\rm I}\mathcal{R}^z$ as we have done in \eqref{eq:gaugeZ2}. Indeed, the $c=\tfrac32$ CFT describing an extended Hamiltonian \eqref{eq:extendedspin1} with $J=J'=D=1$, given in the next subsection, is a particular point on the super-affine line \cite{OBrien20b}. Worth noting is that the $O(3)_1$ CFT (as mentioned below in the context of the Takhtajan-Babujian chain) is another theory on the super-affine line. In our model, we have no guarantee that the velocities of the boson and Ising theories are the same, but the orbifold construction and the definition of sectors apply nonetheless. Note that this combination of a compact boson CFT and Ising CFT matches the discussion in Ref.~\cite{Shiraishi25}. Here we motivate it physically from the vacancy picture, and this approach clarifies that the two CFTs are coupled by an orbifold.

\begin{figure}[ht]
\centering
 \includegraphics[scale=0.4]{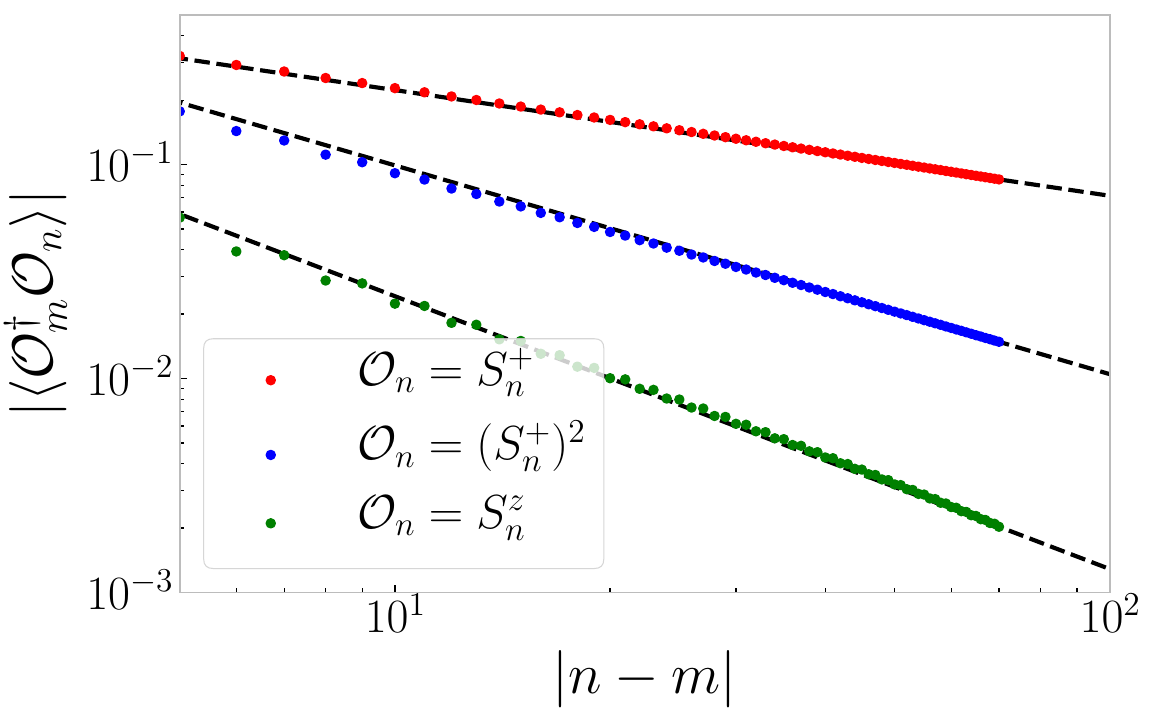}
    \caption{\small  Numerical correlation functions at the multicritical point. The lines are a guide to the eye, using Eq.~\eqref{eq:scalingdim} for $K_m = 0.513$.}
    \label{fig:XXsingleioncorr}
\end{figure}

We can test this field-theory description by numerically computing the two-point functions in the ground state. Two-point functions in a CFT fall off
as $\braket{\mathcal{O}^\dagger_m \mathcal{O}_n}\sim |m-n|^{-2\Delta_{\mathcal{O}}}$. The scaling dimensions of the Ising fields are $\Delta_\sigma=\Delta_\mu=\frac18$, while those of the bosonic fields are given in \eqref{Deltaboson}. With the identifications above, we therefore expect the 
lattice operators to have dimensions
\begin{align} \label{eq:scalingdim}
\left[ S^\pm_n \right] = \frac{1}{16K_{\rm m}} + \frac{1}{8}, \quad \left[ \left(S^\pm_n\right)^2 \right] = \frac{1}{4K_{\rm m}}, \quad \left[ S^z_n + S^z_{n+1}\right] = 1, \quad
\left[ S^z_n -S^z_{n+1}\right] = K_{\rm m} + \frac{1}{8},
\end{align}
where $K_{\rm m}$ is the value of the Luttinger parameter at the multicritical point. 
The numerical results are shown in \cref{fig:XXsingleioncorr}, where we find a good agreement with Eq.~\eqref{eq:scalingdim} for $K_{\rm m} \approx .51$, matching nicely with the theoretical value  $K=\tfrac12$ required by the non-invertible $SU(2)$. At the multicritical point, the operator $\mu\cos(\Theta)$ remains relevant. Thus taking $\delta\ne 0$ here still gaps the system.

What happens when $D$ is decreased past the multicritical point is rather striking. We find an extended critical region, as displayed in \cref{fig:XXsingleionDMRG}. Its existence follows directly from the field-theory analysis. 
In this region, the Ising theory becomes ordered, with a non-vanishing long-range two-point function. 
Using conventional nomenclature, this is the Higgs phase where the gauge symmetry is `spontaneously broken'. The operators $e^{im\Phi}$ with half-integer $m$ are now effectively local. However, since the disorder field is no longer critical, the operator $\mu\cos(\Theta)$ coupling to $\delta$ no longer appears in the low-energy effective theory and cannot gap the system. To probe the situation further, it is convenient to utilize the lattice charge-one bosonic field $\tilde{\Phi}=\Phi/2$ introduced below \eqref{eq:gaugeZ2}, along with its dual $\widetilde{\Theta} =2\Theta$.  Below the multicritical point the Ising degrees of freedom gap out, leaving only the standard free boson \eqref{eq:compactboson} with local operators $e^{i m \tilde\Phi}$ and $e^{i \widehat{m} \tilde\Theta}$. Matching the bosonic fields at the multicritical point tells us that $\tilde{K}=2$ for our new fields. 
As we tune away from the multicritical point, the most relevant symmetry-allowed operator is marginal. The $SU(2)$ symmetry fixes $\tilde{K}$ and the low-energy theory depends on the sign of the coupling to the $SU(2)$-invariant marginal \cite{Affleck_Les_Houches}. This leads to the observed bubble of $c=1$ gapless phase, where the boundary of the bubble corresponds to the marginal coupling changing sign.

Note that we used our SPT at infinite $D$ to anchor our field theory, where in the low-energy theory a unit charge for the free boson corresponds to charge two in the UV. Another approach would be to begin with the bosonic field $\tilde{\Phi}$ where the $U(1)$ acts faithfully. Then to freeze out the unit charges at low energies for large $D$, we need to attach a dressing that ascribes a mass to the odd-charge operators, but not to the even charges. If we call this dressing $\varsigma$, then taking naive operator products of the dressed field  $\varsigma e^{i\tilde \Phi}$ we have that $\varsigma^2 e^{2 i \tilde\Phi}$ is a local operator. Thus we can achieve our goal by fixing $\varsigma^2=1$, i.e., taking an Ising field. The condition that we must dress the unit charge operators can be imposed by exactly the orbifold construction given above! This perspective will be helpful in the next section where we do not have a quotient symmetry to guide us.

\begin{figure}
    \centering
\includegraphics[]{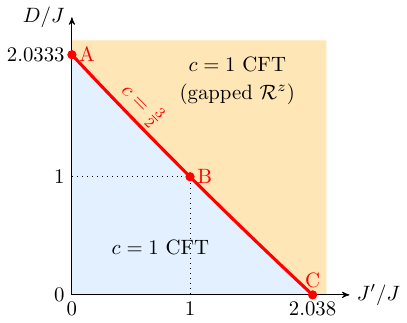}
    \caption{\small Ground state phase diagram of Eq.~\eqref{eq:extendedspin1}. This interpolates our numerically observed central charge $c=\frac{3}{2}$ in the phase diagram \cref{fig:XXsingleionDMRG} for the model Eq.~\eqref{eq:spin1XX} (point A) to an integrable spin-1 chain with central charge $c=\frac{3}{2}$ (point B).}
    \label{fig:appB}
\end{figure}

\subsubsection{Intermezzo: theories with $c=\tfrac32$}
Critical points described by $c=\tfrac{3}{2}$ CFTs often appear in spin-1 antiferromagnetic chains. The best-known example is the Takhtajan-Babujian chain \cite{Takhtajan1982,Babujian1982},  a special point of the $O(3)$-invariant bilinear-biquadratic spin-1 chain. For antiferromagnetic couplings in the $\vec{S}\cdot\vec{S}$ term, its continuum limit is the $O(3)_1$ Wess-Zumino-Witten CFT (aka $SU(2)_2$),  which can be written in terms of three free Majorana-fermion fields. This critical point governs the transition between the Haldane phase and a dimerized region \cite{Affleck1987}. The $c=\tfrac32$ critical behavior survives breaking the $O(3)$ symmetry to $U(1)$ along a particular integrable line of couplings \cite{Zamolodchikov80}. 

One point of the integrable spin-1 XXZ chain is particularly relevant for our analysis. This point, analyzed in depth in \cite{OBrien20,OBrien20b}, describes a self-dual spin-1 chain. It possesses a variety of interesting properties, including critical behavior described by a $c=\tfrac32$ CFT. We can interpolate between this integrable model and the translation-invariant case of our model (Eq.~\eqref{eq:spin1XX}) via
\begin{align} \label{eq:extendedspin1}
H' = - J \sum_n \left( s^+_n s^-_{n+1} + h.c. \right)  - J' \sum_n \left( \left(s^+_n\right)^2 \left( s^-_{n+1} \right)^2 + h.c. \right) - D \sum_n \left( S^z_n \right)^2 \; ,
\end{align}
where we use the convention $s^{\pm} = \frac{S^\pm}{\sqrt{2}} = \frac{S^x \pm i S^y}{\sqrt{2}}$.
The sign of $J$ can be toggled by a unitary transformation; we will set $J>0$ without loss of generality. The case $J'=0$ and $J=1$ corresponds to Eq.~\eqref{eq:spin1XX} with $\delta=0$, while setting $J=J'=D=1$ yields the integrable self-dual model with $c=\tfrac{3}{2}$ criticality. 
We use iDMRG to obtain the ground-state phase diagram of \eqref{eq:extendedspin1}, and display the results in \cref{fig:appB}. We indeed see that the numerically observed $c=\tfrac{3}{2}$ (point A) at $\frac{D}{J} \approx 2.0333$ seems to be smoothly connected to the integrable chain with the same central charge (point B). Our model therefore illustrates the fact \cite{Shi2011} that such critical points occur in non-integrable lattice models.  
We note that this known $c=\frac{3}{2}$ CFT can be described as an orbifold of an Ising CFT and compact boson CFT \cite{Ginsparg1988,Dixon88}, connecting back to the physical arguments for the orbifold description given in the previous subsection.

\subsubsection{Symmetries, emergent anomaly and edge modes}
Now that we understand the field theory underlying the phase diagram in \cref{fig:XXsingleionDMRG}, we can use it to better understand and confirm how the symmetry properties protect topological phenomena.  In particular, as highlighted in \cref{sec:intro}, QSPT transitions exhibit an emergent anomaly. Here we demonstrate this directly, along with the parametrically stable edge modes.

First, the lattice symmetries (with $U(1)$ generated by $\sum_n S_n^z$) act as follows on the fields in the $\sigma$-ordered phase:
\begin{alignat}{2}
\tilde\Phi &\overset{U(\alpha)}{\longrightarrow} \tilde\Phi+ \alpha  \qquad&\tilde\Phi &\overset{\mathcal{R}^x}{\rightarrow} -\tilde\Phi\nonumber\\
\tilde\Theta&\overset{U(\alpha)}{\longrightarrow} \tilde\Theta  &  \tilde\Theta&\overset{\mathcal{R}^x}{\rightarrow}- \tilde\Theta \ .
\end{alignat}
While in the QSPT bosonic theory, which applies at large $D$ where the Ising field is disordered, we have
\begin{alignat}{2}
\Phi &\overset{U(\alpha)}{\longrightarrow} \Phi+ 2\alpha  \qquad&\Phi &\overset{\mathcal{R}^x}{\rightarrow} -\Phi\nonumber\\
\Theta&\overset{U(\alpha)}{\longrightarrow} \Theta  &  \Theta&\overset{\mathcal{R}^x}{\rightarrow}- \Theta \ ,
\end{alignat}
reflecting the fact that charge-one excitations are gapped out. 
We thus recover that the UV $D_4$ symmetry acts as $\mathbb Z_2 \times \mathbb Z_2$ on $\Phi$ and $\Theta$. The $\alpha=\pi/2$ generator of $\mathbb Z_4$ acts as a $\pi$ shift of $\Phi$. Translation in the lattice model acts like an SPT-entangler, which in the CFT manifests as a shift $\Theta \to \Theta +\pi$. Simultaneous $\pi$-shifts of $\Phi$ and $\Theta$ are anomalous (corresponding to a Lieb-Schultz-Mattis (LSM) anomaly in the effective spin chain) \cite{Metlitski18}.
Thus, in the regime where our low-energy theory is described by $\Phi,\Theta$, we have an \emph{emergent} anomaly, preventing any symmetric gapped phase.

\begin{figure}[h]
\centering
\includegraphics[]{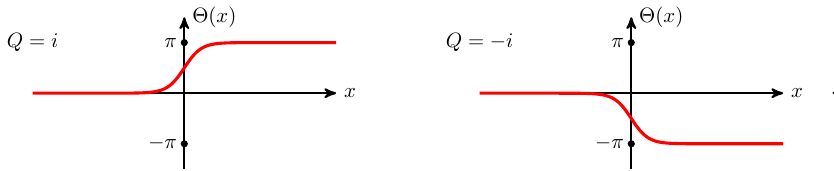}
\caption{\small Interfaces between $\Theta=0$ and $\Theta=\pi$, along with their $\mathbb{Z}_2$ charges. }\label{fig:edge}
\end{figure}

In the same regime, we have that any spatial interface between $\Theta=0$ and $\Theta=\pi$ will host a degenerate edge mode. To see this, note that the $\mathbb Z_2$ symmetry is the shift $\Phi \to \Phi + \pi$. This is generated by the operator $\hat Q = e^{\frac{i}{2} \int_{-\infty}^\infty \partial_x \Theta \mathrm d x}$. Let us consider the basis where $\hat Q$ is diagonal, i.e., the basis of $\Theta(x)$ configurations. Moreover, we only consider configurations with the asymptotic properties $\lim_{x \to -\infty} \Theta(x) = 0 \mod 2\pi$ and $\lim_{x\to +\infty} \Theta(x) = \pi \mod 2 \pi$. Note that this Hilbert space splits into two pieces: $\mathcal H = \mathcal H_+ \oplus \mathcal H_-$, labeled by the eigenvalue $Q = e^{\pm i \pi /2} = \pm i$ (to see this, note that due to the aforementioned asymptotic limits, we have $\int_{-\infty}^{\infty} \partial_x \Theta(x) \mathrm d x = (2m+1)\pi$ for some $m \in \mathbb Z$, implying $Q=i(-1)^m$). Hence, any energy eigenstate belongs either to $\mathcal H_+$ or $\mathcal H_-$, since $Q$ is a well-defined quantum number. Now observe that $\mathcal{R}^x$ toggles the sign of $Q$, i.e., it maps any state of $\mathcal H_+$ to a state in $\mathcal H_-$ and vice versa. {This means that every energy eigenstate (including the ground state) is twofold degenerate for this spatial interface.} Hence, from this field theory perspective, we even have a \emph{strong} zero mode \cite{Fendley16}. \cref{fig:edge} displays configurations which belong to $\mathcal H_+$ and $\mathcal H_-$, respectively.

In this section we have shown that models exhibiting QSPT phenomena naturally give rise to quasi-phases and IR-necessary phase transitions. Moreover, this transition line terminates in a very interesting way, in a $c=\tfrac32$ orbifold CFT at the edge of a bubble described by a free-boson CFT. In the next section we display a rather different model exhibiting RSPT phenomena that has a similar type of phase diagram.

\section{RSPT phenomena in a spin-$\tfrac12$ chain}\label{sec:RSPT}
In this section, we forgo the anomalous low-energy theory and robust quotient structure of the previous section, but see that many of the structures observed above can be seen in the vicinity of an SPT transition. In particular, we explain how RSPT phenomena also give rise to quasi-phases and an IR-necessary phase transition.

Our starting point is the bond-alternating XX chain $H_{\delta XX}$ from \eqref{eq:HXX}, containing two distinct SPT phases separated by a phase transition.  Adding a perturbation breaking its $U(1)$ symmetry down to its $\mathbb{Z}_3$ subgroup causes the topological distinction between the two SPT phases to disappear. 
We described the symmetry fractionalization of $O(2) = U(1)\rtimes \mathbb{Z}_2$ in the ground state in \cref{sec:xxproj}. The two-dimensional representation of the $S_3$ subgroup is generated by 
\begin{align}
V_{2\pi/3} =     \left(
\begin{array}{cc}
 e^{-i \pi/3} & 0 \\
 0 & e^{i \pi/3} \\
\end{array}
\right) \qquad V_\mathcal{C}= \left(
\begin{array}{cc}
 0 & 1  \\
 1 & 0 \\
\end{array} 
\right)\ . \label{eq:S3irrep2d}
\end{align}
We can freely redefine $V_{2\pi/3} \rightarrow - V_{2\pi/3}$, yielding a two-dimensional \emph{linear} irrep of $S_3$ (see \cref{footnote:reps}). Thus $\delta=1$ is in the trivial phase for the $S_3$ subgroup (i.e., the same phase as $\delta=-1$). The non-Abelian linear irrep motivated the RSPT nomenclature in previous work \cite{OBrien20}.

\begin{figure}
\begin{center}
\begin{subfigure}{\textwidth}
\centering
\begin{tikzpicture}
\draw[thick,->] (-4,0) -- (4,0) node[anchor= west]{$\delta$};
\draw[thick,->] (-4,0) -- (-4,1) node[anchor= east]{$\mathcal{J}$};
\node (node1) at (-0,-0.) {$\bullet$}; 
\node (node2b) at (0.,0.5) {$c=1$}; 
\node (node2b) at (0.,-0.5) {$0$}; 
\node (node3) at (2.,0.5) {Trivial phase}; 
\node (node4) at (-2.,0.5) {Trivial phase}; 
\end{tikzpicture}
\caption{\small A hypothetical homogeneous perturbation $\mathcal{J}$ drives us into a gapped phase and so corresponds to a relevant perturbation at the $c=1$ transition. We see  below that this is \emph{not} the phase diagram realized by the model \eqref{eq:HXXJ}. }
\end{subfigure}
\begin{subfigure}{\textwidth}\centering
\begin{tikzpicture}
\draw[thick,->] (-4,0) -- (4,0) node[anchor= west]{$\delta$};
\draw[thick,->] (-4,0) -- (-4,1) node[anchor= east]{$J$};
\draw[very thick,,neworange!90!red] (0,0) -- (0,1);
\node (nodeq) at (-2,1.5) {?}; 
\node (nodeq) at (2,1.5) {?}; 
\node (nodeq) at (0,1.5) {?}; 
\node[neworange!90!red] (node2b) at (0,-.5) {$c=1$}; 
\node (node3) at (2.,0.5) {Quasi-phase 2}; 
\node (node4) at (-2.,0.5) {Quasi-phase 1}; 
\node (node1) at (0,2.5) {}; 
\end{tikzpicture}
\caption{\small  The model \eqref{eq:HXXJ} for small values of $J$. The quasi-phases are locally separated by a transition in the extended phase diagram.}
\end{subfigure}
\end{center}
\caption{\small  Example phase diagrams in the vicinity of $H_{\delta XX}$.}\label{fig:XXJ}
\end{figure}
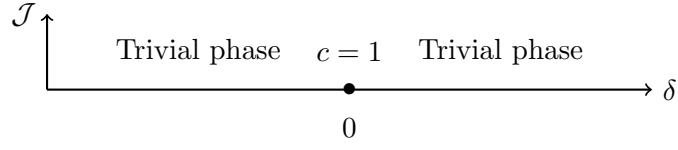
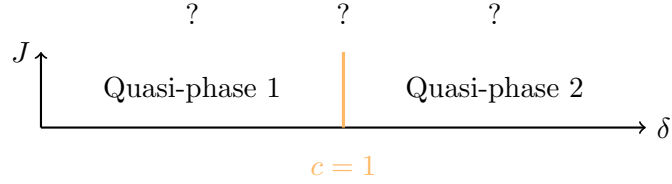
It is precisely this case that is of interest---including a perturbation that breaks $U(1)$ down to $\ZZ_{2m+1}$ removes the distinction between the $\delta=\pm1$ phases. The simplest example, leaving the $\ZZ_3$ subgroup generated by $\mathcal S=\exp(\frac{i \pi}{3}\sum_n Z_n)$, is given by including a charge-3 perturbation
\begin{align}
H = H_{\delta XX} - J \sum_n\left( \sigma^+_{n}\sigma^+_{n+1}\sigma^+_{n+2} + h.c. \right) + \cdots , \label{eq:HXXJ}
\end{align}
where the `$\cdots$' can contain other $S_3$-symmetric terms.  
For $J>0$ the SPT phases either side of $\delta=0$ are no longer topologically distinct, and the simplest conceivable phase diagram would be \cref{fig:XXJ}(a), where we have a single critical point inside the trivial phase. However, this does not occur. We can see this by analyzing the vicinity of the critical point at $\delta=J=0$, as described in \cref{sec:boson}.

 On these fields we have a faithful action of our UV symmetry, and our $S_3$ acts as follows
\begin{alignat}{2}
\Phi &\overset{\mathcal{S}}{\rightarrow} \Phi+ 2\pi/3  \qquad&\Phi &\overset{\mathcal{C}}{\rightarrow} -\Phi\nonumber\\
\Theta&\overset{\mathcal{S}}{\rightarrow} \Theta  &  \Theta&\overset{\mathcal{C}}{\rightarrow}- \Theta \ . \label{eq:bosonS3}
\end{alignat}
As before, translation takes $e^{\pm i\widehat{m}\Theta}\rightarrow (-1)^{\widehat{m}}e^{\pm i\widehat{m}\Theta}$ and leaves $\Phi$ invariant.

From this, we see that at the XX  point the $S_3$ symmetry-allowed primary operators are $\cos(\widehat{m}\Theta)$ and $\cos(3 m \Phi)$ for integer $\widehat{m},m$. The only relevant operator is $\cos(\Theta)$, which couples to $\delta$. This is odd under translation symmetry as needed, and $\Theta=0,\pi$ correspond to the two dimerization patterns.
All charge-three operators in the theory, that will be generated by tuning $J$, are irrelevant, and  so we expect an extended (possibly terminating) critical line.

However, the RG-irrelevance of the perturbation is a red herring. In fact, \cref{fig:XXJ}(a) is excluded even if we tune $K$ to make $\cos(3 \Phi)$ relevant (e.g., by tuning $\Delta$ in \cref{eq:HXX}). This operator drives us into a $\mathbb{Z}_3$ symmetry-breaking phase with three ground states. This phase would itself need to terminate to allow us to connect the two sides of the transition line.
The salient fact is that this CFT and its nearby symmetric gapped phase diagram are \emph{unable} to give a gapped path between the $\delta<0$ and $\delta>0$ regimes. Starting at $\Theta=0$ and taking diagonal $\cos(\widehat{m}\Theta)$ terms we are back in the $U(1)$-symmetric case and we know there is no gapped path to $\Theta=\pi$ either from SPT considerations or the argument given in \cref{sec:boson}. Meanwhile, any path relying on a $\cos(3m\Phi)$ must enter a regime dominated by this term, and this will have a ground state degeneracy, implying a gapless point along the path. 

Because of the parametric stability of a phase transition or intermediate phase separating the $\delta=\pm1 $ regions, we understand that the two sides of the transition realize distinct quasi-phases, as in \cref{fig:XXJ}(b) (we will return to this point with our more precise notion of quasi-phase in \cref{sec:framework}). Nevertheless we know they can be connected, in the remainder of this section we identify an $S_3$-symmetric gapped path and understand how the critical line terminates in the phase diagram. 

\subsection{Building a gapped path between the $S_3$ quasi-phases}\label{sec:gappedS3path}

The spin-$\tfrac12$ chain of \cref{sec:bondalt} exhibits SPT behavior, and so lacks an $O(2)$-symmetric gapped path between the $\delta=\pm1$ regimes. An intuitive reason is that a spatial interface between the two regimes carries a projective irrep of the $O(2)$ symmetry. The smallest such interface is a decoupled qubit that cannot be gapped out while respecting the symmetry. 
If instead there had existed a gapped $O(2)$-symmetric interface, one could use that to sweep across the system and build a gapped path, or, more elegantly, to nucleate pairs of domain walls throughout the system to toggle the regime. 

Now, if breaking $O(2)$ down to $S_3$ makes the phases the same, it should also be possible to gap out the domain wall mode. However, the single decoupled qubit still carries a non-trivial two-dimensional irrep of $S_3$ (described above in \cref{eq:S3irrep2d}) and hence cannot be gapped out. This is a hint that there remains some obstruction to immediately gapping out the transition. The fact that this irrep is not projective is important. If we take three decoupled sites, also a valid interface between $\delta>0$ and $\delta<0$, then the product of three $O(2)$ projective reps remains projective and cannot contain a singlet. However, a product of three two-dimensional linear irreps of $S_3$ \emph{does} contain a singlet, and that allows us to gap out this three-site domain wall.
Indeed, a minimal $S_3$-symmetric domain wall is $\frac{1}{\sqrt{2}}(\ket{\uparrow\uparrow\uparrow}+\ket{\downarrow\downarrow\downarrow})$; we will represent this as a trimer. This is the gapped ground state of $-(\sigma^+_{1} \sigma^+_{2} \sigma^+_{3}+h.c.)$. A natural candidate for a gapped path would be to interpolate from each of the two quasi-phases to a gapped state consisting of a product of these trimers.  

\begin{figure}[h!]
\centering\includegraphics[]{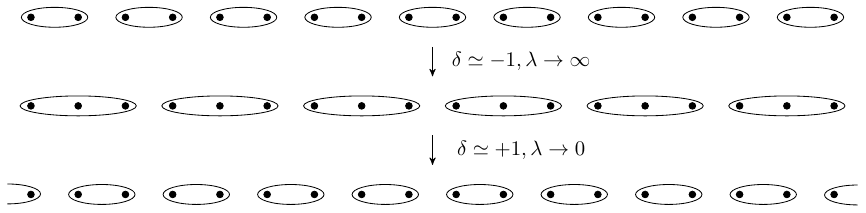}
    \caption{\small  Limiting ground states for the gapped path described in \cref{sec:gappedS3path}.}
    \label{fig:gappedpath}
\end{figure}

This motivates us to consider the model
\begin{align}
H(\delta,\lambda) &= H_{\delta XX} - \lambda \sum_n \left(\sigma^+_{3n+1} \sigma^+_{3n+2} \sigma^+_{3n+3} + h.c. \right) \ .\label{eq:HXXJ2}
\end{align}
For $\lambda \neq 0$, the model is reduced to having an $S_3$ symmetry generated by $\prod_n X_n$ and $e^{\frac{2\pi i}{3} \sum_n Z_n /2}$.
 In the large $\lambda$ limit the Hamiltonian is indeed gapped with a product of trimers as the ground state. The model is particularly simple at $\delta=\pm1 $, where it becomes a chain of decoupled six-site unit cells. A first attempt at (part of) a gapped path would be to simply fix $\delta =-1 $ and tune $0\leq \lambda \leq \infty$. However, this path has a level crossing\footnote{This value is found from diagonalizing the Hamiltonian on a single six-site block---this crossing occurs since within the block, the operators $Z_1Z_2$ and $Z_5Z_{6}$ commute with $H(-1,\lambda)$, and the two limiting states have different quantum numbers.} at $\lambda = 8\sqrt{2}$. This is a fine-tuned issue, and we show in \cref{app:pert} that an infinitesimal shift in $\delta$ will avoid the level-crossing and allow us to connect $\delta =-1 $ to the product of trimers. At large $\lambda$ we can tune $\delta$ to $1$ without a transition and then reduce $\lambda\rightarrow 0$ (this is the same problem as before up to a single-site translation). Thus we have a gapped path between the $S_3$ quasi-phases as described in \cref{fig:gappedpath}.

\subsection{Phase diagram of $S_3$-perturbed XX chain}
Let us now consider the full phase diagram of the Hamiltonian \eqref{eq:HXXJ2}. We have argued the stability of the $c=1$ criticality in the vicinity of $\delta=\lambda=0$, and that we have a gapped path connecting $\delta=\pm1$ in the full phase diagram. The criticality must therefore terminate within this phase diagram. 

\begin{figure}\centering
\includegraphics[]{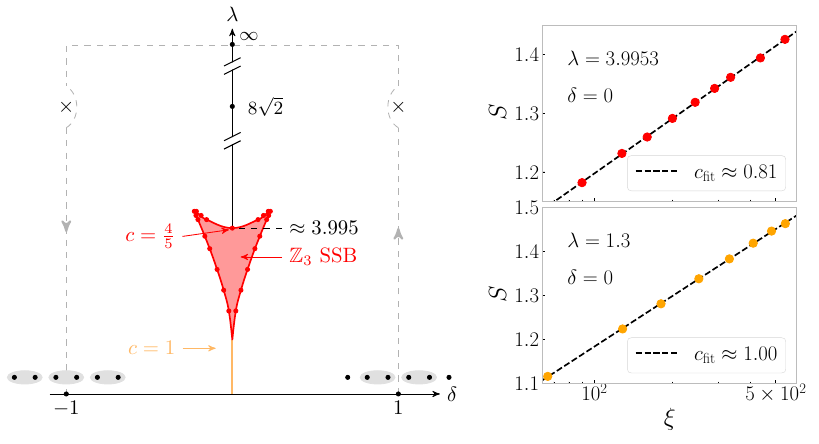}
    \caption{\small Ground state phase diagram for the spin-$\tfrac12$ chain \eqref{eq:HXXJ2}. For small $\lambda$ we have quasi-phases separated by a $c=1$ transition at $\delta=0$. This opens up at a BKT transition into a closed region of $\mathbb{Z}_3$ symmetry-breaking phase, which ends with a Potts transition along the $\delta=0$ line at $\lambda \simeq 3.995$.   The BKT transition is between $\lambda =1.3$ (see inset for $c=1$ entanglement scaling) and $\lambda =1.4$ where we find a symmetry-breaking ground state. The nature of the transition out of the symmetry-breaking phase for $\delta\neq0$ is left to future work. The dashed lines indicate the gapped path between quasi-phases. They avoid points indicated by `$\times$' with infinite degeneracy caused by the level crossing in each unit cell for the decoupled limit. }
\label{fig:RSPTphasediagram}
\end{figure}

Using DMRG we find the ground state phase diagram given in \cref{fig:RSPTphasediagram}. Along the $\delta=0$ line, we observe that for small $\lambda$ we have the expected $c=1$ transition. The system enters a $\mathbb{Z}_3$ symmetry-breaking phase via a BKT transition for $1.3<\lambda<1.4$. We must have a transition out of this phase for consistency with the gapped path at large $\lambda$, and this occurs via a Potts transition at $\lambda \simeq 3.995$. For non-zero $\delta$, the $c=1$ line gaps out into the trivial or RSPT quasi-phase. 
The symmetry-breaking region persists to finite $\delta$ before entering the symmetric phase, where the nature of this transition is left to future work. Of particular interest are the apparent cusps where the boundaries of the symmetry-breaking region meet. For  $\lambda > 3.995$ the symmetry-breaking phase persists for a range of $\delta$ away from zero, and thus we have a re-entrant phase diagram as we tune $\delta$ for a small range of $\lambda$ (we observe regions of symmetry-breaking phase up to $\lambda \simeq 4.4$).

\subsection{Field theory for the full phase diagram}
At $\delta=\lambda=0$ we have the $c=1$ theory at $K=1$, as described above. While this field theory is suitable to describe the phase diagram near $\delta=\lambda=0$, we would like a description that captures the full phase diagram. Recall that the gapped ground state in the large $\lambda$ limit is a product of $S_3$-symmetric trimers (see \cref{fig:gappedpath}). In this state we have that charge-one and charge-two correlators vanish, e.g., $\langle \sigma^+_n \sigma^{-}_m\rangle =\langle \sigma^+_n\sigma^+_{n+1} \sigma^{-}_m\sigma^{-}_{m+1}\rangle= 0$. However, the charge-three correlator has long-range order $$\langle \sigma^+_{3n+1}\sigma^+_{3n+2}\sigma^+_{3n+3} \sigma^-_{3m+1}\sigma^-_{3m+2}\sigma^-_{3m+3}\rangle= 1/4.$$
This indicates that at large $\lambda$, the fields $e^{\pm i\Phi}$ and $e^{\pm 2 i \Phi}$ correspond to infinitely massive excitations, while $\cos( 3 \Phi)$ is condensed. This is not the generic behavior of a compact boson in this phase, rather it is suggestive that the $\mathbb{Z}_3$-charge-one and -two local operators are projected out in the limit. Moreover, we must resolve the tension of the $\cos( 3 \Phi)$-condensed phase having three vacua while our spin chain has a unique vacuum. We will see how this naturally leads us to describe the phase diagram via an orbifolding of the free boson and a $\mathbb{Z}_3$ Potts field.

\subsubsection{Orbifold description of the phase diagram}
To arrive at the construction, we use the following physical intuition that was pre-empted in the discussion of the Ising orbifold above. Charge-one and charge-two operators should have a large mass, while charge-three operators should be light. A simple construction would be to dress our charged operators with another field that gives rise to this mass. If we write our dressed charge-one operator as $\sigma e^{i \Phi}$, naive operator products give us $\sigma^2 e^{2i \Phi}$ and $\sigma^3 e^{3i \Phi}$ as local operators. Then if $\sigma$ ascribes large mass, the simplest way we can have a light charge-three operator is to impose that $\sigma^3 = 1$. I.e., we make $\sigma$ a $\mathbb{Z}_3$ Potts field; this has a natural $\mathbb{Z}_3$ symmetry action. To impose the dressing, we gauge the diagonal $\mathbb{Z}_3$ which effectively glues $\sigma$ to $e^{i \Phi}$ in all local operators. As a byproduct, this gauge symmetry solves our second problem for free---the three vacua of $\cos( 3 \Phi)$ are gauge equivalent and describe the same physical state.

Having motivated the construction, let us unpack it. Define a $\mathbb{Z}_3$ Potts field $\sigma$ with dual field $\mu$, and let $\omega =e^{2\pi i/3}$. We have that $\sigma^3 = \mu^3=1$. We fix a consistent Potts $S_3$ action on these fields as
\begin{alignat}{2}
\sigma &\overset{\mathcal{S}_P}{\rightarrow} \omega^{-1} \sigma  \qquad&\sigma &\overset{\mathcal{C}_P}{\rightarrow} \sigma^\dagger\nonumber\\
\mu&\overset{\mathcal{S}_P}{\rightarrow} \mu  &  \mu&\overset{\mathcal{C}_P}{\rightarrow}\mu^\dagger \ .
\end{alignat}
At criticality, these operators have scaling dimension $\Delta_\sigma=\Delta_\mu = 2/15$. The order-disorder phase transition is tuned by the field $\varepsilon$, with dimension $\Delta_\varepsilon=4/5$; $\varepsilon$ is uncharged under the $S_3$ action. For further details see \cite{Mong14}. 

To describe our phase diagram, as outlined above, we construct an orbifold with the free boson fields by gauging the diagonal $\mathbb{Z}_3$. This acts as
\begin{alignat}{2}
\mathrm{Gauge~symmetry:} \qquad\sigma ~&{\rightarrow} ~\omega^{-1} \sigma \qquad    &
\Phi~ &{\rightarrow} ~\Phi +2\pi/3 \ . \label{eq:gauge}
\end{alignat}
 This leaves mutually local\footnote{Note we implicitly have fixed which twist operator is which so that $\mu e^{i  \Theta/3}$ is local rather than  $\mu^2 e^{i  \Theta/3}$.} operators $V_{m,\widehat{m}}=
\sigma^m e^{i m  \Phi} \times 
\mu^{\widehat{m}}  e^{i \widehat{m} \Theta/3}$, where $m,\widehat{m}$ are integers. The residual lattice $\mathbb{Z}_3$ then acts as $V_{m,\widehat{m}}\overset{\mathcal{S}}{\rightarrow} \omega^mV_{m,\widehat{m}}$ while charge conjugation (which acts diagonally on the Potts and boson) takes 
$V_{m,\widehat{m}}\overset{\mathcal{C}}{\rightarrow} V_{-m,-\widehat{m}}$. Our $S_3$-symmetry allowed operators are hence of the form $(V_{3m,\widehat{m}}+V_{-3m,-\widehat{m}})$. The operators $V_{m,\widehat{m}}$ carry a charge $(-1)^{\widehat{m}}$ under translation by an odd number of sites. The corresponding orbifold CFT has central charge $c=\tfrac{9}{5}$, but this does not appear in our phase diagram, rather various flows from it appear as we discuss now.

To recover our previous free boson description at $\lambda=0$, we suppose our Potts field is in a phase with long-range order $\langle \sigma^\dagger(x) \sigma(y)\rangle$, and exponentially decaying dual correlator $\langle \mu^\dagger(x) \mu(y)\rangle$; i.e., the Higgs phase. This gives us algebraically decaying symmetry-allowed fields of the form  $(V_{3m,\widehat{3m}}+V_{-3m,-\widehat{3m}})$. At $K=1$, the sole relevant operator is $m=0$, $\widehat{m}=1$, giving us back $\cos(\Theta)$ as the operator driving the transition. The two vacua either side of the transition  correspond to $\Theta=0, \pi$ as before. 

Now, as we tune $\lambda$ we add a generic three-site-translation invariant $S_3$-symmetric perturbation that couples to the marginal operator for the free boson and the Potts $\varepsilon$ field. It also couples to $\cos(3\Phi)$ and $\cos(2\Theta)$ (along with higher-order harmonics that are more irrelevant). The simplest possibilities we can expect to see are BKT transitions as we tune $K$, or a Potts order-disorder transition. In our particular phase diagram, as we increase $\lambda$ we tune $K$ and the first transition is where $\cos(3\Phi)$ becomes relevant. Along the $\delta=0$ line, we then have a regime where $\cos(3\Phi)$ flows to three distinct vacua, giving the $\mathbb{Z}_3$ symmetry-breaking phase. Note that as $\sigma$ has long-range order, these vacua are not related by gauge symmetry, due to the aforementioned `Higgsing', and are genuinely distinct\footnote{When both $\sigma=\omega^k$ and $\Phi=2\pi l/3 $ are ordered, we naively get nine states for $0\leq k,l\leq 2$. However, the gauge symmetry \eqref{eq:gauge} relates these in sets of three, to give three physical vacua.}. For $\delta\neq0$ we break the odd-translation-invariance and couple to $\cos(\Theta)$, giving two competing relevant operators. For sufficiently large $\delta$ this drives us out of the symmetry-breaking region back into the trivial gapped phase. The two competing operators, $\cos\Theta$ and $\cos(3\Phi)$, are precisely those that lead to a flow from a free boson to the $c=\tfrac45$ 3-state Potts critical point, once the boson stiffness is tuned appropriately \cite{Lecheminant02}. We thus expect that the edge of the bubble is described by this Potts CFT. Our numerical results are consistent with this expectation, but not definitive; we leave the precise characterization of this transition to future work. 

As we continue to increase $\lambda$, we eventually tune the sign of $\varepsilon$, and enter the Potts disordered phase through a $c=\tfrac{4}{5}$ criticality. This is expected by construction: since $\sigma$ is now exponentially decaying, we have ascribed the mass to our charge-one and -two boson fields. We also have that in the low-energy theory all three vacua are gauge equivalent under \eqref{eq:gauge}, and we return to a unique vacuum. We can also understand the gapped path in this regime. Our perturbation that tunes $\delta$ breaks odd-translation symmetry and so corresponds to $V_{m,\widehat{m}}$ with odd $\widehat{m}$. The dominant symmetric operator generated by this perturbation is thus $(\mu e^{i\Theta/3} +h.c.)$, which as $\mu$ is ordered looks like $\cos(\Theta/3)$. We thus have an effective free boson description in terms of fields $\Phi' = 3\Phi$ and $\Theta'=\Theta/3$, where the gauge redundancy can be absorbed into declaring these fields to be $2\pi $ periodic. Our gapped path then corresponds to the fact recalled above that the $(\cos(\Theta'),\cos(\Phi'))$ phase diagram is gapped away from the origin for any value of $K$ \cite{Lecheminant02}.

\subsubsection{Accessing the $c=9/5$ orbifold on the lattice}
To further validate our field theory analysis we can choose a new perturbation that removes some of the possibilities above. In particular, if we perturb our transition at $\delta=\lambda=0$ by some $U(1)$-symmetric $H'$, then we will never generate vertex operators involving $\Phi$. To exclude the $\Theta$-vertex operators, we use a finite perturbation that shifts $K$ to a regime where these are all irrelevant. 

\begin{figure}
\begin{center}
\includegraphics[scale=0.4]{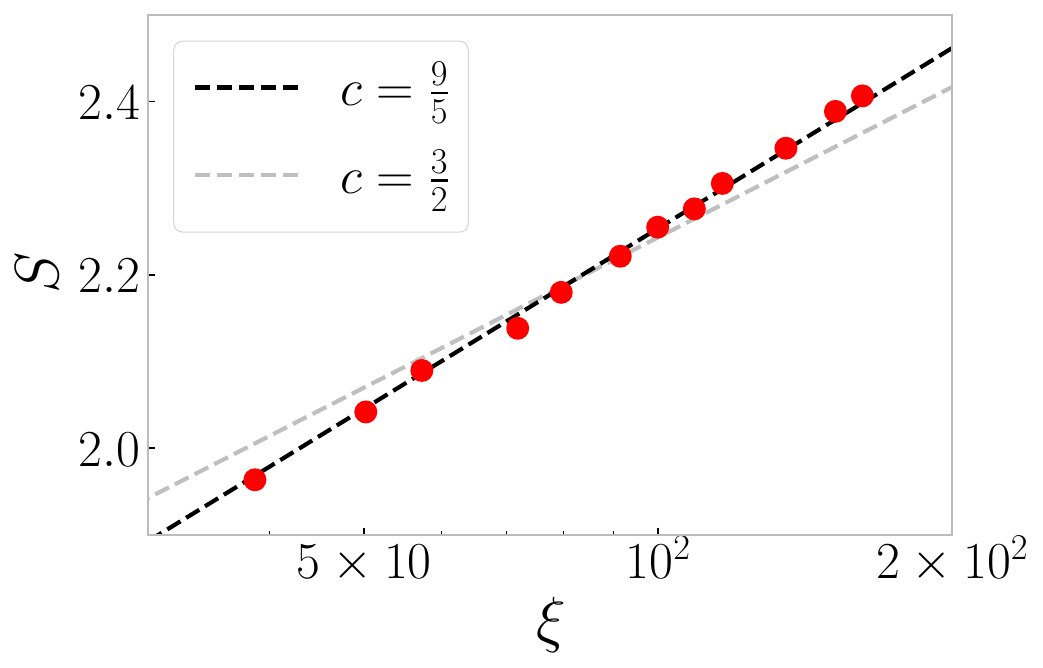}
\caption{\small Finite entanglement scaling for the Hamiltonian \eqref{eq:HXXJ3} at $\mu_c$. The lines $\frac{c}{6} \log \xi $ for $c=3/2$ and $c=9/5$ are given to guide the eye, and we see good agreement with the latter. }
\label{fig:c95}
\end{center}
\end{figure}

We take our two-parameter Hamiltonian as follows
\begin{align}
\tilde{H}(\Delta,\mu) &= -\sum_n(X_nX_{n+1}+Y_nY_{n+1}  -\Delta Z_nZ_{n+1}) -  \mu \sum_n \left(T_{n}^+T_{n+1}^- + h.c. \right) ,\label{eq:HXXJ3}
\end{align}
where $T_n^\pm=\sigma^\pm_{3n+1} \sigma^\pm_{3n+2} \sigma^\pm_{3n+3}$.
We first fix $\Delta =-0.95$ so that all $\Theta$ primaries are irrelevant. Then by tuning $\mu$ we observe a $c=1$ criticality that passes through the above $c=9/5$ multicritical point at $\mu_c \simeq 3.35$. We verified this by finite entanglement scaling in \cref{fig:c95}.

\subsection{Edge modes}
\label{sec:RSPTedge}
In the QSPT case, we saw that edge modes were unavoidable as long as we remained in the effective $c=1$ field theory that described large $D$. Let us now study for the RSPT case whether an interface between $\Theta=0$ and $\Theta=\pi$ always needs to have an edge mode (when one is restricted to the effective $c=1$ Hilbert space where the Potts field is in the Higgs phase).

Our quantum number is now $\hat Q = e^{\frac{i}{3} \int_{-\infty}^\infty \partial_x \Theta \mathrm d x}$. Together with the asymptotic conditions mentioned before (i.e., $\lim_{x \to -\infty} \Theta(x) = 0 \mod 2\pi$ and $\lim_{x\to +\infty} \Theta(x) = \pi \mod 2 \pi$), our Hilbert space now splits into three sectors: $\mathcal H = \mathcal H_1 \oplus \mathcal  H_\omega \oplus \mathcal H_{\bar \omega}$, where the quantum number associated to $\mathcal H_\alpha$ is $Q = -\alpha$. Indeed, using the same profile as above, we now obtain
\begin{align}
Q = e^{i(2m+1)\pi/3} = \left\{ \begin{array}{cl}
- \bar \omega &\textrm{if }m=0\mod 3 \\
-1 &\textrm{if }m=1\mod 3 \\
- \omega &\textrm{if }m=2\mod 3
\end{array}\right. \ .
\end{align}

\begin{figure}
\centering
\includegraphics[]{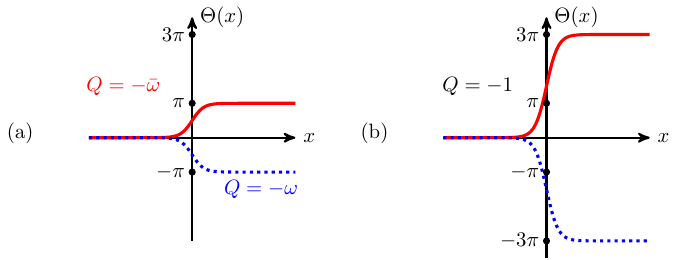}
\caption{\small Interfaces between $\Theta=0$ and $\Theta=\pi\sim 3\pi$, along with their $\mathbb{Z}_3$ charges. (a) forms an $S_3$ doublet, corresponding to an edge degeneracy while (b) forms an $S_3$ singlet with no edge mode.  }\label{fig:RSPTedge}
\end{figure}

Now charge-conjugation has the property $\mathcal{C} Q \mathcal{C} = \bar Q$. Hence, $\mathcal H_\omega \oplus \mathcal H_{\bar \omega}$ transforms under the two-dimensional irrep of $S_3$ (so there are degenerate edge modes for these interfaces), whereas $\mathcal H_1$ is a singlet. At the very least, there are no strong zero modes here. Moreover, if the ground state is in $\mathcal H_1$, then we will not have a necessary edge mode!
Examples of interfaces and their corresponding quantum numbers are given in \cref{fig:RSPTedge}. Note that the last two (which are related by charge-conjugation $\mathcal{C}$) have the same quantum number and can thus hybridize to give a unique symmetric eigenstate. 

We conclude that edge modes can appear for RSPT models, but they depend on the boundary termination. If, for a concrete model, our boundary termination selects a doublet  then we have parametric stability of the doublet until it crosses a boundary transition into the singlet (analogous to the QSPT case). However, we may have a boundary termination that selects the singlet. We will see examples of this choice in a lattice model below. If we consider RG flow in the vicinity of the CFT, we note that the non-degenerate interface mode requires a higher winding, which we expect to be energetically disfavored. This suggests that perturbatively close to the compact boson CFT, one will not flow to this interface. This ties into the more general discussion in \cref{sec:framework}.

\section{Criticality between the Potts and anti-Potts chain within an $S_3$ symmetry-breaking phase}\label{sec:potts}
One of the motivations of the present work is an intriguing phase diagram of Ref. \cite{OBrien20}. The authors considered an $S_3$-symmetric lattice model in the neighborhood of a $U(1)$-symmetric clock model \cite{Vernier19}. This phase diagram appears to contain four phases, two symmetric and two symmetry-breaking, all meeting at the $U(1)$-symmetric multicritical point. In this section we show that the two symmetry-breaking regions, containing the representative fixed-point Potts and `not-$A$' Hamiltonians, are in the same $S_3$-breaking phase, such that the critical point between them is along a terminating critical line. (The symmetric regions are related by Kramers-Wannier duality, and we discuss the corresponding RSPT physics in \Cref{sec:pottsRSPT}.)
To see this, we construct an $S_3$-symmetric gapped path connecting the two models. This is another manifestation of the quasi-phase concept, this time with respect to symmetry-breaking rather than symmetric phases. A similar phenomenon was realized in a symmetry-breaking classical lattice model in Ref.~\cite{Prakash25}.

\subsection{The tale of two Potts models}\label{sec:twoPotts}
We consider Hamiltonians where each site has a three-dimensional Hilbert space for which we have the clock matrices $\sigma$ and $\tau$:
\begin{align}
\sigma = \begin{pmatrix}
0 & 1 & 0 \\
0 & 0 & 1 \\
1 & 0 & 0 \end{pmatrix}
\qquad \textrm{and} \qquad
\tau =  \begin{pmatrix}
1 & 0 & 0 \\
0 & \omega  & 0 \\
0 & 0 & \omega^2\end{pmatrix} \qquad \textrm{where } \omega = e^{2\pi i /3}. \label{eq:sigmatau}
\end{align}
These operators satisfy $\sigma \tau = \omega \tau \sigma$. We will consider systems which have a unitary symmetry $S_3 \cong \mathbb Z_3 \rtimes \mathbb Z_2$ generated by the $\mathbb Z_3$ permutation $\mathcal S$, and the $\mathbb Z_2$ charge conjugation $\mathcal C$
\begin{align}
    \mathcal S &= \prod_n \tau_n & \mathcal C &= \prod_n R^x_n \qquad \textrm{where } \quad R^x
= e^{i\pi  S^x} =  \begin{pmatrix}
-1 & 0 & 0 \\
0 & 0  & -1 \\
0 & -1 & 0\end{pmatrix}.
\end{align}
Notably, they act as
\begin{align}
    \mathcal C \sigma_j \mathcal C &= \sigma_j^\dagger, & \mathcal C \tau_j \mathcal C &= \tau_j^\dagger,\nonumber\\
    \mathcal S^\dagger \sigma_j \mathcal S &= \omega \sigma_j, & \mathcal S^\dagger \tau_j \mathcal S &= \tau_j.
\end{align}
Lastly, we will denote the normalized eigenstates of $\sigma$ as $|A\rangle$, $|B\rangle$, $|C\rangle$, with respective eigenvalues $1,\omega,\bar \omega=\omega^2$. Explicitly,
\begin{align}
    \ket{A} &= \frac{1}{\sqrt{3}}(\ket{0}+ \ket{1}+\ket{2}),\nonumber\\
    \ket{B} &= \frac{1}{\sqrt{3}}(\ket{0}+   \omega \ket{1}+ \bar \omega \ket{2}),\nonumber\\
    \ket{C} &= \frac{1}{\sqrt{3}}(\ket{0}+ \bar \omega \ket{1}+ \omega \ket{2}).
\end{align}
Note that we align notation with \cite{OBrien20}; the alternative notation $X_j = \tau_j^\dagger$ and $Z_j = \sigma_j$ is also frequently used.
\subsubsection{The three-state Potts chain: $H_A$}
The usual Potts chain at its ferromagnetically ordered fixed point has Hamiltonian
\begin{align}
{ H_A = - \sum_n \left( \sigma_n^\dagger \sigma_{n+1}^{\vphantom \dagger} +h.c. \right) } \; . \label{eq:A}
\end{align}
This has three degenerate ground states: $|A\rangle^{\otimes N}$, $|B\rangle^{\otimes N}$ and $|C\rangle^{\otimes N}$. It will also prove to be useful to consider its formulation in terms of spin-1 operators (defined in Eq.~\eqref{eq:spin1operators}). In this notation, we may substitute $\sigma = \left( s^+ \right)^2 + s^-$ to obtain\footnote{For completeness, the dictionary between $\sigma,\tau$ and spin-1 variables is
\begin{align*}
S^z &= \frac{1}{\sqrt{3}i} (\tau - \tau^\dagger) &R^z &= e^{\frac{\pi}{\sqrt{3}} (\tau - \tau^\dagger)} =-e^{\frac{\pi i}{3} (1+\tau + \tau^\dagger)}\\
s^+ &= \frac{S^+}{\sqrt{2}} = \frac{1}{3}(2 - \omega \tau -\bar \omega \tau^\dagger)\sigma^\dagger& (s^+)^2&= \frac{(S^+)^2}{2} =\frac{1}{3}\sigma(1+\omega \tau +\bar \omega \tau^\dagger).
\end{align*}}
\begin{align}
H_A = - \underbrace{ \sum_n \left( s^+_n s^-_{n+1} + \left(s^+_n\right)^2 \left(s^-_{n+1}\right)^2 +h.c. \right)}_{= \textrm{ $O(2)$-symmetric part}} 
- \underbrace{\sum_n \left( s^+_n \left( s^+_{n+1} \right)^2  + \left( s^+_{n} \right)^2 s^+_{n+1} + h.c.\right)}_{= \textrm{ $S_3$-symmetric part}}\ , \label{eq:Aspin}
\end{align}
where the $O(2) \cong U(1) \rtimes \mathbb Z_2$ symmetry is generated by $S^z_{\rm{tot}}=\sum_n S^z_n$ and $\mathcal C$, which contains $S_3$ as its subgroup.

Just as $\sigma$ serves as an order parameter in the Potts chain, one may also use $s^-$ as an order parameter. This is because
\begin{align}
    \bra{A}s^- \ket{A} &=\frac{2}{3}, & \bra{B}s^- \ket{B} &=\frac{2}{3}\omega, & \bra{C}s^-\ket{C} &=\frac{2}{3}\bar\omega \ .\label{eq:order}
\end{align}

\subsubsection{The anti-Potts chain: $H_\notA$}

Ref.~\cite{OBrien20} proposed an intriguing second version of the Potts chain, which seems to be in a different phase of matter. Its ground states are in some sense the complete opposite of the usual Potts chain. More precisely, the ground states of this modified chain are
\begin{align}
\left( \frac{|B\rangle + |C\rangle }{\sqrt{2}} \right)^{\otimes N}, \quad 
\left( \frac{|C\rangle + |A\rangle }{\sqrt{2}} \right)^{\otimes N}, \quad 
\left( \frac{|A\rangle + |B\rangle }{\sqrt{2}} \right)^{\otimes N}. \label{eq:tempnotA}
\end{align}
One might refer to these ground states as `not-$A$', `not-$B$' and `not-$C$', and these are the ground states of the `not-$A$' Hamiltonian in Ref.~\cite{OBrien20}. For later convenience, we consider a Hamiltonian with slightly different ground states. While the states in Eq.~\eqref{eq:tempnotA} are not orthogonal for $N=1$, the following three are
\begin{align}
    \begin{split}
|\bar A \rangle &=R^z\ket{A}= \frac{1}{3}\Big( 2\left(|B\rangle + |C\rangle \right)- |A\rangle\Big)   ,\\
|\bar B \rangle &= R^z\ket{B} = \frac{1}{3}  \Big(2\left(|C\rangle + |A\rangle \right) - |B\rangle  \Big)  ,  \\
|\bar C \rangle &= R^z\ket{C} = \frac{1}{3}\Big( 2\left(|A\rangle + |B\rangle \right)- |C\rangle  \Big).
\end{split}
 \label{eq:notA}
\end{align}
Here
\begin{align}
\label{eq:singlesiteRz}
    R^z = e^{\pi i S^z} =  \begin{pmatrix}
1 & 0 & 0 \\
0 & -1  & 0 \\
0 & 0 & -1\end{pmatrix}
\end{align}
is a $\pi$-rotation around the $z$-axis.

Just as $|A\rangle,|B\rangle,|C\rangle$ are the eigenstates of $\sigma$, we can introduce $\bar \sigma$ as
\begin{align}
\bar \sigma = |\bar A\rangle \langle \bar A| + \omega |\bar B \rangle \langle \bar B | + \bar \omega | \bar C \rangle \langle \bar C | = R^z\sigma R^z  = \begin{pmatrix}
  0 & -1 & 0 \\
0 & 0 & 1 \\
-1 & 0 & 0  
\end{pmatrix}.
\end{align}
We then define the `anti-Potts' chain as follows:
\begin{align}
{ H_\textrm{$\bar A$} = - \sum_n \left( \bar \sigma_n^\dagger \bar\sigma_{n+1}^{\vphantom \dagger} + h.c. \right) } \;. \label{eq:HnotA}
\end{align}
Its three degenerate ground states are $\ket{\bar A}^{\otimes N}$, $\ket{\bar B}^{\otimes N}$ and $\ket{\bar C}^{\otimes N}$. In Appendix~\ref{appendix:pathnotA} we confirm that this model is adiabatically connected, via direct interpolation, to the closely related `not-A' Hamiltonian introduced in Ref.~\cite{OBrien20}. (This also follows indirectly, through the parent-Hamiltonian construction \cite{Schuch11}, from the path of states given in the next section.)

Written as a spin-1 chain we have
\begin{align}
H_\textrm{$\bar A$} = - \underbrace{ \sum_n \left( s^+_n s^-_{n+1} + \left(s^+_n\right)^2 \left(s^-_{n+1}\right)^2 +h.c. \right)}_{= \textrm{ $O(2)$-symmetric part}} 
\color{red} {+} \color{black} \underbrace{\sum_n \left( s^+_n \left( s^+_{n+1} \right)^2  + \left( s^+_{n} \right)^2 s^+_{n+1} + h.c.\right)}_{= \textrm{ $S_3$-symmetric part}}. \label{eq:notAspin}
\end{align}
I.e., compared to Eq.~\eqref{eq:Aspin}, we have simply flipped the sign of the non-$U(1)$-symmetric part.

We may also use $s^-$ as an order parameter to detect the three degenerate ground states. In this case, the order parameter is negated compared to the usual Potts model
\begin{align}
    \bra{\bar A}s^- \ket{\bar A} &=-\frac{2}{3}, & \bra{\bar B}s^- \ket{\bar B} &=-\frac{2}{3}\omega, & \bra{\bar C}s^-\ket{\bar C} &=-\frac{2}{3}\bar\omega \ . \label{eq:ordernot}
\end{align}
Indeed Ref.~\cite{OBrien20} noted that if we cube \cref{eq:ordernot} we get negative values, while \cref{eq:order} gives positive values. Nevertheless, such a sign change does not imply a phase transition in general.
\subsubsection{Criticality between $H_A$ and $H_\notA$ \label{subsec:HAHnotA}}

Consider the direct interpolation
\begin{align}
(1-\beta) H_A + \beta H_\textrm{$\bar A$}\ .
\end{align}
As pointed out by Ref.~\cite{OBrien20} (for the very closely related model), this model is critical at the half-way point $\beta=\tfrac12$ (with the sign of $\langle s_n^- \rangle$ in each symmetry-breaking ground state being different on the two sides). In fact, by comparing Eq.~\eqref{eq:Aspin} and Eq.~\eqref{eq:notAspin}, the critical Hamiltonian is given by the $O(2)$-symmetric part of the two fixed-point Hamiltonians:
\begin{align}
{ \frac{1}{2}\left( H_A + H_\textrm{$\bar A$} \right) = 
- \sum_n \left( s^+_n s^-_{n+1} + \left(s^+_n\right)^2 \left(s^-_{n+1}\right)^2 +h.c. \right) } \; . \label{eq:direct}
\end{align}
Thus, the critical point has $U(1)$ \textit{pivot symmetry}\cite{pivot,pivotDQCP} generated by $S^z_{\rm{tot}}$.
Numerically, we find that it has central charge $c=1$ via entanglement scaling. 
However, as we will now show, criticality can be avoided whilst preserving the symmetry.

\subsection{Paths connecting $H_A$ with $H_\notA$ avoiding the criticality}

\subsubsection{Unitary pivots}

The anti-Potts chain $H_\textrm{$\bar A$}$ (Eq.~\eqref{eq:HnotA}) can be obtained from the Potts chain $H_A$ (Eq.~\eqref{eq:A}) by conjugating by a $\pi$ rotation generated by $S^z_{\rm{tot}}$. Taking a continuous such rotation immediately gives a gapped path connecting the two Hamiltonians.
The $\mathbb{Z}_3$ symmetry-breaking ground states are threefold degenerate at every point along the path; however, the path explicitly breaks charge-conjugation symmetry. Hence, we cannot conclude that we have the same $S_3$ symmetry-breaking phase. 

However, note that $R^z = e^{\frac{\pi}{\sqrt{3}} (\tau - \tau^\dagger)} =-e^{\frac{\pi i}{3} (1+\tau + \tau^\dagger)}$, so we could instead take the rotation generated by $\frac{1}{3}\sum_n (\tau_n + \tau^\dagger_n+\textrm{const})$. This \emph{is} $S_3$-symmetric, but the rotation (except at high-symmetry points) now breaks time-reversal (complex conjugation in the $\tau$ basis). Moreover, in contrast with $S^z_{\rm{tot}}$, \cref{eq:direct} does not have the latter generator as a symmetry.

These pivot paths leave two aspects to be desired:
\begin{enumerate}
\item The latter path connects the $S_3$-breaking phases, but explicitly breaks time-reversal symmetry. Might this symmetry be important? Indeed, we noticed above that the order parameter in the anti-Potts model is negated. Is it the case that if we also preserve complex conjugation symmetry, then the two phases remain distinct?
\item They do not immediately shed light on what happens to the criticality we found above upon interpolating between $H_A$ and $H_\notA$. Apparently we learn the criticality is {unnecessary} in an $S_3$-symmetric phase diagram, but how (un)stable is it? How do we gap it out?
\end{enumerate}

Note the first path is somewhat redundant for the following reason. Pivoting through an angle $2\pi$ generated by $S^z_{\rm{tot}}$ gives us a loop within the $\mathbb{Z}_3$-symmetry-breaking phase. This loop connects $H_A$ back to $H_A$, and passes $H_{\bar A}$ at angle $\pi$. In fact, the generator $\frac{1}{3} S^z_{\rm{tot}}$ has the same effect, since \begin{align} e^{-i \frac{\pi}{3} S^z_{\rm{tot}}}\sigma e^{i \frac{\pi}{3} S^z_{\rm{tot}}}& = \bar\omega \bar\sigma \implies e^{-i \frac{\pi}{3} S^z_{\rm{tot}}} H_A e^{i \frac{\pi}{3} S^z_{\rm{tot}}} = H_{\bar{A}}\nonumber\\
e^{-i \frac{2\pi}{3} S^z_{\rm{tot}}}&=\mathcal{S}^\dagger \implies  e^{-i \frac{2\pi}{3} S^z_{\rm{tot}}} H_A e^{i \frac{2\pi}{3} S^z_{\rm{tot}}}=H_A \ .\end{align}
Thus our original unitary path of states takes us half-way around a closed loop in the space of Hamiltonians, but gets there by traversing the closed loop one and a half times. 

As an aside, since $e^{-i \frac{2\pi}{3} S^z_{\rm{tot}}}=\mathcal{S}^\dagger$ we have that $H_{\bar A}$ is half-way around a $\mathbb{Z}_3$ domain wall pump \cite{Hermele_talk,Kuno24,Manjunath26}. As we will explore in more detail in the Kramers-Wannier dual model below, we can use the results of \cite{Jones25b} to conclude that the half-way point \cref{eq:direct} has a $U(1)$ symmetry (as we already observed directly). This implies a phase transition along the line connecting $H_A$ to $H_{\bar{A}}$, since it is not consistent to break the $\mathbb{Z}_3$ subgroup of $U(1)$ without breaking the full group. Moreover, we also learn something about the neighborhood of this critical point in any nearby $S_3$-symmetric phase diagram. In the Kramers-Wannier dual model we have a charge pump and can use the symmetry arguments around \cref{eq:bosonS3} to rule out any $S_3$-symmetric gapped path in the neighborhood of the critical point; by duality the same applies here. We will see this for a particular perturbation below.

In \cref{sec:path}, we construct a path of product states which preserves both charge conjugation and complex conjugation symmetry. In \cref{sec:halfway} we consider a gapped Hamiltonian path with all these symmetries. Moreover, it sheds light on the criticality, and we see that we tune through a $c=\tfrac32$ multicritical point to terminate the $c=1$ line.

\subsubsection{A path of product states \label{sec:path}}
Consider a family of Hamiltonians with threefold degenerate ground states $\ket{A_\lambda}^{\otimes N}$, $\ket{B_\lambda}^{\otimes N}$, and $\ket{C_\lambda}^{\otimes N}$ where
\begin{align}
|A_\lambda\rangle &=  \sqrt{ \frac{3}{4(2+\lambda^2)}} \left( (1-\lambda)|A\rangle - (1+\lambda) |\bar A\rangle \right)\\
&\propto  (2-\lambda)\ket{A} - (1+\lambda) \left( \ket{B} + \ket{C} \right)  . \label{eq:MPSstate}
\end{align}
The symmetry-related partners are $|B_\lambda\rangle ^{\otimes N}= \mathcal S |A_\lambda\rangle^{\otimes N}$ and $|C_\lambda\rangle^{\otimes N} = \mathcal S |B_\lambda\rangle^{\otimes N}$.
Charge conjugation preserves $|A_\lambda\rangle^{\otimes N}$ and exchanges $|B_\lambda\rangle^{\otimes N}$ with $|C_\lambda\rangle^{\otimes N}$. 
While the above state and all the formulas we derive for it are well-defined for any real value of $\lambda$, we will focus on the window $-1 \leq \lambda \leq 1$, where this path connects the $A$ state ($\lambda=-1$) to the $\bar A$ state ($\lambda=1$).
The family of states has the property that $\mathcal R^z$ maps $\lambda \to - \lambda$, i.e., $ \mathcal R^z|A_\lambda\rangle = -|A_{-\lambda}\rangle$. 

The above path is normalized ($\langle A_\lambda|A_\lambda \rangle = 1$) and
\begin{align}
\langle A_\lambda | B_\lambda \rangle = \langle A_\lambda | C_\lambda \rangle= \langle B_\lambda | C_\lambda \rangle = \frac{\lambda^2-1}{\lambda^2+2}\ . \label{eq:overlap}
\end{align}
This shows that for large enough regions (more precisely, at least two sites), the three states remain linearly independent (at least for \emph{finite} $\lambda$), such that this is a physical path with a threefold ground state degeneracy (i.e., there will be \emph{some} gapped, symmetric parent Hamiltonian \cite{Schuch11}). Indeed, let us consider the following family of cat states of product states:
\begin{align}
\ket{\psi(\alpha)} \propto \left( 1 + \prod_n \tau_n + \prod_n \tau_n^\dagger \right) \bigotimes_{n=1}^N \left(  \ket{B}_n + \ket{C}_n -\alpha \ket{A}_n \right) 
\end{align}
This relates to Eq.~\eqref{eq:MPSstate} with $\alpha = \frac{2-\lambda}{\lambda+1}$. Indeed, we can see $\ket{\psi(\infty)}$ is the conventional Potts cat state, $\ket{\psi(\tfrac12)}$ is the anti-Potts cat state, and $\ket{\psi(0)}$ is the not-$A$ cat state. 
We can straightforwardly rewrite this as:
\begin{align}
\ket{\psi(\alpha)} \propto \bigotimes_{n=1}^N \left( \tau_n + \tau_n^\dagger - \alpha \right) \left( \ket{A}^{\otimes N} + \ket{B}^{\otimes N} + \ket{C}^{\otimes N} \right). \label{eq:path}
\end{align}
This form makes it trivial to write down an MPS description (of the form \cref{eq:MPSform}), from which one can derive that the path is `gapped' as it tunes between the aforementioned three cases. By this we mean that while the dominant transfer-matrix eigenvalue is threefold degenerate (due to the cat state nature of the state), there is a gap to the subleading eigenvalue, implying that its parent Hamiltonian would remain gapped throughout.

Returning to the product state description, one can also straightforwardly calculate various observables:
\begin{align}
  \langle A_\lambda| \sigma_n | A_\lambda \rangle &= \frac{1-2\lambda}{2+\lambda^2}\nonumber\\
    \langle A_\lambda| \bar \sigma_n | A_\lambda \rangle &= \langle A_\lambda| \mathcal R^z \sigma_n \mathcal R^z | A_\lambda \rangle = \langle A_{-\lambda}| \sigma_n | A_{-\lambda }\rangle = \frac{1+2\lambda}{2+\lambda^2}\ .  
\end{align}
For any (finite) value of $\lambda$, at least one of these two order parameters is nonzero!
Moreover, this also implies that 
\begin{align}
\langle A_\lambda| s^\pm_n | A_\lambda\rangle = \frac{-2 \lambda}{2+\lambda^2} \qquad \textrm{and} \qquad 
\langle A_\lambda|\left( s^\pm_n \right)^2 | A_\lambda \rangle = \frac{1}{2+\lambda^2} \ .
\end{align}
Note that $\langle s^\pm_n \rangle$ vanishes for $\lambda=0$, a consequence of the $\mathcal{R}^z$ symmetry of this point! (Note $\mathcal{R}^z s^\pm_n\mathcal{R}^z =-s^\pm_n.$)

Unlike the simple pivot paths in the previous subsection, this path of product states explicitly preserves both charge conjugation and time reversal symmetry. Hence, enforcing this additional symmetry does not separate the phases.
It does not, however, tell us what happens in the vicinity of the critical point in Eq.~\eqref{eq:direct}. To study this terminating criticality, we turn to the study of Hamiltonians.

\begin{figure}
    \centering
    \includegraphics[]{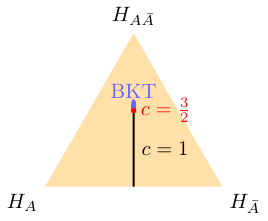}
    \caption{\small  The terminating criticality between $H_A$ and $H_\notA$.}
    \label{fig:phasediagramA}
\end{figure}

\subsubsection{Gapping out the criticality with $H_\AnotA$  and the orbifold}
\label{sec:halfway}
We have seen that in the spin-1 notation, $H_A$ and $H_{\bar A}$ consist of the same $U(1)$-symmetric piece (which determines the $c=1$ criticality we found in Eq.~\eqref{eq:direct}) and have opposite signs for the $\mathbb Z_3$-symmetric part. At the half-way interpolation point we gain $\mathcal R^z$ symmetry. The challenge to terminate the criticality is to break $U(1)$ down to $\mathbb Z_3$ without violating $\mathcal R^z$. Since $H_A$ and $H_{\notA}$ are distinguished by the sign of $\langle s^- \rangle$, we can include a $\mathbb{Z}_2$ Ising field $\varsigma$ that corresponds to this sign, and the gapped path where the sign changes will correspond to a disordered regime for this field.

One Hamiltonian in this disordered regime is given by
\begin{align}
\label{eq:HAnotA}
H_\AnotA
= 
&-\sum_n \left( \left(s^+_n\right)^2 \left(s^-_{n+1}\right)^2 + \left(s^+_{n-1}\right)^2 \left(s^+_{n}\right)^2 \left(s^+_{n+1}\right)^2 + h.c.\right).
\end{align}
The expectation $\langle s^- \rangle$ vanishes if we remove the spin zero on each site from the support of our ground state (e.g., by fixing $R_n^z=-1$). In our Hamiltonian, we have kept only those terms which commute with ${R}^z_n$, even on a single site. The Hamiltonian annihilates spin zero ($R_n^z=1$) and the ground state will locally satisfy ${R}^z_n =- 1$. 
On the effective spin-$\tfrac12$ with ${R}^z_n =- 1$ that remains at low energy, this is an XX model perturbed by an $S_3$-symmetric term (as in \cref{eq:HXXJ}). We have numerically confirmed that its ground state is in the Potts phase. In fact, its correlation length $\xi \approx 1$ and it monotonically decreases as we tune to either $H_A$ or $H_\notA$. 

Let us now consider $H = \alpha H_A + \beta H_\AnotA + \gamma H_\notA$ (with $\alpha,\beta,\gamma \geq 0$). The numerically obtained phase diagram is shown in barycentric coordinates $(\alpha,\beta,\gamma)$, in \cref{fig:phasediagramA}. This is reminiscent of the phase diagram \cref{fig:XXsingleionDMRG}, and the field theory description is closely related. Indeed, we combine the description of the $c=1$ free boson for \cref{eq:direct} with the Ising $\varsigma$ field that describes the sign of $\langle s^- \rangle$. Gauging the diagonal $\mathbb{Z}_2$ in the field theory we have local operators
\begin{align}
    \varsigma^m e^{i m \Phi/2},\quad \mu^{\widehat{m}} e^{\pm i\widehat{m}\Theta}\ .
\end{align}
Here $\Phi/2$ is $2\pi$ periodic, while a $\pi$ shift together with toggling the sign of $\varsigma$ is a gauge symmetry. Our lattice $\mathbb{Z}_3$ symmetry acts as $\Phi/2 \rightarrow \Phi/2+2\pi/3$, while $\mathcal{C}$ toggles the sign of $\Phi$ and $\Theta$. Considering the effective spin-$\tfrac12$ XX model limit $H_\AnotA$, where $\mu$ has long-range order, we see that translation takes $\mu^{\widehat{m}}e^{\pm i\widehat{m}\Theta}\rightarrow (-1)^{\widehat{m}}\mu^{\widehat{m}}e^{\pm i\widehat{m}\Theta}$ and leaves $\Phi$ invariant.

At the line connecting $H_A$ and $H_{\notA}$, the $\varsigma$ field has long-range order and we tune off criticality by $\varsigma \cos(3 \Phi/2)$, giving us the two symmetry-breaking phases. As we tune $\beta$, we drive the Ising degrees of freedom through a transition while we are in a regime of $1/2<K<9/8$ where the boson remains critical. After passing through the $c=\tfrac32$ point, $\varsigma \cos(3 \Phi/2)$ is exponentially decaying, and our dominant symmetry-allowed perturbations are $\cos(3 \Phi)$ and $\cos(2\Theta)$ (note $\mu \cos(\Theta)$ breaks translation symmetry, else it would dominate). There is a regime (the bubble) where both of these operators are irrelevant; we then pass through a BKT transition at $K=9/8$ into the $\cos(3 \Phi)$ dominated phase. Naively this has twice as many vacua as $\cos(3 \Phi/2)$, but our gauge symmetry means that the effective radius of $\Phi$ is $2\pi$ in this regime, so we have exactly three vacua, as required by the phase diagram.

\section{RSPT phenomena in the dual Potts--anti-Potts phase diagram} \label{sec:pottsRSPT}
We will now perform a Kramers-Wannier transformation on $H_A$ and $H_\notA$ in order to find two Hamiltonians both belonging to the trivial symmetric phase, that nevertheless feature a critical point when interpolating between them. This resolves the question of whether or not the two symmetric phases from Ref.~\cite{OBrien20} can be connected.

\subsection{Hamiltonians and ground states}

The Kramers-Wannier mapping for the $\ZZ_3$ subgroup of $S_3$ can be defined as
\begin{align}
    \tau_n &\rightarrow \sigma_n^\dagger \sigma_{n+1}^{\vphantom \dagger} &\sigma_n^\dagger \sigma_{n+1}^{\vphantom \dagger} &\rightarrow \tau_{n+1}\ .
\end{align}
Under this mapping, the dual Hamiltonians, which we will denote with tildes, are
\begin{align}
    \tilde H_A = - \sum_n \left( \tau_n +h.c. \right)\ , \qquad \qquad
    \tilde H_{\bar A} =-\sum_n \left( K_{n-1,n,n+1} \; \tau_n +h.c. \right)
\end{align}
where
\begin{align}
K_{n-1,n,n+1} &=e^{\frac{\pi i}{3} ((1- \omega)(\sigma_{n-1} \sigma_n^\dagger + \sigma_{n}^\dagger \sigma_{n+1}) +h.c. )}\ .
\end{align}

It is useful to compare the ground state wavefunctions of $\tilde H_{A}$ and $\tilde H_{\bar A}$. The ground state of $\tilde H_{A}$ is the usual fixed-point paramagnetic wavefunction
\begin{align}
    \ket{\tilde\psi_A} =\bigotimes_n\ket{0}_n = \frac{1}{\sqrt{3}^N}\sum_{s_n = A,B,C} \ket{s_1 s_2 \ldots s_N}
\end{align}
i.e., it is an equal-weight superposition of all $\ket{A}$, $\ket{B}$, and $\ket{C}$ configurations. To identify the ground state of $\tilde H_{\bar A}$, we can use the fact that 
\begin{align}
    \tilde H_{\bar A} = \tilde{\mathcal R^z} \tilde H_{A} \tilde{\mathcal R^z} \ ,
\end{align}
where 
\begin{align}
\label{eq:RSPTpivot}
    \tilde{\mathcal R}^z =\prod_{n} \tilde R^z_n = \prod_n e^{\frac{\pi}{\sqrt{3}} \left(\sigma_n^{\dagger}\sigma^{}_{n+1}-\sigma_n^{}\sigma_{n+1}^\dagger \right) } =  \prod_n -e^{\frac{\pi i}{3} \left(1+\sigma_n^{\dagger}\sigma_{n+1}+\sigma_n^{}\sigma_{n+1}^\dagger \right) }
\end{align}
is the dual of the $\mathcal R^z$ under the Kramers-Wannier transformation. Importantly, in the $\sigma$-basis $\{ \ket{s_n} \}_{s_n\in \{A,B,C\}}$, we can write
\begin{align}
    -e^{\frac{\pi i}{3} \left(1+\sigma_n^\dagger\sigma_{n+1}+\sigma_n\sigma_{n+1}^\dagger \right) } = -(-1)^{\delta_{s_n,s_{n+1}}} \ . \label{eq:Rzidentity}
\end{align}
In words, this means that the operator $\tilde{\mathcal R}^z$ assigns a $-1$ sign to the wavefunction for every pair of neighboring spins that \textit{does not} align (i.e., for every domain wall). Indeed, this expression is clear from the fact that it is the dual of $R^z$ in Eq.~\eqref{eq:singlesiteRz}, which negates the spin if it is either $\ket{1}$ or $\ket{2}$.

The effect of $\tilde{\mathcal R}^z$ turns out to count the winding (mod 2) of the spins as we go around the chain. To see this, note that any sequence of spins that does not wind around the clock must change its configuration an even number of times (i.e., what goes up must come down). On the other hand, the shortest sequence of spins that winds around the clock once (e.g., $...ABCA...$) needs to change its configuration 3 times, resulting in a phase $(-1)^3 = -1$, and any equivalent sequence differs by an even number of changes. To conclude, the wavefunction is
\begin{align}
    \ket{\tilde\psi_{\bar A}}= \tilde {\mathcal R}^z \ket{\tilde\psi_{A}} = \frac{1}{\sqrt{3}^N}\sum_{s_n = A,B,C} (-1)^{W(\{s_n\})} \ket{s_1 s_2 \ldots s_N}
\end{align}
where $W(\{s_n\})$ counts the winding of the spins around the chain. From the dual picture of the previous section, we have that these two Hamiltonians belong to the same (trivial) phase, we see this explicitly below. The midpoint between these two Hamiltonians $\frac{1}{2}(\tilde H_A+\tilde H_{\bar A})$ is a critical point, with a $U(1)$ symmetry generated by the dual of the pivot Hamiltonian.
\subsection{A path of MPS ground states}\label{sec:RSPTpath}
We can find an $S_3$-symmetric path connecting $\ket{\tilde\psi_{A}}$ to $\ket{\tilde\psi_{\bar{A}}}$ by Kramers-Wannier transforming the path of cat states in \cref{eq:path}. We find:
\begin{align}
\ket{\tilde \psi(\alpha)} \propto \bigotimes_{n=1}^N \left( \sigma_n^\dagger \sigma_{n+1}^{} +  \sigma_n^{} \sigma_{n+1}^\dagger - \alpha \right) \ket{0}^{\otimes N} \ . \label{eq:mpspath}
\end{align}
In particular, we see that for $\alpha=\tfrac12$ this assigns a minus sign whenever two spins are non-aligned in the $\sigma$-basis (up to global normalization), i.e., we find $\ket{\tilde\psi_{\bar A}}$. For $\alpha \to \infty$ it is $\ket{\tilde\psi_{A}}$. For $\alpha = 0$ it is equal to the RSPT state found in \cite{OBrien20}.

Recall that a translation invariant matrix-product state (MPS) on a chain of length $M$ is of the form
\begin{align}
\ket{\psi}= \sum_{j_1,\dots j_L} \mathrm{tr}\left( \mathcal{A}_{j_1}\cdots\mathcal{A}_{j_M}\right) \ket{j_1,\dots,j_L} \ . \label{eq:MPSform}
\end{align}
Here $\mathcal{A}_{j}^{\alpha,\beta}$ is a three-index tensor; for fixed site-index $j$ we have a $\chi\times\chi$ matrix, where $\chi$ is the bond dimension. From \cref{eq:mpspath} we can read off the MPS expression for this whole parameterized path. In particular, in the $\{ \ket{A}, \ket{B}, \ket{C} \}$ basis, our $\chi=3$ MPS tensor $\mathcal A$ is:
\begin{align}
\mathcal A_{A} = \left( \begin{array}{ccc} 2-\alpha & 0 & 0 \\
-1-\alpha &0 & 0 \\
-1 - \alpha & 0 & 0\end{array} \right), \quad
\mathcal A_{B} = \left( \begin{array}{ccc} 0& -1-\alpha & 0 \\
0 &2-\alpha  & 0 \\
0 &-1 -\alpha  & 0\end{array} \right),\quad
\mathcal A_{C} = \left( \begin{array}{ccc} 0 & 0 & -1-\alpha \\
0 & 0 & -1 - \alpha \\
0 & 0 & 2-\alpha \end{array} \right)\ .\label{eq:MPS}
\end{align}
From this MPS we can derive the entanglement spectrum $(\alpha^2+2)^{-1}(1,1,\alpha^2)$. The correlation length is given by $\xi^{-1}= -\log \lvert (1-2\alpha)/(2+\alpha^2)\rvert$. 

For $\alpha=0$, we have a rank-two entanglement spectrum in a $\chi=3$ MPS representation; the MPS tensor above can be related by a gauge transformation to the $\chi=2$ tensor given in \cite{OBrien20}. Note that $\alpha=0$ has a finite correlation length. At our modified RSPT point $\alpha = \tfrac12$ we find an entanglement spectrum of $\left( \frac{4}{9},\frac{4}{9},\frac{1}{9}\right)$, and zero correlation length. Just like in the SSB case, this path adiabatically connects these two RSPTs to the trivial product state. For $\alpha<1$ the entanglement doublet is dominant, and the entanglement transition occurs at $\alpha=1$ (corresponding to $\lambda =\tfrac12$). 

 If an MPS state is symmetric under an on-site symmetry $u$, then we have a local symmetry fractionalization of the form $\sum_{j'} u_{j,j'}\mathcal{A}_{j'} = V^{\dagger}\mathcal{A}_{j} V $, where we leave the bond indices implicit \cite{Chen11,Schuch11,Cirac21}. (Altogether, the $V$ form a projective representation of the symmetry group, and the cohomology class of this representation is the SPT invariant.)
For all $\alpha$, we have the fixed symmetry-fractionalization (interpreting bond indices in \eqref{eq:MPS} in the $\sigma$-basis)
\begin{align}
    \sum_{a'}\tau_{a,a'} \mathcal{A}_{a'}=\tau  \mathcal{A}_{a}\tau^\dagger \qquad 
\sum_{a'}\mathcal{C}_{a,a'} \mathcal{A}_{a'} =\mathcal{C} \mathcal{A}_{a}\mathcal{C} \ . \label{eq:MPSsymmfrac}
\end{align}
In the bond space, in the $\tau$-basis, this decomposes as a singlet $\tau\ket{0} = \mathcal{C}\ket{0} = \ket{0}$ and a doublet
\begin{align}
V_\tau =     \left(
\begin{array}{cc}
 \omega^2  & 0 \\
 0 & \omega  \\
\end{array}
\right) \qquad V_\mathcal{C}= \left(
\begin{array}{cc}
 0 & 1  \\
 1 & 0 \\
\end{array} 
\right)\ .
\end{align}
(With conventions fixed below \cref{eq:MPSform}.) The two-dimensional representation dominates for $\alpha<1$, motivating the name RSPT.

\subsection{Field theory description of a gapped path}\label{sec:RSPTfieldtheory2}
Applying Kramers-Wannier duality to the models in \cref{fig:phasediagramA}, we have a phase diagram containing a gapped path between $\tilde{H}_A$ and $\tilde{H}_{\bar A}$. While the precise form of the perturbing dual Hamiltonian $\tilde{H}_{A\bar A}$ is not particularly enlightening, it is instructive to analyze the field theory description. 

First, the lattice Kramers-Wannier corresponds to gauging the $\mathbb{Z}_3$ symmetry $\Phi/2 \rightarrow \Phi/2+2\pi/3$. Following this gauging, the lattice $\mathbb{Z}_3$ acts on the $\Theta$ field, so to align with conventions above, we relabel both $\Phi/2 \leftrightarrow \Theta$ and $\varsigma \leftrightarrow\mu$. The local fields are then
\begin{align}
    \varsigma^m e^{i m \Phi/2},\quad \mu^{\widehat{m}} e^{\pm i\widehat{m}\Theta}\ .
\end{align}
Here $\Phi/2$ is $2\pi$ periodic and the lattice $\mathbb{Z}_3$ symmetry acts as $\Phi/2 \rightarrow \Phi/2+2\pi/3$. 
(Note that the gauge symmetry that we used to glue the Ising and bosonic fields would be a $\pi$-shift in $\Theta$ together with toggling the sign of $\mu$. This is equivalent to a $\pi$-shift in $\Phi/2$ together with toggling the sign of $\varsigma$, as can be seen by the final local field content.)

With these variables, the $c=1$ transition between $\tilde{H}_A$ and $\tilde{H}_{\bar A}$ is driven by $\mu \cos(\Theta)$. Tuning beyond the Ising, transition we have a bubble with no relevant operators, while the transition into the gapped symmetric phase is driven by $\cos(2\Theta)$. Note that in this regime,  $\cos(2\Theta)$ has a unique ground state by the gauge symmetry. 
\subsection{RSPT edge modes depend on boundary termination}
Let us now show explicitly that $\tilde H_{\bar A}$ can exhibit a two-dimensional edge mode at the boundary, exhibiting the same phenomena as the RSPT Hamiltonian of Ref.~\cite{OBrien20}. However, as we will see shortly, there exists a different choice of boundary Hamiltonian which will give rise to a unique ground state.

The degeneracy in this model happens if we naively truncate the Hamiltonian, throwing away all terms that lie beyond the cut. Let us consider a half-infinite chain with $n \ge 0$. To obtain the Hamiltonian at $n=0$, we first expand $K_{n-1,n,n+1}$ as a sum
\begin{align}
K_{n-1,n,n+1} &=\frac{1}{9}\left[1-(2\omega \sigma^\dagger_{n-1}\sigma_{n}^{\vphantom \dagger}+2\omega \sigma_n^{\vphantom \dagger}\sigma^\dagger_{n+1}-4\sigma_{n-1}^{\vphantom \dagger}\sigma^\dagger_{n+1} -4 \omega\sigma_{n-1}^{\vphantom \dagger}\sigma_{n}^{\vphantom \dagger}\sigma_{n+1}^{\vphantom \dagger}+h.c.)\right] \ .
\end{align}
Thus, considering $K_{-1,0,1}$ and throwing all terms that involve the spin at $n=-1$ away, we arrive at
\begin{align}
    K^\text{trunc}_{0,1} = \frac{1}{9}(1-2\omega \sigma_0 \sigma_1^\dagger -2\bar \omega \sigma_0^\dagger \sigma_1)
\end{align}
and therefore the Hamiltonian with boundary reads
\begin{align}
   \tilde H_{\bar A,n\ge 0} &= - (K_{0,1}^\text{trunc} \tau_{0} + h.c.)  -\sum_{n=1}^\infty \left( K_{n-1,n,n+1} \; \tau_n +h.c. \right).
\end{align}
To understand the effect of this boundary Hamiltonian, we may perform an $S_3$-symmetric basis transformation $U = \prod_{n=0}^\infty \tilde R^z_n$, i.e., \cref{eq:RSPTpivot} truncated to act on the half-infinite chain.
Under this transformation, we find
\begin{align}
   U\tilde H_{\bar A,n\ge 0}U^\dagger &= 
   \frac{1}{3}(\tau_{0}^{} + \tau_{0}^\dagger)  -\sum_{n=1}^\infty \left( \tau_n^{} +\tau_n^\dagger \right)\ .
\end{align}
Thus, in this basis, the Hamiltonian is completely decoupled, and we may analyze each site separately. In the basis $\ket{0}$, $\ket{1}$ and $\ket{2}$, the $S_3$ symmetry block diagonalizes into the trivial representation (the singlet) spanned by $\ket{0}$ and the two-dimensional representation (the doublet) spanned by $\ket{1}$ and $\ket{2}$. Now, for each site in the bulk ($n \ge 1$), the ground state is the singlet, which minimizes $-(\tau_n^{} + \tau_n^\dagger)$. On the other hand, for the boundary site $n=0$, because of the positive coefficient, the term $\frac{1}{3}\left(\tau_0^{} + \tau_0^\dagger\right)$ penalizes the singlet, and therefore we are left with a twofold degeneracy given by the doublet, illustrating the feature of the RSPT.

From this perspective, it is now clear that the degeneracy is but an artifact of choosing a particular truncation of the boundary Hamiltonian. Instead, we can simply gap out the boundary of $\tilde H_{\bar A}$ if we reverse the sign of the boundary term. Namely, the Hamiltonian
\begin{align}
   {\color{red}+} (K_{0,1}^\text{trunc} \tau_{0} + h.c.)  -\sum_{n=1}^\infty \left( K_{n-1,n,n+1} \; \tau_n +h.c.\right), \label{eq:RSPTedge}
\end{align}
has a unique gapped ground state.
\subsection{$\tilde{H}_{\bar{A}}$ is half-way around a charge pump}\label{sec:rsptpump}
Charge pumps are non-contractible loops in the space of gapped $\tilde G$-symmetric Hamiltonians (or their ground states), classified by the $\tilde G$-charge pumped to the boundary when the loop is traversed on a semi-infinite system \cite{Thouless83,Hsin20,Shiozaki22,Kitaev_talk,Hermele_talk,Bachmann24,Jones25b,Prakash24}. For our purposes, unitary pivot loops are charge pumps when $U(2\pi)=\mathbb{I}$  with periodic boundaries, but $U(2\pi)=\mathcal{O}_L\mathcal{O}_R$ for $G$-charged operators $\mathcal{O}_{L/R}$ when taking open boundaries. 

In the next section, we will discuss some general connections between charge pumps and terminating critical lines. For now we simply note that the examples of quasi-phases encountered above all had a charge pump (or a domain-wall pump in the SSB case) in the vicinity of the critical point. Other examples of RSPT physics have also been shown to be connected to charge pumps \cite{Jones25b}. The purpose of this subsection is twofold: (1) we show that a charge pump exists in the neighborhood of the critical point along the line $\lambda \tilde{H}_A+ (1-\lambda)\tilde{H}_{\bar A} $, and (2) we investigate connections between the charge pump and RSPT phenomena: i.e., entanglement spectrum degeneracy and edge modes.  
\subsubsection{Pivot loops}
Above we give two equivalent forms for the $\mathbb{Z}_2$ `RSPT entangler' \eqref{eq:RSPTpivot}. We can write these as pivot loops
\begin{align}
\label{eq:RSPTpivot2}
    \tilde{\mathcal R}^z(t) =  \prod_n -e^{\frac{ i t}{3} \left(1+\sigma_n^{\dagger}\sigma_{n+1}+\sigma_n^{}\sigma_{n+1}^\dagger \right) } \qquad \tilde{\mathcal S}^z(t) =  \prod_n e^{\frac{t}{3\sqrt{3}} \left(\sigma_n^{\dagger}\sigma^{}_{n+1}-\sigma_n^{}\sigma_{n+1}^\dagger \right) }
\end{align}
where, as we show in \Cref{app:onsager}, $  \tilde{\mathcal R}^z=   \tilde{\mathcal R}^z(\pi) =  \tilde{\mathcal S}^z(3\pi)= \tilde{\mathcal S}^z(\pi)$. Hence, $\tilde{\mathcal S}^z(\pi)$ is also an RSPT entangler, but the operators $\tilde{\mathcal R}^z(t)$ and $\tilde{\mathcal S}^z(t)$ generate different loops in the space of Hamiltonians (indeed $\tilde{\mathcal S}^z(t)$ breaks $\mathcal{C}$ except at $t=0, \pi$).

In fact, $\tilde{\mathcal S}^z(2\pi)$ acts as a charge pump \cite{Hsin20,Shiozaki22,Hermele_talk}. We show in \Cref{app:onsager} that  \begin{align}
\tilde{\mathcal S}^z(2\pi) = \begin{cases} \mathbb{I} \qquad & \mathrm{periodic}\\
\sigma_1^{^\dagger} \sigma_L^{} \qquad &\mathrm{finite~ region} \end{cases} \ . \label{eq:Szpump}
\end{align}
This follows from the identity $e^{\frac{2\pi}{3\sqrt{3}} \left(\sigma_n^{\dagger}\sigma^{}_{n+1}-\sigma_n^{}\sigma_{n+1}^\dagger \right)}=\sigma_n^\dagger \sigma_{n+1}$. 
By making a connection to the Onsager ferromagnet, we can moreover show that we have a strict circular loop
\begin{align}
\tilde{\mathcal S}^z(t) \tilde{H}_A \tilde{\mathcal S}^z(t)^\dagger  = \frac{1}{2}(\tilde{H}_A +\tilde H_{\bar A})  +\frac{1}{2}(\tilde{H}_A -\tilde H_{\bar A})\cos(t) + H' \sin(t) \ , \label{eq:strict}
\end{align}
for a $\mathcal{C}$-breaking Hamiltonian $H'$ given in \Cref{app:onsager}. Using the arguments given in \cite{Jones25b} we have an anomalous $\mathbb{Z}_3 \times U(1)^{\mathrm{pivot}}$ symmetry at
$\tilde{H}_A +\tilde H_{\bar A}$, which hence cannot be short-range entangled. This anomaly implies a phase transition along the line connecting the two Hamiltonians, which, as we know, is realized at the $U(1)$-symmetric point in \cref{fig:phasediagramA}. 

Moreover, once identified as a $c=1$ point, the existence of this non-trivial pump in the vicinity of the critical point tells us that our loop locally corresponds to the $U(1)$ charge pump of the free boson, generated by $\partial \Theta/2\pi$. In particular, the $U(1)$ action must act as $\Phi \rightarrow \Phi +\alpha$. This means that after breaking $U(1)$ down to $\mathbb{Z}_3$, the low-energy action continues to $\Phi \rightarrow \Phi +2\pi/3$. This
is enough to exclude a gapped $S_3$-symmetric path in the neighborhood of the critical point, by the same reasoning as in \cref{sec:RSPT}. 

\subsubsection{Pump phenomena for a family of MPS and entanglement spectrum}
In this subsection we explore the connection between the pump and the presence of a dominant entanglement doublet (that corresponds to the two-dimensional irrep of $S_3$). This degeneracy was used to characterize RSPTs `deep in the phase' in Ref.~\cite{Jones25} (although as we have argued here, it is not a robust characterization). We know from the symmetry fractionalization of \cref{eq:MPSsymmfrac} that the ground state of 
$\tilde{H}_\notA$ has the RSPT-associated leading entanglement spectrum degeneracy, but here we show that this is not a consequence of the existence of a pump. Note that it is well known that charge pumps imply a spectral flow in the entanglement spectrum (see e.g., Ref.~\cite{Hayward18}), and thus if we begin with a dominant singlet, we must have a level crossing with a dominant doublet as we go around the loop. We are interested in a leading degeneracy that occurs due to a representation of a symmetry, and this would only occur at $S_3$-symmetric points on the loop (since $\mathbb{Z}_3$ has only one-dimensional irreps). Here we show that these two can be decoupled, i.e., that the dominant doublet need not occur at an $S_3$-symmetric point.

To see this, let us first
write the MPS ground state around the pivot loop as  
\begin{align}
\mathcal{B}_a^{n,n'}(t) = {\frac{1}{\sqrt{3}}\omega^{a(n-n')}}~\underbrace{\frac{1}{3}\Big(1+ 2 \cos \left(\frac{t+2n'\pi}{3}\right)\Big) }_{\Lambda_{n'}(t)}\label{eq:MPSCPTparameter}
\end{align}
where the index $a$ is in the $\sigma$ diagonal basis
(at $t=\pi$ this is a canonical form of the MPS \cref{eq:MPS} for $\alpha =\tfrac12$ and can be found directly as in \cite{Jones25}), and $0\leq n,n'\leq 2$. We can read off the entanglement spectrum $\Lambda_{n'}^2(t)$, and at $t=2\pi k$ a singlet dominates while at $t=\pi+2\pi k$ we see a dominant doublet corresponding to the $S_3$ symmetry fractionalization. As we vary $t$ we see a $6\pi$-periodic spectral flow in the entanglement eigenvalues. 

As we traverse the loop, the MPS has a consistent $\mathbb{Z}_3$ symmetry fractionalization 
\begin{align}
\sum_{a'}\tau_{a,a'} \mathcal{B}_{a'}(t)=\tau \mathcal{B}_{a}(t)\tau^\dagger \ ,\end{align}
while cycling the loop acts as 
\begin{align}
    \mathcal{B}_{a}(t+2\pi)=\sigma \mathcal{B}_{a}(t)\sigma^\dagger \ .\end{align} This gives another way to identify the charge pump: it manifests as the non-commutativity of $\sigma$ and $\tau$ \cite{Inamura24,Jones25b}.

The spectral flow exchanges the doublet and the singlet, but this is not universal. Consider now the following $\alpha$-parameterized deformation of the above family of MPS
\begin{align}
\mathcal{A}_a^{n,n'}(t,\alpha) = \frac{1}{\sqrt{3}}\omega^{a(n-n')} ~{\sqrt{\frac{2}{3(2 \alpha^2 + 1)}}\Big(\alpha+ \cos(t/3+2n'\pi/3 )\Big)}\ .\label{eq:MPSCPT2}
 \end{align}
For all choices of $\alpha$ we see spectral flow as above (and the families coincide for $\alpha=\tfrac12$), while the symmetry fractionalization is $\alpha$ independent. Nevertheless, for $0<\alpha<1/4$ at $t=\pi$ the singlet dominates the doublet\footnote{The limit of $\alpha=0$ is in fact a sequence of two charge-two pumps with period $\pi$, and we see dominant doublets half-way round the primitive pump at $t=\pi/2$. Finite $\alpha$ removes this accidental feature. }. The transfer matrix has $t$-independent non-zero eigenvalues $\lambda_1=1$ and $\lambda_2=\lambda_3=1-\frac{3}{4 \alpha ^2+2}$. We thus have a corresponding family of gapped parent Hamiltonians, and despite the non-trivial $S_3$ fractionalization at $t=0,\pi$, in neither case does the doublet dominate.

The pump does leave a subtle imprint on the entanglement spectrum at $S_3$-symmetric points. In \cref{app:pumpspectrum} we show that the number of doublets must equal the number of singlets at $S_3$-symmetric points, whenever we have a $\mathbb{Z}_3$ pump.

\subsubsection{Edge modes and the charge pump}
We have seen that considering the fixed Hamiltonian $\tilde H_{\bar A}$, we can choose a boundary termination so that there is no degeneracy (this is not possible for SPT phases). However, from general considerations \cite{Hsin20,Shiozaki22,Jones25b}, charge pumps give rise to boundary transitions, and perhaps this could lead to a guaranteed RSPT boundary degeneracy for a boundary termination compatible with a non-trivial pump loop. More precisely, since the loop \eqref{eq:strict} of bulk Hamiltonians is a $\mathbb{Z}_3$ charge pump, we have that any periodic and $\mathbb{Z}_3$-symmetric boundary termination of these Hamiltonians must have a degeneracy at some point around the loop. In fact, if we take an $S_3$-symmetric boundary Hamiltonian, the loop \eqref{eq:strict} will be $S_3$-equivariant \cite{Jones25b}, and the boundary transitions will be symmetric under reflection $t \rightarrow -t$. There is, however, no guarantee on general grounds that the boundary transition would correspond to the $S_3$-symmetric bulk (i.e., $t=0$ or $t=\pi$), and we confirm below that a simple choice of boundary termination is gapped for both of the $S_3$-symmetric bulk Hamiltonians.  

Indeed, consider the following $S_3$-symmetric and $2\pi$-periodic boundary termination
\begin{align}
H(t) = \sin^2(t/2) (K_{0,1}^\text{trunc} \tau_{0} + h.c.) -\cos^2(t/2)\frac{1}{3}(\tau_0 +\tau_{0}^\dagger) 
-\tilde{\mathcal S}^z(t) \sum_{j=1}^\infty (\tau_j +\tau_{j}^\dagger) \tilde{\mathcal S}^z(t)^\dagger    \ . 
\end{align}
For $t=0$ this is simply $\tilde{H}_A$ (up to a normalization on the first site), with trivial unique ground state. For $t=\pi$ it is \cref{eq:RSPTedge}, again with a unique ground state. 
By diagonalizing the boundary Hamiltonian, we find that the required boundary transition occurs at the $\mathcal{C}$-related pair $t=\pi/2$ and $t=3\pi/2$. 
Finally, we note that this is a pump loop at finite distance from the critical point. It would be interesting to study the same problem in the vicinity of the critical point.
\section{Quasi-phases: broader framework}\label{sec:framework}
\subsection{Quasi-phases via interfaces}
\subsubsection{Definitions}
In light of the examples of different quasi-phases considered above, we expect quasi-phase physics to be a property of the vicinity of a particular critical point, rather than of the gapped phases themselves. Indeed, this is evidenced by the fact that in all the examples, the critical line did not simply end but required a bigger theory to describe the nearby gapped path. 

We give the following definition, relative to a particular, not necessarily isolated, critical point:
\begin{shaded}
\noindent\textbf{Quasi-phases.} A critical point hosts distinct nearby gapped quasi-phases if
\begin{enumerate}
    \item there are relevant perturbations that flow to at least two gapped regimes;
    \item there is no perturbation of the critical point that flows to an invertible interface between these regimes.
\end{enumerate}
In the presence of a $G$-symmetry, one can further require perturbations to be symmetric.
\end{shaded}
\noindent 
We discuss the meaning of an `invertible interface' in the following subsection (\ref{sec:invinterface}) but in the special case where the two gapped phases are SRE, such an invertible interface is equivalent to the \emph{absence} of an edge mode, i.e., we have a short-range entangled unique interface. Although the above definition is in terms of interfaces, the following subsection discusses how it, in fact, connects to the (non-)existence of a nearby gapped path. The definition thus intuitively serves two roles: it makes precise in which sense edge modes (or rather, interface modes) are universal properties near the critical point, and how describing a nearby gapped path requires a bigger parent theory.

We allow the perturbations\footnote{We restrict to small couplings for both bulk homogeneous perturbations as well as inhomogeneous perturbations that can create interfaces, and we take the IR limit for arbitrarily small but \emph{fixed} couplings.} mentioned in the above definition to be spatially inhomogeneous. This means that bulk irrelevant perturbations should be taken into account---while they vanish in the bulk scaling limit, they can lift interface degeneracies, as we will show in \cref{sec:irrelevant}, and they thus make the notion of quasi-phase more restrictive than one might a priori expect. Note that under this definition, critical points separating distinct gapped phases (for example, the Ising transition separating $\mathbb{Z}_2$-trivial from $\mathbb{Z}_2$-symmetry-breaking, or transitions between distinct SPT phases) would be said to host distinct quasi-phases. 
We use the term \emph{quasi-phase} to particularly emphasize those cases where some pair of gapped regimes are globally connected in the phase diagram (i.e., are in the same phase), yet distinguished near the critical point, as in the examples above. 

Indeed, the models we studied in the preceding sections naturally fit into this framework. For instance, in the QSPT case of \cref{sec:QSPT}, a flow to a non-degenerate interface between quasi-phases---fixing spin-zero rather than the effective spin-1/2 on a single site---requires a perturbation of order $D\gg 1$ (so, in particular, is not a small perturbation). Such a small perturbation cannot exist from the perspective of the original compact boson CFT, as the quotient symmetry does not support the necessary representations for a unique invertible interface. In the RSPT setting of \cref{sec:RSPT}, the mechanism is more subtle, but we expect that any RG-flow at small coupling will select the $\pm \pi$ interfaces that are degenerate and non-invertible, rather than the invertible superposition of $\pm 3\pi$ that is available at high energies (see the discussion in~\cref{sec:bosoninterface} and \cref{fig:RSPTedge}).

 \subsubsection{Invertible interfaces and gapped paths}
\label{sec:invinterface}
\begin{figure}[h]
\includegraphics[]{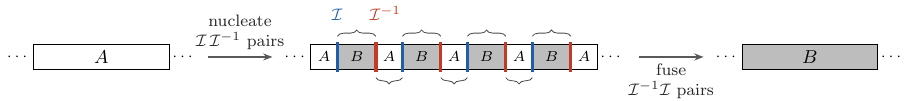}
\caption{\small Invertible interfaces can be used to construct a gapped path between gapped systems $A$ and $B$ \cite{Gaiotto_2019}.}\label{fig:interface}
\end{figure}
 We first recall the definition: an invertible interface $\mathcal{I}$ from phase $A$ to phase $B$ is a codimension-one defect with a corresponding interface $\mathcal{I}^{-1}$ from $B$ to $A$ such that if both appear close to each other, the combination flows to the identity defect under RG (i.e., one can `heal' the original interface by adding another one). More concretely, in the case of gapped phases, it means that the finite region which contains interfaces $\mathcal{I}$ and $\mathcal{I}^{-1}$  can be deformed, without closing the gap, to a trivial defect within a single phase \cite{Kitaev_talk,Gaiotto_2019}. Conversely, starting from a pure $A$ regime, one can nucleate pairs  $\mathcal{I}\mathcal{I}^{-1}$ without closing the gap. Furthermore, in the special case of short-range entangled phases of spin chains, an invertible interface hosts no protected degeneracy; for example, interfaces between two different SPTs necessarily have zero-energy edge modes, and this means they cannot be invertible. However, this concept is useful in more general settings, including long-range entangled and gapless systems \cite{Kitaev_2012,Davydov_2011,Gaiotto_2019,Prembabu:2025qvi}. This allows us to apply our definition far more broadly. 

At first sight, relating phases of matter to the interfaces between them seems like a secondary result rather than a defining property. In fact, the interface picture is deeply related to other definitions, including connectivity along paths in some parameter space.  Indeed, in analyzing the usual gapped phases of matter, the existence of an invertible interface is equivalent to the standard notion of gapped symmetric paths between two phases $A$ and $B$ \cite{Kitaev_talk,Gaiotto_2019}. 
To see this, if we have such a gapped path then we can implement it slowly in space between two semi-infinite bulk regions in phases $A$ and $B$, respectively. This flows to an invertible interface, where the inverse corresponds to the reverse path. Conversely, if we have an invertible interface then we can nucleate pairs of them in space and then annihilate each interface with its other neighbor to give a gapped path, as in \cref{fig:interface}. 

Implicit in this discussion is a hierarchy of scales. The alternating $A$ and $B$ regions should be at a mesoscopic scale, where the size of each region is much larger than the correlation lengths of the gapped system and the interfaces, but much smaller than the system size. This allows us to treat the interfaces as independent and locally fuse them.

This gapped path is not obviously in the vicinity of the critical point. However, we argue that if the critical point with $A$ and $B$ as two neighboring gapped phases admits a perturbation that flows to an invertible interface between them, then, even infinitesimally close to said critical point, one can still adiabatically connect $A$ and $B$. To see this, let $\lambda$ be the perturbation strength for some relevant perturbation that can flow to $A$ or $B$ (e.g., depending on the sign of the perturbation). By the definition of quasi-phases above, we know we can make $\lambda$ arbitrarily small, but for a given choice of $\lambda$, the gapped phases have a correlation length $\xi \sim \lambda^{-\nu}$ (for some critical exponent $\nu$). Hence, for each choice of $\lambda$, we can find a finite length-scale $\xi_{\rm meso}(\lambda) \gg \xi$. This means that we can construct a new operator whose unit cell is order $\sim \xi_{\rm meso}(\lambda)$ which has the property that perturbing the original critical point with strength $\sim \lambda$ will flow to a gapped phase with alternating regions of $A$ and $B$, with gapped interfaces $\mathcal I$ and $\mathcal I^{-1}$. At this point, we can repeat the usual reasoning in \cref{fig:interface}. Note that the existence of this gapped path can still be consistent with the original critical point being of codimension one: the operator that builds the gapped bridge is non-uniform and its unit cell diverges as one takes $\lambda \to 0$.

In the other direction, given a gapped path $A\rightarrow B$ constructed from small perturbations of our critical point, subject to usual adiabatic spatial-gluing assumptions this can be implemented slowly in space to give us an interface. The couplings are of the same order as those in the gapped path, so we remain in the vicinity of the critical point.
The inverse interface comes from the reverse path. Altogether we have a physically motivated equivalence between gapped paths and invertible interfaces in the vicinity of the critical point, characterizing equivalence of quasi-phases. 

It is helpful to clarify the connection between these ideas and the existing terminology of (generic) unnecessary criticality. As outlined in \cref{sec:conceptual}, an unnecessary criticality in the sense of Refs.~\cite{BiSenthil19,Jian20,Prakash23} has codimension one in parameter space but is not implied by the global phase diagram. Here we point out that in the presence of quasi-phases, such a codimension-one critical point can be IR-necessary even if it is unnecessary according to the global classification of phases. However, we also stress that it is not the case that all such criticalities are IR-necessary. In the next subsection we see that even if there are no homogeneous relevant operators connecting the two regimes of some relevant operator, an inhomogeneous irrelevant operator can enable a path between them. While the example has codimension larger than one, we think this mechanism makes it likely that generic unnecessary criticalities that are not IR-necessary exist in theories beyond the free boson.

\subsubsection{Example: irrelevant operators and interfaces}\label{sec:irrelevant}
Here we unpack, via an example, our claim that irrelevant operators can give rise to gapped paths in the vicinity of a critical point. 
We take the $c=1$ free boson, with conventions defined in \cref{sec:boson}. We enforce only the unitary charge conjugation symmetry, $(\Phi,\Theta) \rightarrow_{\mathcal{C}} -(\Phi, \Theta) $. The symmetric primary vertex operators are then $\cos(m \Phi)$ and $\cos(\widehat{m}\Theta)$.
We fix the Luttinger parameter $K<1/8$ so that $\cos(\Phi)$ and all its higher harmonics are irrelevant. In this regime $\cos(\widehat{m}\Theta)$ is relevant for $\widehat{m}^2<2/K$ which gives at least four primary operators. This is hence \emph{not} a codimension-one critical point (and so cannot be an unnecessary critical point in the sense of \cite{BiSenthil19,Jian20,Prakash23}). Nevertheless, it will serve to illustrate our point that exploring the relevant directions only does not suffice to understand all of the gapped paths in the vicinity of the critical point.

Perturbing the critical point by $\lambda \cos(\Theta)$ drives us to a unique vacuum $\Theta=\pi$ (positive coupling) or $\Theta=0$ (negative coupling). There are no non-trivial SPT classes for $\mathbb{Z}_2$ and so these regimes can be connected in the broader $\mathcal{C}$-symmetric phase diagram. Any homogeneous relevant perturbation depends only on $\cos(\widehat{m}\Theta)$ and, as discussed above, cannot continuously connect the two vacua. 
Using the homogeneous irrelevant perturbation $\cos(\Phi)$ to explore the space $\lambda \cos(\Theta) + \mu \cos(\Phi)$ will not give rise to a gapped path. For any non-zero $\lambda$ we are driven to the corresponding $\Theta$ vacuum, and at $\lambda=0$ we remain critical. However, we can use this operator to construct an \emph{inhomogeneous} gapped path.

\begin{figure}[h]
\centering
\includegraphics[]{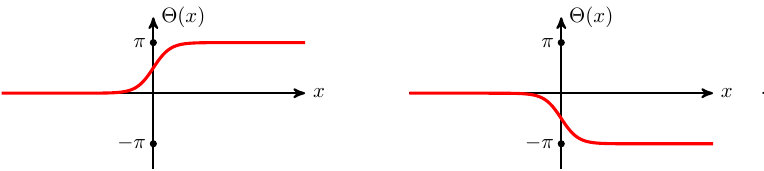}
\caption{\small `Minimal' interfaces between $\Theta=0$ and $\Theta=\pi$; $\mathcal{C}$ exchanges them. }\label{fig:edge2}
\end{figure}
To see this, introduce a smooth interface between the $\Theta=0$ and $\Theta=\pi$ vacua by perturbing the CFT with $\lambda(x) \cos(\Theta)$. In doing so, we maintain the emergent\footnote{This symmetry together with $\mathcal{C}$ would mean $\Theta=0$ and $\pi$ were two different SPT classes for $O(2)$, if preserved, but we will explicitly break the $U(1)$ with an irrelevant operator.} $U(1)$ symmetry with charges $Q=\tfrac{1}{2\pi}\int \partial_x\Theta~ \mathrm d x$.
Suppose, for example, that the RG flow energetically selects the minimal $\pm\pi$-shift interfaces as in \cref{fig:edge2}, with $Q=\pm 1/2$. 
Now, if we add the irrelevant perturbation $\cos(\Phi)$, for any finite coupling we break the emergent $U(1)$ and couple the $Q=\pm1/2$ interfaces: our lowest-energy interface becomes unique (it is one of the two $\mathcal{C}$-symmetric superpositions). We can then construct a gapped path as described in the previous subsection. Note that the same discussion applies for any charge $\pm Q$ pair of interfaces---these will be split by an infinitesimal $\cos(\Phi)$.

We contrast this with, say, the $S_3$ RSPT case of \cref{sec:RSPTedge}. Then any $S_3$-symmetric operator, relevant or irrelevant, cannot couple the $Q=\pm1/2 $ interfaces and, if these are at the lowest energy, then our RG flow selects a degenerate (non-invertible) interface. Of course, if the $Q=\pm3/2 $ interfaces are the endpoint of the flow then we can connect them with an irrelevant $S_3$-symmetric perturbation.

\begin{figure}
\centering
\includegraphics[]{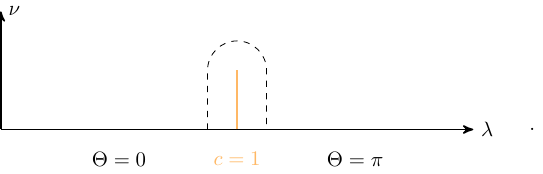}
\caption{\small A terminating $c=1$ critical line that does \emph{not} separate distinct quasi-phases. At $\lambda=0$, we fix the couplings to all relevant $\cos(\widehat{m} \Theta)$ to zero, $\lambda$ tunes the coupling to $\cos(\Theta)$ only. As we tune $\nu$ we increase the Luttinger parameter $K$ and eventually $\cos(\Phi)$ becomes relevant. We can then connect $\Theta=0$ to $\Theta=\pi$ by tuning through the $\cos(\Phi)$ dominated regime \cite{Lecheminant02}. Note that even for small $\nu$, $\cos(\Phi)$ can generate a gapped path via interfaces, as argued in the main text.}\label{fig:terminating}
\end{figure}

With a similar set-up, we can generate a terminating critical line, as in \cref{fig:terminating}. While the regions $\Theta=0$ and $\Theta=\pi$ do not constitute distinct quasi-phases, it is also not a generic unnecessary criticality, since for $\nu=0$ we have codimension at least four and we need to tune these operators to reach criticality (see also \cite{Prakash25}). Nevertheless, as stated above, we expect that in other settings we will be able to identify generic unnecessary critical points that do not separate distinct quasi-phases. Indeed, since we have argued that distinct quasi-phases forbid a gapped path connecting them arbitrarily close to the original critical point, this suggests that the broader phase diagram (which does include a gapped path) \textit{requires a bigger parent theory} for its description, unlike the example in \cref{fig:terminating}. This is consistent with the various rich examples of multicriticality we found in the previous sections.

\subsubsection{Lowest-energy interfaces in the free boson CFT}\label{sec:bosoninterface}
We thus see that the lowest energy interface is a key feature of our critical point. Determining the fate of the RG flow in a given lattice model is non-trivial. We give two reasons that the $Q=\pm1/2$ interfaces are natural for the free boson. First, semiclassically we can consider the energy $E[\Theta] = \int \Big((\partial_x \Theta)^2 + \sum_n t_n \cos(n\Theta)\Big) \mathrm{d} x$. Then, since the potential is reflection invariant about $n\pi$, any smooth configuration can, by reflection, be mapped onto a configuration on $[ 0,\pi]$ with the same energy. Although the image configuration will no longer be smooth in general, we can smooth it out, which we expect to decrease the energy. Second, at the free-fermion point $K=1$ with free-fermion perturbations, these interfaces are the Jackiw-Rebbi solitons and there is a unique zero mode \cite{Jackiw:1975fn}. Charge $Q=\pm3/2$ interfaces include additional charged fermionic excitations, which will increase the energy. Including interacting perturbations at $K=1$, or shifting $K$, requires further analysis, but perturbatively we expect some regime where this ordering of interfaces is stable in the quantum model. 
\subsection{Framework for constructing examples}
\label{sec:frameworkconstruct} In the previous subsection we sharpened our definition of quasi-phases, but a natural question is how to \textit{construct} examples of IR-necessary critical points that host such (distinct) quasi-phases. Here we use the interface picture to give intuition for building examples, along the lines of the intuitive discussion in \cref{sec:gappedS3path}. This unifies the disparate set of models we have seen above into a common framework. 

In particular, we propose that a natural set-up for exploring IR-necessary criticality and quasi-phases in 1+1D is given by spin chains where we have sites that transform under a representation $\rho$ or $\bar \rho$ for an internal group $G$ in an alternating fashion, with an additional site-centered inversion symmetry $\mathcal P$---see \cref{fig:interfacechain}.
In cases where $\rho$ is a non-trivial linear irrep, we might expect that natural classes of Hamiltonians might exhibit IR-necessary criticality: indeed, below we will argue that the examples above do indeed fit into this context. We first review how the usual SPT transitions also fit into this framework, where $\rho$ is projective rather than linear.

\subsubsection{Effective spin chain at a direct transition and LSM constraints}
\begin{figure}
\centering
\includegraphics[]{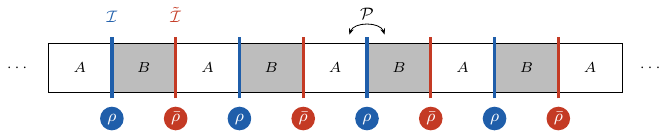}
    \caption{\small A model for a direct transition between gapped $d$-dimensional SRE phases $A$ and $B$. 
    Interfaces on codimension-one surfaces $A\rightarrow B$ ($B\rightarrow A$) are denoted by $\mathcal I$ ($\tilde{\mathcal{I}}$). We focus on $d=1$ spin chains: for sufficiently wide slabs of gapped phase, any modes localized on the interfaces form an effective spin chain with local symmetry representation $\rho(g)$ or its conjugate. We enforce a site-centered inversion symmetry, $\mathcal{P}$, on the effective chain (not the microscopic model). When $\rho$ is projective, LSM constraints can forbid an SRE ground state for the effective chain (and hence the system).}
    \label{fig:interfacechain}
\end{figure}

Suppose we have two gapped SRE phases $A$ and $B$. A candidate direct transition is tuning to the balance point between periodically nucleating bubbles of $B$ phase inside a background of $A$ and vice versa. If they are in the same phase, this tuning could give rise to a gapped path from a uniform chain of $A$ phase to a uniform chain of $B$ phase.
At the `balanced' half-way point, within the mesoscopic regions of $A$ and $B$ the chain is gapped, and the interesting low-energy physics occurs at the interfaces $\mathcal I$ between $A$ and $B$ and $\tilde{\mathcal{I}}$ between $B$ and $A$; see \cref{fig:interfacechain}. 

The question is whether this half-way point can host a unique gapped (SRE) ground state. Given our UV symmetry group, $G$, the state on the effective spin chain should be $G$-symmetric. The $G$ action is determined by the degrees of freedom at the interface. In particular, $\mathcal{I}$ hosts a projective representation of $G$ with class $\omega$ while  $\tilde{\mathcal{I}}$ hosts a projective representation of $G$ with class $-\omega$.
Moreover, we demand (effective) site-centered inversion symmetry, which exchanges $A$ and $B$, so that there is no bias towards either the $A$ or $B$ phase.

Given this set-up we may have LSM constraints that imply we cannot have an SRE ground state. In particular, Ref.~\cite{Ogata21} shows that if the projective class $\omega \neq 2\nu$ then we \emph{cannot} have an inversion and $G$-symmetric SRE state on our effective chain. 
As an example, consider the bond-alternating XXZ chain of \cref{sec:bondalt}. There the interface between trivial and SPT hosts a non-trivial projective representation of $\mathbb{Z}_2\times\mathbb{Z}_2$ that cannot be decomposed as a square (since there is only one non-trivial class for this group). Hence, the corresponding ground state cannot be SRE---in fact the tuning we describe is exactly the bond-alternation where the midpoint gives us the gapless phase transition point.

\subsubsection{Two routes to a trivial singlet and orbifold theories}
If $\omega=2\nu$, and in particular when $\omega=0$, there is no obstruction to a gapped symmetric ground state coming from LSM. Nevertheless, as we have seen, this does not mean that it is always simple to find such a state (or parent Hamiltonian). In the QSPT case, analyzed in \cref{sec:QSPT}, the UV $D_4$ symmetry acts on the IR degrees of freedom as an effective projective $\mathbb{Z}_2\times \mathbb{Z}_2$ (it is natural to restrict to the faithful symmetry group acting at low energies, else all SPT classes become trivial \cite{Anfuso07,MoudgalyaPollmann15,WangWenWitten18,Verresen21}). There is thus an emergent LSM obstruction to having an SRE state. To reach a gapped singlet we must include additional high-energy degrees of freedom (in this case the frozen $\ket{0}$ states) that allow us to detect that we have a trivializable linear representation of $D_4$. Once this degree of freedom is brought down to low energy, we can put our chain into a $D_4$-symmetric singlet. In fact we can reach a product state, supported on the unfrozen local singlets. 

In the $S_3$ RSPT cases discussed above we have no such emergent obstruction. At low energies we always have a linear representation; there are no quotients supporting projective irreps. We can have the following, seemingly mild, barrier to a symmetric product state though: when the `simplest' interface is in a two-dimensional irrep of $S_3$, by Schur's lemma we cannot gap out this interface into a symmetric product state. In \cref{sec:RSPTfieldtheory2} we gap out the critical point by tuning the available low-energy degrees of freedom (the gapped regime of the Ising variable), similar to the QSPT case. As exemplified in \cref{sec:gappedS3path}, another option is available in the RSPT case: we can construct trivial interfaces by constructing new $A\rightarrow B$ interfaces by considering a sequence of elementary interfaces.
 For the $S_3$ case we take three sites, then $\mathcal{I}\otimes \tilde{\mathcal{I}}\otimes\mathcal{I}\simeq \mathbf{2} \otimes \mathbf{2}  \otimes \mathbf{2}  \simeq \mathbf{1} \oplus \mathbf{1'}  \oplus \mathbf{2} \oplus \mathbf{2}  \oplus \mathbf{2} $, where $\mathbf{2} $ is the doublet, $\mathbf{1} $ is the trivial singlet and $\mathbf{1'} $ is the sign singlet. Thus by energetically favoring the composite interface, we have an on-site singlet available in our effective spin chain, and can construct a trivial symmetric product state.
This is a new feature for RSPTs: for (Q)SPTs $\mathcal{I}\otimes \tilde{\mathcal{I}}\otimes\mathcal{I}$ has projective class $\omega-\omega+\omega$ and so the LSM constraint remains for any composite interface.

Abstracting these ideas, consider a CFT with one symmetry-allowed relevant direction that takes us into SRE phases $A$ and $B$, and where the low-energy $A\rightarrow B$ interfaces do not host symmetric singlets. Then $A$ and $B$ are distinct quasi-phases, and we can attempt to construct phase diagrams that connect them by either:
\begin{enumerate}
    \item Bringing down new low-energy degrees of freedom that transform in a $G$-singlet. 
    \item Energetically favoring composite interfaces that contain $G$-singlets. 
\end{enumerate}
Both of these naturally give rise to orbifold constructions. In the first case we add a field where we tune between gapped regimes that have low-energy singlets (un)available. This is the mechanism for the $c=3/2$ (Ising $\times$ free boson) orbifold above (applicable to both QSPT and RSPT). The orbifolding corresponds to `labeling' operators that we want to be at high energy in some regime, while composite operators can remain at low energy. In the second option we instead use the orbifold to attach a field that can push individual lowest-energy interfaces to high energy, while allowing composites to remain at low energy. This is the mechanism we saw in the $c=9/5$ (Potts $\times$ free boson) orbifold. Altogether we expect that generalizations of our constructions beyond free boson examples will be possible using these ideas. 

\subsection{Charge pumps and terminating critical lines}\label{sec:frameworkpump}
\begin{figure}
\includegraphics[]{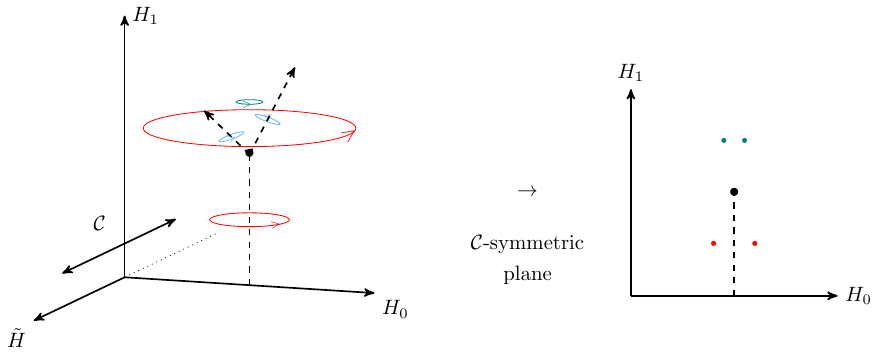}
\caption{\small Possible bifurcation of a diabolical critical line to leave the $\tilde{G}\rtimes\mathcal{C}$-symmetric space of models. The critical lines obstruct us from trivializing the red loops of $\tilde{G}$-charge $q$. The $\mathcal{C}$ symmetry, and consistency with the red loop, means that above the bifurcation point the two blue loops each pump equal charge $q'=q/2$ (if there is no such $q'$ then the bifurcation cannot occur, implying the critical line cannot terminate in the $\tilde{G}\rtimes\mathcal{C}$-symmetric space \cite{Jones25b}). The upper green loop is trivializable and thus pumps no charge. Projecting this into the  $\tilde{G}\rtimes\mathcal{C}$-symmetric space we see a terminating critical line. The bifurcation point $\bullet$ is constrained since it must be consistent as a junction of $\tilde{G}$-symmetric criticalities, and also with the respective charge pumps. While we display the gapless locus as lines, we can have extended regions obeying the same constraints, see also Ref.~\cite{Berg11}. }\label{fig:frameworkpump}
\end{figure}
In this final subsection, we provide a complementary perspective on quasi-phases which might provide a promising starting point for further investigation. The quasi-phase phenomena studied in this work correspond to non-trivial structure within the same phase of matter. Taking the trivial phase for definiteness, one simple probe of the structure beyond the phase diagram of adiabatically connected gapped Hamiltonians comes from loops of gapped Hamiltonians within this phase. As described in \cref{sec:rsptpump}, non-contractible such loops correspond to generalized Thouless charge pumps \cite{Thouless83,Hsin20,Cordova20,Hermele_talk,Shiozaki22,Wen23,Bachmann24}. For $\tilde{G}$-symmetric Hamiltonians in 1+1D, equivalence classes of loops are classified by the $\tilde{G}$-charge pumped through the bulk as we traverse the loop \cite{Bachmann24}; this charge may be observed at a boundary of the system. 

As soon as we identify a non-trivial $\tilde{G}$-charge pump within some phase diagram, we necessarily have a `diabolical locus'---this is a non-SRE region of the phase diagram that obstructs us from contracting the corresponding loop---and since it obstructs loops it must be of codimension at least two \cite{Hsin20}.  In a two-dimensional $\tilde{G}$-symmetric phase diagram containing a non-contractible loop, this appears generically as a gapless point, but can appear as some gapless region or intermediate phase. As emphasized in \cite{Prakash24}, this naturally gives rise to multicritical points, that require tuning at least two parameters, but not to extended codimension-one critical lines. One way that such a stable criticality could arise is when we are actually considering a larger symmetry group $G$, and we pump charge for a subgroup, $\tilde{G}$. Then it is possible that, by restricting to $G$-symmetric Hamiltonians, we fine-tune one of the couplings so that the criticality remains in this space and becomes codimension one. One mechanism which guarantees this is when the $\tilde{G}$-charge pump implies a relative $G$-SPT between two $G$-symmetric Hamiltonians around the loop \cite{Jones25b}---then the criticality is implied by the $G$-SPT classification. In this work, we have analyzed cases where we have a single $G$-trivial phase either side of an IR-necessary critical line, and so the pump \emph{does not} imply an SPT. 

The models analyzed above have symmetry group $G=\tilde{G}\rtimes\mathbb{Z}_2^C$ comprising a $\mathbb{Z}_2^C$ charge conjugation symmetry along with a $\tilde{G}$ for which we pump a charge (in our examples $\tilde{G}=\mathbb{Z}_k$ for $2\leq k\leq4$). Suppose we have a $\tilde{G}$-charge pump consisting of a curve between two $G$-symmetric Hamiltonians ($H_A$ and $H_B$) and its $\mathcal{C}$ reflection (e.g., these could be at the red bullets in the right-hand side of \cref{fig:frameworkpump} as part of the red pump loop on the left-hand side). If the pumped $\tilde{G}$-charge, $q$, cannot be decomposed into two identical smaller charges, then there must be a critical point on any $G$-symmetric path between $H_A$ and $H_B$---they are in different $G$-SPT phases. On the other hand, suppose $q$ can be written as the sum of two identical charges, $q=2q'$. Then if $H_A$ and $H_B$ are separated by a $G$-symmetric critical point, this critical point can disappear by connecting to both a $\tilde{G}$-symmetric gapless locus and its $\mathbb{Z}_2^C$ mirror image, where a loop around each such gapless locus pumps charge $q'$. At a high level, by analyzing such junctions of gapless loci we could learn something about the stability of the $G$-symmetric critical point. (Note that the existence of the $G$-symmetric critical point is non-trivial and \emph{not} implied by a pump without some additional structure: for example, a deformation of a genuine SPT transition, or where the pump is a strict circular loop as in \cref{eq:strict} \cite{Jones25b}.)

For pump loops to give rise to codimension-one terminating critical surfaces in this fashion, the Hamiltonians around the loop should \emph{not} be $G$-symmetric (the loop may, however, be $G$-equivariant). Then a codimension-two locus in the $\tilde{G}$-symmetric phase diagram can appear to be codimension one in the $G$-symmetric slice, and we have stability so long as the diabolical point remains $G$-symmetric. If we instead take a $G$-symmetric loop, then the stability of the pump is a priori unconnected to the codimension-one criticality. Indeed, according to the classification \cite{Bachmann24}, we can deform it in a $G$-symmetric way to a reference pump loop. Thus our examples above should be contrasted with the examples given in Ref.~\cite{Prakash24}, where the pump loop exists in the $G$-symmetric subspace. 

We observe that the QSPT and RSPT models considered above are all in the vicinity of charge pumps, and the corresponding phase diagrams are thus compatible with the presence of the pump. We note that related ideas on using charge pumps to constrain the topology of the phase diagram near an SPT transition appeared in Ref.~\cite{Berg11}.
In the QSPT case this reduces to the usual connection between pumps and SPTs so long as we remain in the regime where charges outside the quotient are at high energies. In \cref{eq:spin1XX}, for large $D$, we have that a loop enclosing the critical point in an $\mathcal{R}^x$-breaking direction pumps a unit charge \emph{for the low energy theory}. This means that so long as our low-energy theory cannot resolve `fractional charges', our critical line remains pinned in the $\mathcal{R}^x \sim \mathcal{C}$-symmetric plane.
 Once we leave the quotient SPT regime, we realize that what we identified as a unit $U(1)$ charge was in fact charge-two for our UV $U(1)$ symmetry. This charge is allowed to bifurcate into two charge-one obstructions, and indeed these must appear in the $\mathcal{R}^x$-breaking direction to prevent us trivializing the loop. It would be interesting to identify these as perturbations of our $c=\tfrac32$ theory.
In the RSPT model studied in \cref{sec:RSPT}, our phase diagram is in the vicinity of a genuine SPT transition---the bond-alternating XX model (as in the `failed SPT' of Ref.~\cite{Prakash23}). This SPT transition comes along with a $U(1)$ charge pump as described in \cref{app:pumpxx}, and so we can argue the IR-$\mathbb{Z}_3$ symmetry action and stability as in \cref{sec:rsptpump} (this is of limited value in this particular example, since we know the IR-symmetry action independently). We defer further study of the constraints arising from non-trivial charge pumps that do not imply SPTs, but they appear to play an important role in this phenomenology.

\section{Outlook}
In this paper we have introduced the conceptual framework around quasi-phases of quantum matter, defined in the neighborhood of IR-necessary critical points, and explored associated phenomena. We analyzed a series of lattice models that host rich quasi-phase physics, including both SRE and symmetry-breaking phases. The phase diagrams we studied included regimes where two quasi-phases are separated by a transition, but are globally connected beyond a termination of the IR-necessary criticality. In each case, the phase diagram was explained by analyzing flows from a larger orbifold field theory. Following the examples, we rationalized them as part of a broader framework, and gave a local definition of quasi-phases in terms of interfaces in the vicinity of a critical point. This suggested a method to construct further examples of lattice models with similar phenomenology. We moreover connected to other literature on unnecessary criticality and charge pumps.

There are numerous exciting directions to pursue based on this work. The proposed definition of quasi-phases given in \cref{sec:framework} can be applied to general spacetime dimension, and it would be of great interest to identify examples of IR-necessary criticality and quasi-phases in higher dimensions. Known unnecessary criticalities in higher dimensions would be a natural place to look \cite{BiSenthil19}. The interface picture moreover applies in the gapless setting \cite{Prembabu:2025qvi}, and it would be most interesting to identify gapless analogues of quasi-phases and their transitions. 
Even in the 1+1D spin chain setting there remain interesting problems. Our examples centered around the $c=1$ free boson CFT. It would be interesting to find phase diagrams where the IR-necessary criticality is in a different universality class. As outlined above, the expected connection between charge pumps and quasi-phase phenomenology, particularly those pumps in the vicinity of an IR-necessary critical point, requires clarification.

\section*{Acknowledgements}
We are grateful to Michael Levin and Abhishodh Prakash for helpful discussions and motivating questions. 
We acknowledge use of AI tools (ChatGPT 5 and 6 and Claude Opus 5.5) for assistance in preparing numerical code in \Cref{sec:RSPT}, preparing Figures 11 and 14, and with the proof of certain identities in \cref{app:AKLT}. The authors validated any output and take full responsibility for the content of the manuscript. The work of P.F.\ was supported in part by the UK Engineering and Physical Sciences Research Council grant EP/X030881/1. 
\\
\textit{Note added:} in the final stages of preparing this manuscript, we became aware of independent work \cite{Chanda26} that analyzes an adiabatic path between two symmetry-breaking regimes in a classical model, akin to the models in \cref{sec:potts}.

\bibliography{arxiv.bbl}

\nolinenumbers

\begin{appendix}
\crefalias{section}{appendix}

\section{Charge pump phenomena in the XX model} \label{app:pumpxx}
\begin{figure}
\centering
\includegraphics[]{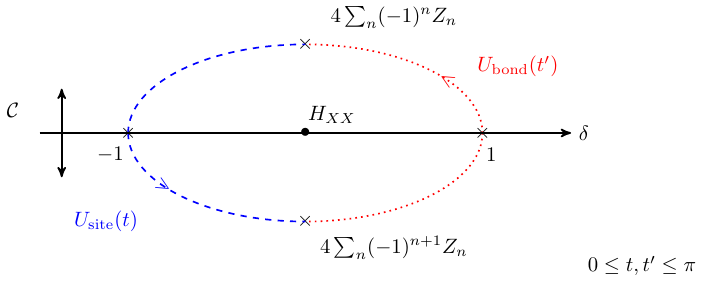}
\caption{\small A $U(1)$-pump cycle for the XX model (the two-dimensional phase diagram is dual to the Rice-Mele model \cite{Rice82}). This cycle is $\mathcal{C}$-equivariant, with fixed points at  $H_{\delta XX}$ for $\delta=\pm1$. Note that the isospectral unitary equivalence given applies only to the particular Hamiltonians indicated by $\times$; other points in the plane are gapped away from the origin. This loop pumps a unit $U(1)$ charge, and thus we conclude that $\delta=\pm1$ are in different $U(1)\rtimes\ZZ_2$ SPT classes. This in turn implies they are in different $\ZZ_2\times\ZZ_2$ SPT classes.}\label{fig:pumpXX}
\end{figure}
We can prove that $H_{\delta XX}$ for $\delta=\pm1 $ are in two different $\ZZ_2\times\ZZ_2$ SPT phases using a charge pump argument, as described in \cref{sec:frameworkpump}. This is not merely another perspective on the projective representations analyzed in \cref{sec:xxproj}. The pump invariant has greater stability than the SPT phase and we can try to utilize this stability to understand the phase diagram after explicit symmetry breaking. 

To derive a pump loop, first define $U_{n,n+1}(t)=e^{-i \frac{t}{4} \left(Y_n X_{n+1}-X_n Y_{n+1}\right)}$. Note that this commutes with the full $U(1)$ symmetry.
We then construct $U_\mathrm{site}(t)= \prod_k U_{2k-1,2k}(t) $ and 
$U_\mathrm{bond}(t)= \prod_k U_{2k,2k+1}(t) $, which are both products of two-site rotations.  Now, let $H_0=4 \sum_n (-1)^n Z_n$. Conjugating by $U_\mathrm{site}(t)$, we reach the trivial $\delta=-1$ model at $t=\pi/2$ and $-H_0$ at $t=\pi$. The Hamiltonian $-H_0$ is the same as $H_0$ shifted by half of a two-site unit-cell, so conjugating by $U_\mathrm{bond}(t)$ we continue and find a closed loop in the space of Hamiltonians that passes the $\delta=1$ model; this is illustrated in \cref{fig:pumpXX}. 

Concretely let \begin{align}
  W(t)= \begin{cases}  U_\mathrm{site}(t) \qquad &0\leq t \leq \pi\\
     U_\mathrm{bond}(t-\pi)U_\mathrm{site}(\pi)\qquad &\pi\leq t \leq 2\pi \end{cases}\ .
\end{align} 
The loop $H(t)=W(t)H_0W^\dagger(t)$ is $\mathcal{C}$-equivariant; in particular
$\mathcal{C} H(t) \mathcal{C} = H(\pi-t)$ for $0\leq t\leq \pi$ and $\mathcal{C} H(t) \mathcal{C} = H(3\pi-t)$ for $\pi\leq t\leq 2\pi$. Thus $\mathcal{C}$ fixes $H_{\delta XX}$ and otherwise mirrors the path in the $\delta$ axis.

To see that this loop pumps a unit $U(1)$ charge, it is simplest to compare the actions of the unitary loop on open or periodic boundaries as applied to the ground state of $H_0$. This ground state is given by $\ket{\psi_0} =\ket{\uparrow\downarrow\uparrow\downarrow \dots\uparrow\downarrow}$. Applying $U_\mathrm{site}(\pi)$ we get $\ket{\downarrow\uparrow\downarrow\uparrow \dots\downarrow\uparrow}$, independent of the boundary termination. Now, if we have a periodic system, we can apply $U_\mathrm{bond}(\pi)$ to return to $\ket{\psi_0}$. If we have open boundaries, we leave out $U_{L,1}$ and apply $U_\mathrm{bond}^{\mathrm{OBC}}(t)= \prod_{k=1}^{L/2-1} U_{2k,2k+1}(t) $. For $t=\pi$ this gives $\ket{\downarrow\downarrow\uparrow\downarrow \dots\uparrow\downarrow \uparrow\uparrow} = X_1 X_L \ket{\psi_0}$. We have thus pumped a unit charge operator to each boundary, and in one-dimensional systems this rigorously characterizes the pump \cite{Bachmann24}. This analysis of the pump applies for the isospectral loop at finite radius in \cref{fig:pumpXX}. However, since the phase diagram is gapped away from the origin, we can deform it to as close as we like to the critical point (indeed, the two directions correspond to $\cos(\Theta)$ and $\sin(\Theta)$ in the notation of \cref{sec:boson}). As an aside, note that in the vicinity of the critical point we have a loop corresponding to the low-energy dual $U(1)$, described on the lattice in Ref.~\cite{Pace25}. However, this lattice dual $U(1)$ does not commute with the on-site $U(1)$ symmetry and so will leave this phase diagram.

Now, from Result 4 in Ref.~\cite{Jones25b} (since $\mathcal{C}$ acts as a charge conjugation), we have that $\delta=\pm1$ are in different $O(2)=U(1)\rtimes \ZZ_2$ SPT classes. This is because a unit charge is not the sum of two charges in this group. If we consider perturbations that respect any subgroup $\ZZ_N\rtimes \ZZ_2$, we conclude that $\delta=\pm1$ are in different classes for $N$ even, but not for $N$ odd. This is because for $N$ odd, we can write $1=2\times (N+1)/2 \mod{N}$. Nevertheless, the loop is a non-trivial charge pump for odd $N$, and this can constrain the nearby phase diagram.

\section{Effective Hamiltonians from degenerate perturbation theory}\label{app:pert}

\subsection{Large $D$ regime in the spin-1 QSPT model}
In the large $D \gg 1$ regime of the spin-1 chain in Eq.~\eqref{eq:spin1XX}, it is useful to decompose the Hamiltonian $H = H_0 + V$ with
\begin{align}
H_0 = - D \sum_n  \left( S^z_n \right)^2 \qquad \textrm{and} \qquad V = - \sum_n \left( 1 + (-1)^n \delta \right)\left( s^+_n s^-_{n+1} + {\rm h.c.} \right) .
\end{align}
For convenience we expressed $V$ in terms of the $s^\pm = \frac{S^\pm}{\sqrt{2}}$ operators. The space of ground states  of $H_0$ is $\sim 2^{N}$ degenerate, with effective spin-$\tfrac12$ Pauli operators $Z_n = S^z_n$ and $\sigma^+_n = (s^+_n)^2$. If $P = \prod_n Z_n^2$ is the projector onto this degenerate ground state subspace, we observe that $PVP=0$. Thus, at leading order in perturbation theory,
\begin{align}
    H_\textrm{eff} = E_0 + PVGVP,
\end{align}
where $G = \frac{1}{E_0 - H_0} (1-P)$ and $E_0 = -ND$ is the ground state energy of $H_0$.

This can be straightforwardly computed as follows:
\begin{align}
PVGVP = -\frac{1}{2D} PV^2P = -\frac{1}{2D} \sum_n \left( 1 + (-1)^n \delta \right)^2 \; P \left( s^+_n s^-_{n+1} + {\rm h.c.} \right)^2 P
\end{align}
where
\begin{align}
P \left( s^+_n s^-_{n+1} + {\rm h.c.} \right)^2 P &= 
P \left( (s^+_n)^2 (s^-_{n+1})^2 + (s^+_n s^-_n)(s^-_{n+1} s^+_{n+1}) + [n \leftrightarrow n+1] \right) P \\
&= \sigma^+_n \sigma^-_{n+1} + \sigma^-_{n} \sigma^+_{n+1} + \frac{1+Z_n}{2} \; \frac{1-Z_{n+1}}{2} + \frac{1-Z_n}{2} \; \frac{1+Z_{n+1}}{2} \\
&= \sigma^+_n \sigma^-_{n+1} + \sigma^-_{n} \sigma^+_{n+1} +\frac{1-Z_n Z_{n+1}}{2}.
\end{align}

Putting everything together, we thus obtain the following effective spin-$\tfrac12$ Hamiltonian at second order in degenerate perturbation theory:
\begin{align}
H_{\rm eff} = - ND - \frac{1}{4D} \sum_n \left( 1 + (-1)^n \delta \right)^2 \; \left( X_n X_{n+1} + Y_n Y_{n+1} - Z_n Z_{n+1} + 1 \right) .
\end{align}
This agrees with Eq.~\eqref{eq:Heff} in the main text, up to an irrelevant constant. See also Ref.~\cite{Chen03}.
\subsection{Near the first-order transition in the RSPT gapped path}
Consider the model \cref{eq:HXXJ2} at $\delta=-1$. We have
\begin{align}
H(-1,\lambda) &= -4\sum_{n}\left(X_{2n-1}X_{2n}+Y_{2n-1}Y_{2n}\right) - \lambda \sum_n \left(\sigma^+_{3n+1} \sigma^+_{3n+2} \sigma^+_{3n+3} + h.c. \right) \ . 
\end{align}
This consists of decoupled six-site terms that we can diagonalize. The ground-state level crossing occurs at $\lambda =8\sqrt{2}$.

For $\delta=-(1+\epsilon)$ we have
\begin{align}
H(-(1+\epsilon),\lambda) \propto H(-1,\lambda(1+\epsilon/2)^{-2}) -\frac{\epsilon^2}{(1+\epsilon/2)^{2}}\sum_{n}\left(X_{2n}X_{2n+1}+Y_{2n}Y_{2n+1}\right)  \ .
\end{align}
Fixing $\lambda$ so we are at the first-order transition, we reduce to degenerate perturbation theory for 
\begin{align}
H_0 + \tilde{\epsilon}~ V = {H(-1,8\sqrt{2})} - \tilde{\epsilon}\sum_{n}\left(X_{2n}X_{2n+1}+Y_{2n}Y_{2n+1}\right)\ .    
\end{align}
$H_0$ is the decoupled model that can be diagonalized with an overall ground state degeneracy of $2^{n_{\mathrm{blocks}}}$; we will view this as an effective spin-$\tfrac12$ chain. For each block we have ground states labeled $\ket{\uparrow}$ and $\ket{\downarrow}$; these are eigenstates of $Z_1Z_2$ with eigenvalue $+1$ and $-1$ respectively. To first order, the perturbation has no effect on the low-energy space. At second order, we generate couplings within and between neighboring blocks as follows:
\begin{align}
H_{\mathrm{eff}} &= \textrm{const} - \tilde{\epsilon}^2\sum_j \left(h_X \sigma^x_j +h_Z\sigma^z_j+J \sigma^z_j\sigma_{j+1}^z \right) \\
h_X &\simeq 0.0589\qquad h_Z \simeq 0.148 \qquad J \simeq 0.00176 \ .
\end{align}
Since this effective Ising chain with a non-zero longitudinal field has a unique gapped ground state, we avoid the level crossing in the $H(\delta,\lambda)$ phase diagram by taking $\delta=-(1+\epsilon)$. 
\section{RSPT phenomena with AKLT states}
\label{app:AKLT}
Spin-$S$ AKLT-type models give a simple family of spin chains where the ground state has non-trivial symmetry fractionalization of an $SO(3)$ symmetry. These models are constructed with a virtual spin $s\geq S/2$ that realizes another representation of $SO(3)$. For integer virtual spin, $s$, such chains are in the trivial $SO(3)$-SPT class, while for half-integer $s$ the SPT class is non-trivial.

Nevertheless, for integer $s$ these models have RSPT edge modes. These will be removable without a phase transition, as expected on general grounds. It remains interesting to consider explicit paths. Examples of such paths are given in Ref.~\cite{Pollmann12}, demonstrating that the $S=2$, $s=1$ AKLT state is adiabatically connected to a product state (which explicitly breaks $SO(3)$ symmetry but preserves $\mathbb Z_2 \times \mathbb Z_2$) and to a product state of singlets (which preserves $SO(3)$ but explicitly breaks translation symmetry).

In this appendix we give a general path that interpolates, for fixed $S$, between different virtual spins $s_1$ and $s_2$. This path preserves all symmetries, \emph{including translation symmetry}, and we conjecture that when $s_1 = s_2 \mod 1$ then all states along the path have a finite correlation length. Note that when the transfer matrix has a unique largest eigenvalue, we can use the correlation length to bound the spectral gap from below \cite{Fannes90}.
If $s_1 \neq s_2 \mod 1$ then there must be a phase transition, examples of such paths appear in Refs.~\cite{Zang10,Jiang10,Zheng11,Gozel19}.

\subsection{AKLT states for a spin-$S$ chain}
Consider a spin-$S$ chain with $S$ integer. The AKLT-type state where we have singlets of virtual spin-$s$ (on each site $s\otimes s$ is projected into $S$) is given by the MPS tensor \cite{Fannes90,Tu09}
\begin{align}
A_{n}^{\alpha,\beta} = \langle s,\alpha-s,s,s-\beta;S,S-n \rangle \times (-1)^{\alpha} \times \sqrt{\frac{2s+1}{2S+1}}, \label{eq:AKLT}
\end{align}
where $\langle j_1,m_1,j_2,m_2;J,M\rangle$ are the Clebsch-Gordan coefficients. In the above expression $\alpha,\beta=0,1,\cdots,2s$ and $n=0,1,\cdots,2S$, and the MPS tensor is in canonical form. From standard fusion of SU(2) representations, the state is well-defined (i.e., nonzero) only if $s \geq S/2$. The spectrum of the transfer matrix (which has a simple form in terms of Wigner $6j$ symbols) and parent Hamiltonians are given in \cite{Fannes90}.

By construction, we have
that this AKLT state is spin-rotation symmetric. In particular, using the conventional definition of the spin matrices, we obtain: 
\begin{center}
\begin{tikzpicture}
\node at (0,0){
    \begin{tikzpicture}[inner sep=1mm]
    \node[R] (R) at (0, 0) {$A$};
    \node[U] (U) at (0, -1) {$R_\gamma(\theta)$};
    \draw[-] (R) -- (U) -- (0, -1.7);
    \draw[-] (R) -- (0.7,0);
    \draw[-] (R) -- (-0.7,0);
    \end{tikzpicture}
    };
\node at (1.5,0.1){$=$};
\node at (4,0.5){
    \begin{tikzpicture}[inner sep=1mm]
    \node[R] (R) at (0, 0) {$A$};
    \node[V] (Vr) at (1.25, 0) {$R_\gamma(\theta)$};
    \node[V] (Vl) at (-1.25, 0) {$R_\gamma(\theta)^\dagger$};
    \draw[-] (R) -- (0, -0.7);
    \draw[-] (R) -- (Vr)-- (2,0);
    \draw[-] (R) -- (Vl)-- (-2.2,0);
    \end{tikzpicture}
    };
\node at (8,0.1){for $\gamma=x,y,z$.};
\end{tikzpicture}
\end{center}

\subsection{Interpolating between AKLT states with distinct virtual spin}
Here we define a path of MPS for a spin-$S$ chain which interpolates between an AKLT state with virtual spin $s_1$ and an AKLT state with virtual spin $s_2$. To do this, our MPS will have a virtual spin $s_1 \oplus s_2$ (i.e., bond dimension $\chi = (2s_1+1)+(2s_2+1)$) defined as:
\begin{align}
A_n = \left( \begin{array}{cc} \cos(\varphi) \; B_n(s_1,s_1) & \sqrt{\sin(\varphi)\cos(\varphi)} \; B_n(s_1,s_2) \\
\sqrt{\sin(\varphi) \cos(\varphi)} \; B_n(s_2,s_1) & \sin(\varphi) \; B_n(s_2,s_2) \end{array} \right) \label{eq:AKLTpath}
\end{align}
where $n=0,1,\cdots,2S$ labels the physical index. Here, for each $n$, $B_n(j_1,j_2)$ is a $(2j_1+1)\times(2j_2+1)$-dimensional matrix defined as:
\begin{align}
B_n(j_1,j_2)^{\alpha,\beta} = \langle j_1,\alpha-j_1,j_2,j_2-\beta;S,S-n \rangle \times (-1)^{\alpha} \times \sqrt{\frac{\sqrt{(2j_1+1)(2j_2+1)}}{2S+1}} 
\end{align}
where $\alpha = 0,1,\cdots,2j_1$ and $\beta = 0,1,\cdots,2j_2$.
We have verified that the above MPS is spin-rotation symmetric. 

Note that for $\varphi=0$, we recover the spin-$S$ AKLT state with virtual spin $s_1$, whereas for $\varphi=\pi/2$ we recover the case with virtual spin $s_2$.
If $s_1$ and $s_2$ are in distinct projective classes (i.e., one is integer and the other half-integer), then the mixed terms $B_n(s_1,s_2)$ will vanish such that this path will just give a simple first-order transition for $\varphi = \pi/4$. In fact, even for the same projective class this problem occurs if $S$ is not contained in the tensor product of $s_1$ and $s_2$.  However, we conjecture that when all the $B_n(s_i,s_j)$ are non-vanishing, then this gives a symmetric finite-correlation-length path interpolating between them.

The $B_n(s_i,s_j)$ are non-vanishing when certain triangle inequalities and integral constraints are satisfied \cite[Sec.~34.2]{NIST:DLMF}. 
These reduce to the following conditions that we impose on the path:
\begin{itemize}
\item[(i)] $s_1,s_2 \geq S/2$, and $S$ is an integer, meaning that the limiting AKLT states are well-defined.
\item[(ii)] $s_1 \equiv s_2 \mod 1$ (i.e., both virtual spins belong to the same projective class), and $\lvert s_1-s_2\rvert\leq S$.

\end{itemize}
With these conditions, one can show that the above MPS is automatically in middle-canonical form. In particular, its transfer matrix has its largest eigenvalue equal to $1$ and the corresponding left- and right-eigenvector is
\begin{align}
\Lambda = \left( \begin{array}{cc}
\frac{\cos(\varphi)}{\sqrt{2s_1+1}} \mathbb I_{2s_1+1} & 0 \\
0 & \frac{\sin(\varphi)}{\sqrt{2s_2+1}} \mathbb I_{2s_2+1} \end{array} \right).
\end{align}
This follows from writing out the equation for the left- and right-eigenvector and using the following orthogonality conditions on the $B_n$ (these reduce to standard $3j$ symbol orthogonality conditions \cite[Sec.~34.3]{NIST:DLMF}) 
\begin{align}
\sum_n B_n(s_a,s_c)B_n^T(s_b,s_c) &= \delta_{s_a,s_b}\sqrt{\frac{2s_c+1}{2s_a+1}} \mathbb{I}_{2s_a+1}  \nonumber\\
\sum_n B_n^T(s_c,s_a)B_n(s_c,s_b) &= \delta_{s_a,s_b}\sqrt{\frac{2s_c+1}{2s_a+1}} \mathbb{I}_{2s_a+1}  \ .
\end{align}
As a sanity check on the eigenvector, $\textrm{Tr}(\Lambda^2) = 1$. This also gives the entanglement entropy along our path:
\begin{align}
S = - (2s_1+1) p \log p - (2s_2+1) q \log q \qquad \textrm{with } p = \frac{\cos^2(\varphi)}{2s_1+1} \textrm{ and } q= \frac{\sin^2(\varphi)}{2s_2+1}.
\end{align}
In particular, the entanglement multiplets intersect when $\tan^2(\varphi) = \frac{2s_2+1}{2s_1+1}$.

We now illustrate our conjecture that this gives a well-defined gapped path connecting different virtual representations for certain values of the physical spin-$S$.

\begin{figure}
    \centering
    \begin{tikzpicture}
\node at (0,0) {(a)};
 \node at (3.5,0) {\includegraphics[width=0.4\textwidth]{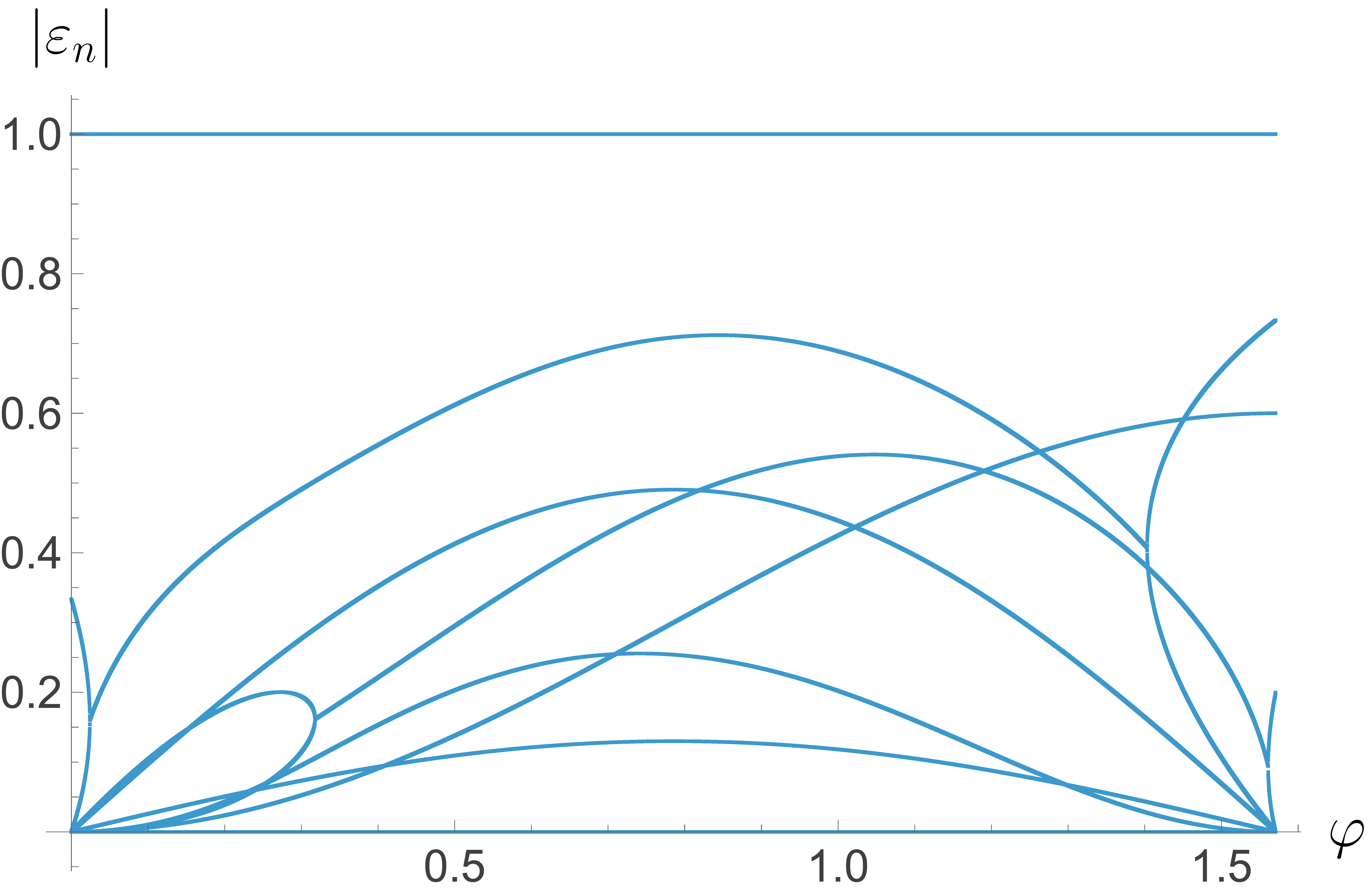}};
 \node at (7,0) {(b)};
\node at (10.5,0) {\includegraphics[width=0.4\textwidth]{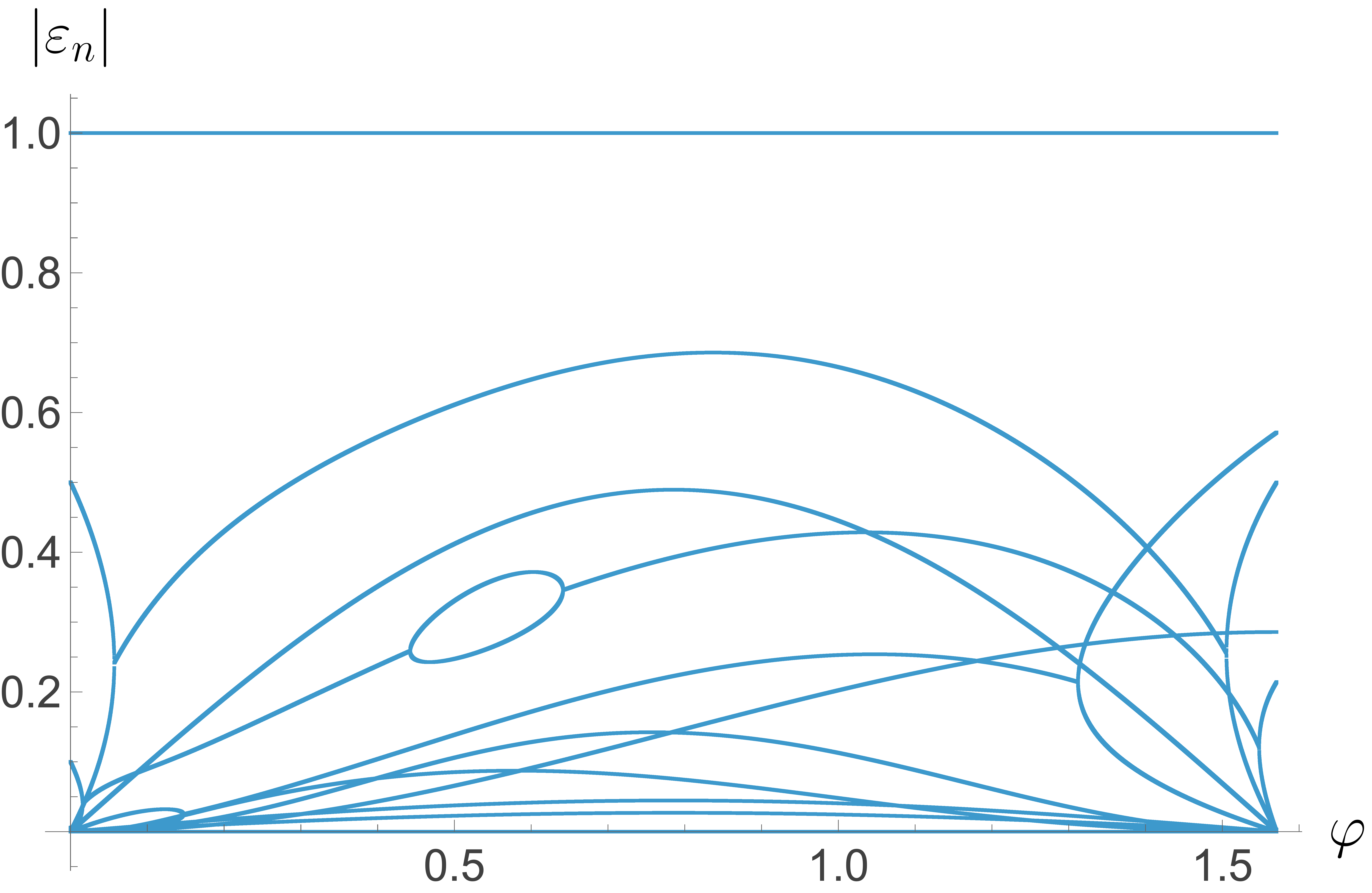} };
    \end{tikzpicture}
    \caption{\small  (a) Interpolating a spin-1 chain from the AKLT state with virtual spin-$\tfrac12$ ($\varphi=0$) to a case with virtual spin-$\tfrac32$ ($\varphi=\pi/2$).  (b) Interpolating a spin-2 chain from the AKLT state with virtual spin-$1$ ($\varphi=0$) to a case with virtual spin-$2$ ($\varphi=\pi/2$). We are plotting the spectrum, $\varepsilon_n$, of the transfer matrix (obtained using Mathematica \cite{mathematica}); the fact that there is always a unique largest eigenvalue implies that the correlation length remains finite throughout the interpolation. }
    
    \label{fig:interpolation}
\end{figure}

\subsubsection{Spin-1 chain: from virtual spin-\texorpdfstring{$\tfrac12$}{1/2} to virtual spin-\texorpdfstring{$\tfrac32$}{3/2}}

The paradigmatic example of the AKLT state is of course the case of the spin-1 chain with a virtual spin-$\tfrac12$. According to the SPT classification, the case with a virtual spin-$\tfrac32$ is in the same phase of matter. We are not aware of a study in the literature where it was explicitly shown that these two are indeed connected (although if we allow blocking of sites, one could use the construction in \cite{Schuch11}). Here, we show that one can connect them whilst preserving all symmetries, including both $SO(3)$ and translation symmetry. Indeed, we take the path defined by Eq.~\eqref{eq:AKLTpath} with physical spin $S=1$ and virtual spins $s_1=\tfrac12$ and $s_2=\tfrac32$. This is arguably the simplest setting where one can study the AKLT-interpolation, yet it already seems to straddle the line of solvability: using Mathematica we can diagonalize the transfer matrix, but the eigenvalues are specified as roots of certain polynomials with degrees as large as seven. We plot the corresponding solutions in \cref{fig:interpolation}(a). The key take-away message is that the second-largest eigenvalue is always strictly smaller than $1$, i.e., the correlation length $\xi$ remains finite throughout this interpolation, confirming that these two AKLT states belong to the same phase of matter. In this case, it seems that the largest correlation length along the path is achieved at the endpoint $\varphi = \pi/2$, i.e., the spin-1 AKLT state with virtual spin-$\tfrac32$.

\subsubsection{Spin-2 chain: from virtual spin-1 to virtual spin-2}

Unlike the spin-1 chain with spin-$\tfrac12$ edge modes, the spin-2 chain with `fractionalized' spin-1 edge modes is not a robust SPT phase. As before, while gapped paths have been considered, to the best of our knowledge we do not have a trivialization preserving both $SO(3)$ \emph{and} translation symmetry. However, this is a subtle question: what does one \emph{mean} by a trivial state with $SO(3)$ and translation symmetry? Surely, a product state is incompatible with these properties (indeed any translation-invariant spin-rotation symmetric MPS must have $\chi\geq 2$).

As a first step to show that the spin-$2$ chain with spin-$1$ edge mode is not a distinct phase even with $SO(3)$ and translation symmetry, we show that it can be adiabatically connected to the spin-$2$ AKLT state with virtual spin-$2$. Since here the virtual spin is the same as the physical spin, it might already be clear that this property is not special. In \cref{fig:interpolation}(b), we show the spectrum of the transfer matrix for Eq.~\eqref{eq:AKLTpath} with $S=2$, $s_1=1$ and $s_2=2$. This was obtained numerically. Again we see that the correlation length remains finite throughout, the maximal value being $\xi \approx 2.65$.

\subsubsection{Other interpolations for the spin-2 chain}

The above interpolation shows directly that the spin-2 AKLT states with virtual spin-1 and with virtual spin-2 are in the same phase of matter. Still one might object: this path did not remove the edge mode, in fact, it made it larger! Indeed, the final AKLT Hamiltonian will have a fivefold (spin-2) degeneracy per edge. Just like the spin-1 edge mode, this spin-2 edge mode is perturbatively stable. However, we can eventually remove it by also including a virtual spin-0 sector. Indeed, we can consider Eq.~\eqref{eq:AKLTpath} with a physical spin $S=2$, a virtual spin $s_1=2$ and a virtual spin $s_2=0$. In this case, we cannot go all the way to $\varphi=\pi/2$ since the AKLT state with virtual spin-$0$ is not well-defined ($B_n(0,0)=0$). Nevertheless, the MPS path is well-defined for all values $\varphi < \pi/2$ (albeit it will no longer automatically be in canonical form). 

Taking \begin{align}
\Lambda = \left( \begin{array}{cc}
\alpha \mathbb I_{5} & 0 \\
0 & \beta \end{array} \right).
\end{align} 
as an ansatz for the dominant right/left-eigenvector of the transfer matrix, we find a leading eigenvalue $\lambda = \tfrac{1}{2}\cos(\varphi)(\cos(\varphi)+\sqrt{1+3\sin^2(\varphi)})$. Then define $\mu=\alpha/\beta =\lambda/( \sin(\varphi)\cos(\varphi)\sqrt{5})$; and we find that the entanglement spectrum corresponds to a fivefold degenerate $\lambda_2^2 = \mu^2/(1+5\mu^2) $ and a singlet $\lambda_0^2 = 1/(1+5\mu^2) $. At $\tan(\varphi)=\sqrt{5}/4$ we cross over from a dominant fivefold degenerate Schmidt value to a dominant entanglement singlet. In the limit $\varphi\rightarrow\pi/2$ we find an equal total weight on the spin-0 and spin-2 sectors ($\lambda_0^2=1/2$ and the fivefold degenerate $\lambda_2^2=1/10$).

\section{Connecting $H_\notA$ to the not-$A$ Hamiltonian} \label{appendix:pathnotA}
In this appendix, we justify the claim in \Cref{sec:twoPotts} that $H_\notA$ is connected to the not-$A$ Hamiltonian in Ref.~\cite{OBrien20}. There it was observed that the product states $\left(\frac{  |B\rangle + |C\rangle}{\sqrt{2}} \right)^{\otimes N}$ (and their symmetry-related partners) are the ground states of a Hamiltonian $H_{\text{not-}A} = H_0 - H_A$ where $H_A$ is the Potts model defined in Eq.~\eqref{eq:A} and $H_0$ is the $U(1)$-symmetric model introduced in Ref.~\cite{OBrien20} as:
\begin{align}
H_{0}= 3 \sum_n \left(  \left(s^+_n\right)^2 \left(s^-_{n+1}\right)^2 - s^+_n s^-_{n+1} +h.c. \right) - \sum_n \left( \tau_n + \tau_n^\dagger \right).
\end{align}
In fact, we can write
\begin{align}
\begin{split}
H_{\text{not-}A} = &- \underbrace{\sum_n \left( 2s^+_n s^-_{n+1}  - 4\left(s^+_n\right)^2 \left(s^-_{n+1}\right)^2 + \tau_n +h.c. \right)}_{= \textrm{ $O(2)$-symmetric part}}\\
&+ \underbrace{\sum_n \left( s^+_n \left( s^+_{n+1} \right)^2  + \left( s^+_{n} \right)^2 s^+_{n+1} + h.c.\right)}_{= \textrm{ $S_3$-symmetric part}}
\end{split}
\label{eq:HnotAFendley}
\end{align}
which can be compared to our $H_{\bar A}$ model in Eq.~\eqref{eq:notAspin}: the $S_3$-symmetric term is the same (with the same sign), but the $O(2)$-symmetric part is different.

\begin{figure}
    \centering
    \includegraphics[scale=0.33]{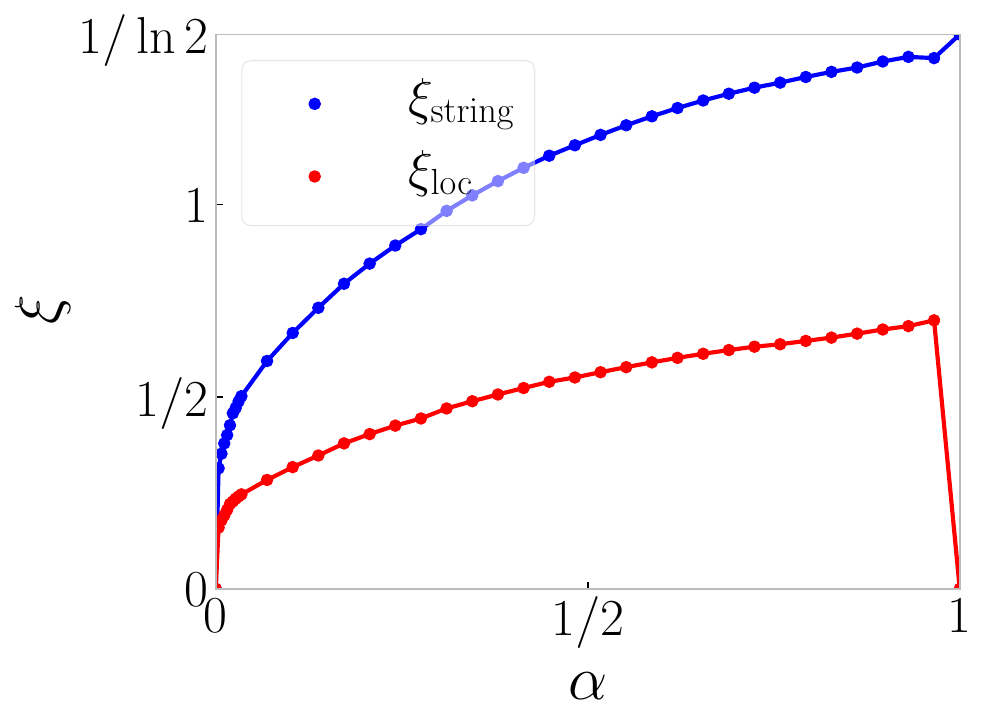}
    \includegraphics[scale=0.33]{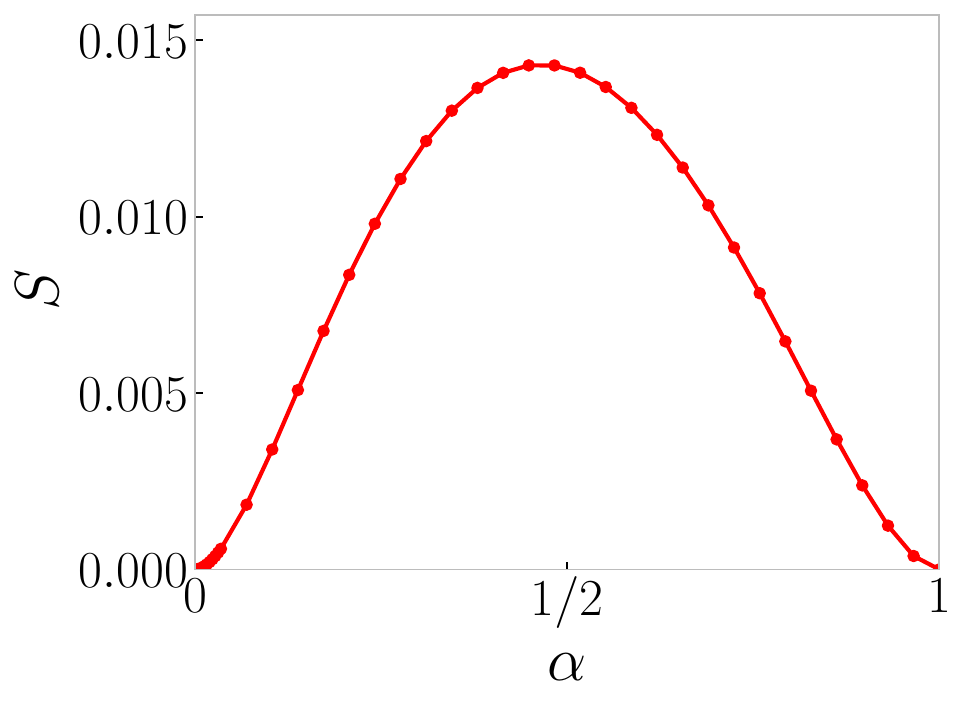}
    \includegraphics[scale=0.33]{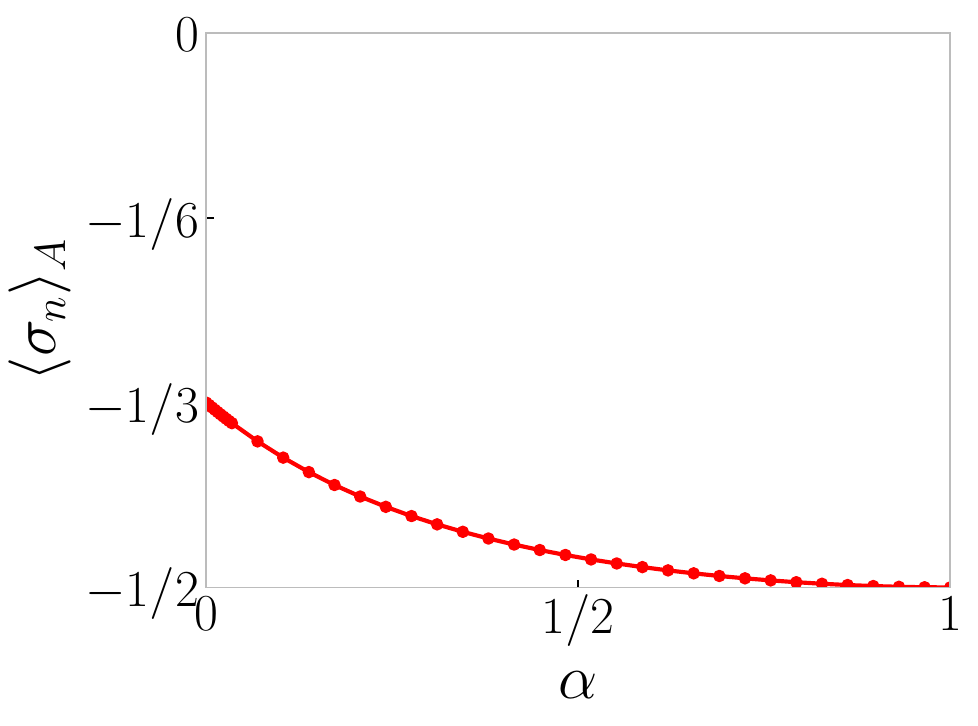}
    \caption{\small  Interpolating from the $H_{\bar A}$ Hamiltonian \eqref{eq:HnotA} ($\alpha=0$) to the not-$A$ Hamiltonian in Ref.~\cite{OBrien20} and reproduced in Eq.~\eqref{eq:HnotAFendley} ($\alpha=1$). We show the correlation length (both for local and nonlocal observables), entanglement entropy, and expectation value of $\sigma_n$ in the charge-conjugation-preserving ground state.}
    \label{fig:notApath}
\end{figure}

We now show that $H_{\text{not-}A}$ is adiabatically connected to $H_{\bar A}$ via direct interpolation:
\begin{align}
(1-\alpha)H_{\bar A} + \alpha H_{\text{not-}A}.
\end{align}
The correlation length $\xi$, entanglement entropy $S$, and expectation value $\langle \sigma_n \rangle_A$ (with respect to the unique $\mathcal C$-preserving state) are shown in \cref{fig:notApath}. The entanglement entropy remains very small throughout. The correlation length for local observables jumps when we reach the model of Ref.~\cite{OBrien20}. This can be understood as follows: at that point, the ground state is a product state, hence $\xi = 0$ for local observables. But this limit need not be continuous, despite the path being gapped. Indeed, if we have a correlation function $\sim c \;e^{-n/\xi}$, then if $c$ vanishes continuously, then the correlation length can appear to jump. This is due to the asymptotic nature of how a correlation length is defined. As an aside, we also plot the correlation length $\xi_\textrm{string}$ which determines the exponential decay of the disorder operator for the broken $\mathbb Z_3$ symmetry $\mathcal S$. This turns out to be continuous and equals $1/\log 2$ for the product state at $\alpha=1$ (this also becomes the local correlation length of the RSPT MPS after a Kramers-Wannier transformation, see \cref{sec:RSPTpath}); this nonzero value is possible since the different product states are not orthogonal for finite regions. In contrast, the states at $\alpha=0$ are orthogonal even on a single site, since they are rotations of the original states of the Potts model (see Eq.~\eqref{eq:notA}). Hence, we would argue that the $\alpha=0$ point (i.e., the Hamiltonian $H_{\bar A}$) defines a natural fixed-point limit of `not-$A$'.
\section{Charge pumps and Onsager-integrable clock models}\label{app:onsager}
\subsection{The Onsager ferromagnet generates a pump}
The Onsager-integrable chiral clock Hamiltonians are given by 
\begin{align}
A_0 & =-\frac{1}{N} \sum_{j} \sum_{m=1}^{N-1} \alpha_m \tau_j^{m} \qquad
&A_1 &=-\frac{1}{N}\sum_{j} \sum_{m=1}^{N-1} \alpha_m \sigma_j^{-m}\sigma_{j+1}^{m} \qquad &\alpha_m = \frac{1}{1-{\omega}^{m}}; \label{eq:A0A1explicit}
\end{align}
where $\omega =e^{ 2\pi i/N}$, and $\sigma$, $\tau$ are the standard $\mathbb{Z}_N$ clock operators generalizing \cref{eq:sigmatau}.
For our purposes we are particularly interested in $N=3$, but we can leave $N$ general for now. (Note that $A_0$ is the $\mathcal{T}$ conjugate of the Onsager paramagnet in \cite{Jones25}; since $A_1$ is invariant under $\mathcal{T}$, these Hamiltonians obey the usual Onsager algebra.)

For the analysis in \Cref{sec:rsptpump}, we first conjugate these Hamiltonians by the dipolar transformation $D=\prod_j \tau_j^{j}$. We find that 
\begin{align}
A_0 & =-\frac{1}{N} \sum_{j} \sum_{m=1}^{N-1} \alpha_m \tau_j^m \qquad
&\hat{A}_1 &=-\frac{1}{N}\sum_{j} \sum_{m=1}^{N-1} \omega^{-m}\alpha_m \sigma_j^{-m}\sigma_{j+1}^{m};
\end{align}
satisfy the Onsager algebra. 

Note that $\hat{A}_1$ has a $\mathbb{Z}_2^T$ symmetry (complex conjugation in the $\sigma$ basis, which conjugates $\sigma$ and leaves $\tau$ invariant). This symmetry takes $A_0 \rightarrow \overline{A}_0$, given by $-\frac{1}{N} \sum_{j} \sum_{m=1}^{N-1} \alpha_{-m} \tau_j^m$. 
Then\footnote{This, and the next lines, are analogous reasoning to Section VIII of \cite{Jones25b}.}, by symmetry we have that $\overline{A}_0$ and $\hat{A}_1$ also satisfy the Onsager algebra. 

Since $\alpha_m + \alpha_{-m}=1$, the Hamiltonian $H_{\mathrm{Potts}}=-\frac{1}{N} \sum_{j} \sum_{m=1}^{N-1} \tau_j^m=A_0+\overline{A}_0 $. By linearity, $H_{\mathrm{Potts}} $ and $\hat{A}_1$ satisfy the Dolan-Grady relation
\begin{align}
  [H_{\mathrm{Potts}},\hat{A}_1]=\Big[\big[[H_{\mathrm{Potts}},\hat{A}_1],\hat{A}_1 \big],\hat{A}_1\Big] \ .\label{eq:DG}
\end{align}
Hence, pivoting $H_{\mathrm{Potts}}$ by $\hat{A}_1$ gives a strict circular loop in Hamiltonian space. (Using the results of \cite{Jones25b} the two Hamiltonians do not satisfy an Onsager algebra for $N>2$.)

Now, to identify this pivot loop as a charge pump, let $U_{A_1}(t) = e^{-i t A_1}$.
We can write $U_{A_1}(t)=\prod_{j} A_{j,j+1}(t)$ where the individual gate is given by \begin{align}A_{j,j+1}(t)= \exp\left(\frac{i t}{N} \sum_m {\alpha_m\sigma_j^{-m}\sigma_{j+1}^m}\right).\end{align}
Using standard identities \cite{Jones25b}, we find $A_{j,j+1}(2\pi)=\omega^{\frac{N-1}{2}} \sigma_j^{-1} \sigma_{j+1}^{}$. For a periodic system we thus find $U_{A_1}(2\pi)\propto \mathbb{I}$, while it acts as a pump applied to a finite region with boundaries. Now consider $U_{\hat{A}_1}(t) = D U_{A_1}(t) D^\dagger$. Decomposing into gates, we then have $\hat{A}_{j,j+1}(2\pi)=\omega^{\frac{N-3}{2}} \sigma_j^{-1} \sigma_{j+1}^{}$, giving analogous charge pump behavior for loops generated by $\hat{A}_1$.

\subsection{Connection to the $N=3$ RSPT}
Restricting to $N=3$ we have that 
\begin{align}
H_{\mathrm{Potts}}& = \frac{1}{3}\tilde{H}_A=-\frac{1}{3} \sum_{j}(\tau_j+\tau_j^\dagger) \qquad
&\hat{A}_1 &=\frac{i}{3\sqrt{3}}\sum_{j}(\sigma_j^\dagger \sigma_{j+1}^{}-\sigma_j^{}\sigma_{j+1}^{\dagger}) \ .
\end{align}
From \cref{eq:RSPTpivot2} 
\begin{align}
\label{eq:RSPTpivot3}
    \tilde{\mathcal R}^z(t) &=  \prod_n -e^{\frac{ i t}{3} \left(1+\sigma_n^{\dagger}\sigma_{n+1}+\sigma_n^{}\sigma_{n+1}^\dagger \right) } =(-1)^L e^{it L/3}e^{-i H_A t/3}\\\tilde{\mathcal S}^z(t) &=  \prod_n e^{\frac{t}{3\sqrt{3}} \left(\sigma_n^{\dagger}\sigma^{}_{n+1}-\sigma_n^{}\sigma_{n+1}^\dagger \right) }= e^{-i \hat{A}_1 t} \ .
\end{align}
In particular, for $N=3$, $\tilde{\mathcal S}^z(t) = U_{\hat{A}_1}(t)$, and so we conclude that $ \tilde{\mathcal S}^z(t)$ has the charge pump behavior given in \cref{eq:Szpump}. We also notice that $\tilde{R}^z(2\pi)$ is identity (up to a phase) for any boundary condition, due to \cref{eq:Rzidentity}.
Finally, since 
\begin{align}
A_1 = \frac{1}{2} \left( \frac{1}{3}H_A- \hat{A}_1\right) \ ,
\end{align}
 on periodic boundaries we have that $U_{A_1}(2\pi) = \mathbb{I} =\tilde{\mathcal R}^z(\pi) \tilde{\mathcal S}^z(-\pi) $. Hence $ \tilde{\mathcal S}^z(\pi)=  \tilde{\mathcal R}^z(\pi)$, and so $\tilde{\mathcal S}^z(\pi)$ entangles the RSPT.

Now, using \Cref{eq:DG} we have that $\hat{A}_1$ and $\tilde{H}_A$ satisfy the Dolan-Grady relation, and so pivoting  $\tilde{H}_A$ with $ \tilde{\mathcal S}^z(t)$ generates a strict circular loop. In particular:
\begin{align}
\tilde{\mathcal S}^z(t) \tilde{H}_A \tilde{\mathcal S}^z(t)^\dagger  = \frac{1}{2}(\tilde{H}_A +\tilde H_{\bar A})  +\frac{1}{2}(\tilde{H}_A -\tilde H_{\bar A})\cos(t) + i [ \tilde{H}_A,\hat{A}_1 ]  \sin(t) \ .\label{eq:strictappendix}
\end{align}
Following \cite{Jones25b} we can conclude there must be a phase transition along the line $(1-\lambda)\tilde{H}_A+\lambda\tilde H_{\bar A}$, as observed in the numerics.

\subsection{A CPT-symmetric phase diagram}\label{app:CPT}
The pivot Hamiltonian $\hat{A}_1=\sum_j\frac{i}{3\sqrt{3}} \left(\sigma_j^{\dagger}\sigma^{\vphantom\dagger}_{j+1}-\sigma_j^{\vphantom\dagger}\sigma_{j+1}^\dagger \right)$ has $\ZZ_3$-symmetry breaking ground states ($D=\prod_j \tau_j^{j}$ applied to the simple ferromagnetic states). Note that $\hat{A}_1$ and $\tilde{H}_A$ both have a $\ZZ_2^{\mathrm{CPT}}$ symmetry, combining $\mathcal{C}$, spatial inversion and complex conjugation in the $\sigma$ basis. This implies that $\tilde H_{\bar A}$ has the same symmetry and that we can construct a $\ZZ_3\rtimes\ZZ_2^{\mathrm{CPT}} $ phase diagram  $H = \alpha \tilde H_A + \beta \hat{A}_1 + \gamma \tilde{H}_\notA$ (with $\alpha,\beta,\gamma \geq 0$). In this phase diagram, the pump acts as a strict circular loop everywhere. Indeed, for each point in the phase diagram, there is a strict circular loop that pumps charge and intersects the phase diagram at this point and its mirror image under  $ \tilde H_A\leftrightarrow\tilde{H}_\notA $. Hence, we either have that neither point is short-range entangled (as in the SSB phase), or we have a critical point separating the two (as in the trivial phase). 
Due to the simplicity of \cref{eq:strictappendix}, this is a very strongly constrained phase diagram, and hence we can argue the presence of a phase transition between $ \tilde H_A$ and $\tilde{H}_\notA $ despite the two being in the same phase.

We also have that 
$\tilde{H}_\notA$ has the RSPT-associated leading entanglement spectrum degeneracy for $S_3^{\mathrm{CPT}}=\mathbb{Z}_3 \rtimes \ZZ_2^{\mathrm{CPT}}$.  The MPS is given in \cref{eq:MPSCPTparameter} and 
at the $\mathrm{CPT}$ symmetric point $t=\pi\sim3\pi\sim5\pi$ we have a non-trivial fractionalization. The exact form will depend on $t$, but will always be non-trivial. For example 
\begin{align}
\mathcal{B}_a^{n,n'}(3\pi) = \underbrace{\frac{1}{\sqrt{3}}\omega^{a(n-n')}(-1)^{\delta_{n',0}}}_{\Gamma_a^{n,n'}} ~\underbrace{\frac{1}{3}(2-\delta_{n',0})}_{\Lambda_{n'}}\label{eq:MPSCPT}
\end{align} and 
\begin{align}
\sum_{a'}\mathcal{C}_{a,a'}{\Gamma}_{a'}^\dagger = \mathfrak{C}\Gamma_{a}\mathfrak{C} \qquad  \mathfrak{C}= \left( \begin{array}{ccc} -1 & 0 & 0 \\
0 &0& 1 \\
0& 1 & 0\end{array} \right)\ .
\end{align}
This gives one one-dimensional and one two-dimensional irrep of $S_3^{\mathrm{CPT}}$, corresponding to the two entanglement blocks. Given the entanglement spectrum found above, we see that the 2d irrep is dominant, and this feature is parametrically stable.

\section{Charge pumps and the entanglement spectrum}\label{app:pumpspectrum}
In this appendix we continue the theme of \cref{sec:rsptpump} and explore the connection between the presence of a non-trivial pump loop, and degeneracies in the entanglement spectrum at high-symmetry points. 

In the case where the pump implies a non-trivial SPT at one of the high-symmetry points \cite{Jones25b}, then we will necessarily see entanglement spectrum degeneracy there implied by the non-trivial SPT projective symmetry fractionalization. Of interest to us is when the pump does not imply a non-trivial SPT at a high-symmetry point. We will focus on the case where we have a family of MPS, and then we generically have the charge pump appearing in the algebra of the fractionalized symmetry and the monodromy gauge transformation coming from traversing the loop \cite{Inamura24,Jones25b}. We show that generically for $S_3$-symmetric points on a $\mathbb{Z}_3$ pump loop, there are the same number of doublets and singlets. (See Ref.~\cite{Jones25b} for how this can fail at `removable singularities' around the MPS loop. In this case, we should then interpret a doublet or a singlet that participates in the pump's spectral flow as appearing at zero entanglement eigenvalue. This is fine-tuned, but we note that it does occur when we pivot starting with a product state.)

Let us fix the $\mathbb{Z}_N\rtimes \mathbb{Z}_2^C$ case for definiteness, and suppose we have a $2\pi$-periodic family of $\mathbb{Z}_N$-symmetric states, with MPS representation $\mathcal{A}(t)$. Let the $\mathbb{Z}_N$ be generated by $\prod_j \tau_j$. 
We can pick a gauge where \begin{align}
\sum_{a'}\tau_{a,a'} \mathcal{A}_{a'}(t)=V(\tau) \mathcal{A}_{a}(t)V(\tau)^\dagger \ ,\end{align}
while cycling the loop acts as 
\begin{align}
    \mathcal{A}_{a}(t+2\pi)=W \mathcal{A}_{a}(t)W^\dagger \ ,\end{align}
    (up to unimportant overall phases).
    A charge-$k$ $\mathbb{Z}_N$ pump manifests as $V(\tau)W=\omega^k WV(\tau)$. Moreover, $V(\tau)^N=W^N = \mathbb{I}$.

    For $N=3$, a non-trivial charge pump means that 
    $V(\tau)W=\omega^{\pm1} WV(\tau)$. This means that for any fixed $t$, the bond Hilbert space supports a non-trivial projective representation of $\mathbb{Z}_3\times\mathbb{Z}_3$. This has a unique irrep (up to fixing a basis) \cite{Else12} generated by $\sigma$ and $\tau$, and thus the bond space decomposes as a direct sum of $m$ copies of this three-dimensional irrep. One can see this directly: $V(\tau)$ has eigenvalues $\{1,\omega,\omega^{-1}\}$, and $W$ cycles them, so each eigenspace has the same degeneracy. 

    At high-symmetry points where we have an additional charge conjugation symmetry, this will fractionalize as 
    \begin{align}
\sum_{a'}\mathcal{C}_{a,a'} \mathcal{A}_{a'}(t)=V(\mathcal{C}) \mathcal{A}_{a}(t)V(\mathcal{C})^\dagger \ .\end{align}
We can fix phases so that $V(\mathcal{C})^2=\mathbb{I}$, and define $\tilde{V}(\tau) =\omega^\kappa V(\tau)$ for some $\kappa\in\{0,\pm 1\}$ (preserving $\tilde{V}(\tau)^3=\mathbb{I}$). Then
$\tilde{V}(\tau)$ and $V(\mathcal{C})$ form a linear representation of $S_3$, with $V(\mathcal{C})\tilde{V}(\tau)V(\mathcal{C})=\tilde{V}(\tau)^\dagger $. $S_3$ has one-dimensional and two-dimensional irreps, but we are constrained by the non-trivial $\mathbb{Z}_3$ action. Let us fix a basis by the eigenvalue of $\tilde{V}(\tau)$ and an index $j$ that tracks which of the $m$ irreps of $\mathbb{Z}_3\times \mathbb{Z}_3$ we are in. Then for each $j$ we have one $S_3$ doublet (coupling $\ket{\omega, j}$ and $\ket{\omega^{-1},j}$) and one $S_3$ singlet ($\ket{1,j}$). We thus find a non-trivial interplay between the presence of a charge pump and high-symmetry points, but this does not imply a \emph{dominant} entanglement doublet at any high-symmetry point. As in our discussion of boundary physics, it would be interesting to analyze this for a pump loop in the vicinity of a critical point.
\end{appendix}
\end{document}